\documentclass[acmsmall,nonacm]{acmart}

\usepackage{array}
\usepackage{hyperref}
\usepackage{booktabs}
\usepackage{multirow}
\usepackage{graphicx}
\usepackage{makecell}
\usepackage{colortbl}
\usepackage{longtable}
\usepackage{enumerate}
\usepackage{enumitem}
\usepackage{pifont}
\usepackage{tabularx}
\newcolumntype{L}{>{\raggedright\arraybackslash}X}

\newcolumntype{C}[1]{>{\centering\arraybackslash}m{#1}}

\newcommand{\mycomment}[1]{}

\AtBeginDocument{%
  \providecommand\BibTeX{{%
    \normalfont B\kern-0.5em{\scshape i\kern-0.25em b}\kern-0.8em\TeX}}}

\setcopyright{acmcopyright}
\copyrightyear{2023}
\acmYear{2023}
\acmDOI{}

\acmJournal{TOSEM}
\acmVolume{X}
\acmNumber{X}
\acmArticle{}
\acmMonth{8}

\begin{document}

\title{AIOps Solutions for Incident Management: Technical Guidelines and A Comprehensive Literature Review}

\author{Youcef Remil}
\email{yre@infologic.fr}
\affiliation{%
  \institution{University of Lyon, INSA Lyon}
  \streetaddress{CNRS UMR 5205, FR-69621}
  \city{Lyon}
  \country{France}
}
\affiliation{%
  \institution{Infologic}
  \streetaddress{99 avenue de Lyon, FR-26500}
  \city{Bourg-L{\`{e}}s-Valence}
  \country{France}
}

\author{Anes Bendimerad}
\email{abe@infologic.fr}
\affiliation{%
  \institution{Infologic}
  \streetaddress{99 avenue de Lyon, FR-26500}
  \city{Bourg-L{\`{e}}s-Valence}
  \country{France}
}

\author{Romain Mathonat}
\email{rma@infologic.fr}
\affiliation{%
  \institution{Infologic}
  \streetaddress{99 avenue de Lyon, FR-26500}
  \city{Bourg-L{\`{e}}s-Valence}
  \country{France}
}

\author{Mehdi Kaytoue}
\email{mka@infologic.fr}
\affiliation{%
  \institution{University of Lyon, INSA Lyon}
  \streetaddress{CNRS UMR 5205, FR-69621}
  \city{Lyon}
  \country{France}
}
\affiliation{%
  \institution{Infologic}
  \streetaddress{99 avenue de Lyon, FR-26500}
  \city{Bourg-L{\`{e}}s-Valence}
  \country{France}
}

\renewcommand{\shortauthors}{Remil et al.}

\ccsdesc[500]{General and reference~Surveys and overviews}
\ccsdesc[500]{Computing methodologies~Artificial intelligence}
\ccsdesc[500]{Computer systems organization~Maintainability and Maintenance}

\keywords{}


\begin{abstract}

The management of modern IT systems poses unique challenges, necessitating scalability, reliability, and efficiency in handling extensive data streams. Traditional methods, reliant on manual tasks and rule-based approaches, prove inefficient for the substantial data volumes and alerts generated by IT systems. Artificial Intelligence for Operating Systems (AIOps) has emerged as a solution, leveraging advanced analytics like machine learning and big data to enhance incident management. AIOps detects and predicts incidents, identifies root causes, and automates healing actions, improving quality and reducing operational costs. However, despite its potential, the AIOps domain is still in its early stages, decentralized across multiple sectors, and lacking standardized conventions. Research and industrial contributions are distributed without consistent frameworks for data management, target problems, implementation details, requirements, and capabilities. This study proposes an AIOps terminology and taxonomy, establishing a structured incident management procedure and providing guidelines for constructing an AIOps framework. The research also categorizes contributions based on criteria such as incident management tasks, application areas, data sources, and technical approaches.  The goal is to provide a comprehensive review of technical and research aspects in AIOps for incident management, aiming to structure knowledge, identify gaps, and establish a foundation for future developments in the field. 

\end{abstract}

\maketitle

\section{Introduction}

\subsection{Context and Motivation}

IT environments of today are constantly becoming larger and more complex as new technologies emerge and new work methods are adopted. They encounter challenges in ensuring efficiency and reliability. Many organizations are transitioning from product delivery to service releases, shifting from traditional static infrastructures to dynamic blends of on-premises, managed, private, and public cloud environments. Due to factors like device mobility, evolving runtime environments, frequent updates, upgrades, and online repairs, these systems are increasingly vulnerable to failures~\cite{chen2020towards, dang2019aiops, notaro2021survey, salfner2010survey}. According to~\citet{lin2018predicting}, Microsoft's Azure cloud system experiences failures in approximately 0.1\% of its server nodes daily. Such failures can lead to reduced system availability, financial losses, and negative user experiences~\cite{chen2019outage}. Surveys conducted by the International Data Corporation (IDC) reveal that application downtime can cost businesses up to \$550,000 per hour~\cite{elliot2014devops, cappuccio2013ensure, huff2015breaking}. These significant losses trigger the need for autonomic and self-managing systems to address the root causes of failures and enhance the quality and responsiveness of IT services~\cite{bogatinovski2021artificial, farshchi2018metric, ren2019time}.

Traditional IT management solutions, relying on expert systems and rule-based engines, frequently encounter shortcomings in adaptiveness, efficiency, and scalability~\cite{prasad2018market, levin2019aiops}. These solutions often overlook the real-time state of a system, leading to inaccurate predictive analysis based on its current condition. Additionally, they are rooted in a conventional engineering mindset that emphasizes manual execution of repetitive tasks and individual case analysis, often relying on bug reproduction steps or detailed logs~\cite{dang2019aiops}.

These factors have sparked interest in replacing multiple conventional maintenance tools with an intelligent platform capable of learning from large volumes of data to proactively respond to incidents. Accordingly, organizations are turning to AIOps to prevent and mitigate high-impact incidents. The term AIOps was initially introduced in 2017 by Gartner to address the AI challenges in DevOps~\cite{prasad2018market}. Initially, AIOps stemmed from the concept of IT Operations Analytics (ITOA). However, with the growing popularity of AI across various domains, Gartner later redefined AIOps based on public opinion, characterizing it as Artificial Intelligence for Operating systems~\cite{shen2020evolving}. AIOps involves the application of big data and machine learning techniques to intelligently enhance, strengthen, and automate various IT operations~\cite{dang2019aiops, notaro2021survey, prasad2018market}. AIOps learns from a diverse range of data collected from services, infrastructures, and processes. Subsequently, it autonomously takes initiatives to detect, diagnose, and remediate incidents in real-time, leveraging this acquired knowledge~\cite{prasad2018market, chen2020towards, dang2019aiops}.

\begin{figure}
	\centering
	\includegraphics[width=0.9\linewidth]{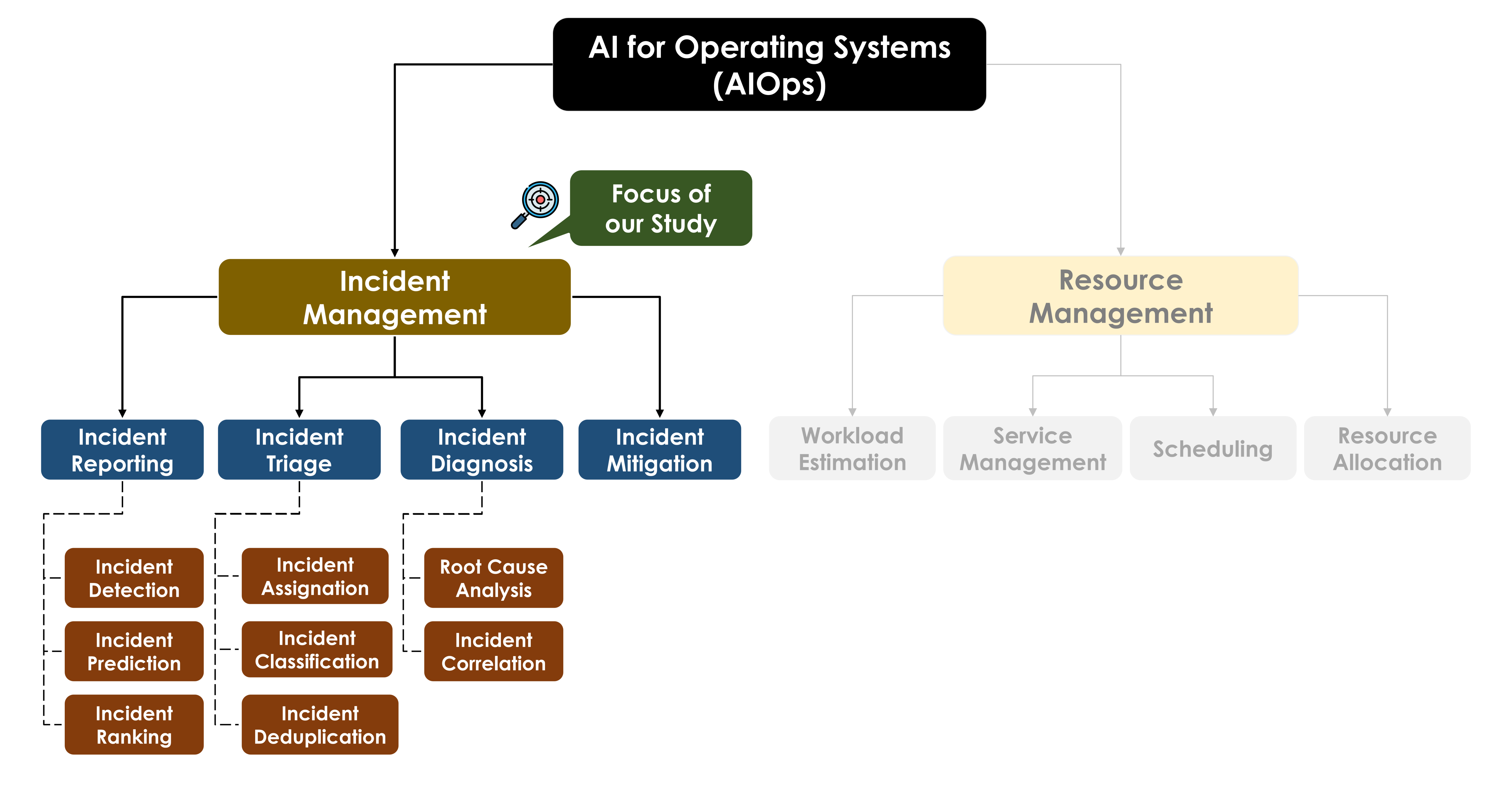}
    \captionsetup{font=footnotesize}
	\caption{Exploring the research landscape of AIOps subareas with a focus on Incident Management.}%
	\label{fig:taxo_aiops}
\end{figure}

To date, a universally accepted formal definition of AIOps is yet to emerge due to its novelty and its expansive scope bridging research and industry. While some definitions focus solely on the capabilities and benefits of AIOps, without delving into its conceptual framework and operational processes, several research efforts have taken the initiative to propose a comprehensive definition (refer to Table~\ref{tab:definitions_aiops}). 
These definitions commonly converge on two key points~\cite{notaro2021survey}. Firstly, AIOps entails the application of artificial intelligence to enhance, fortify, and automate a wide range of IT operating systems. Secondly, AIOps emphasizes the provision of complete visibility, control, and actionable insights into the past, present, and potentially future states of the system. Other definitions also emphasize the importance of various aspects such as robust data collection, data ingestion and effective querying capabilities, scalable infrastructure, and real-time operations. It is highly important to note that AIOps extends beyond the management of incidents and the automation of maintenance processes. We concur with the findings of~\citep{notaro2021survey} that AIOps includes two primary subareas: incident management and resource management, as depicted in Figure~\ref{fig:taxo_aiops}. While the resource management procedure covers techniques for optimal allocation and utilization of resources for IT operations, our study specifically focuses on exploring the capabilities of AIOps to assist the incident management procedure. Our objective is to redesign and categorize the entire maintenance workflow routines related to incident management. This is achieved by considering the diverse array of contributions made in AIOps and aligning them with the industrial needs through clearly defined phases. 

\begin{table}[t]
    \captionsetup{font=footnotesize}
    \caption{Available AIOps definitions with corresponding capabilities.}
    \label{tab:definitions_aiops}
    \centering
    \resizebox{\textwidth}{!}{%
    \begin{tabular}{>{\raggedright\arraybackslash}p{1cm} >{\raggedright\arraybackslash}p{11cm} >{\raggedright\arraybackslash}p{5cm}}    \toprule
    \textbf{Work} & \textbf{Provided Definition} & \textbf{Capabilities} \\ \midrule 
    \cite{prasad2018market} & "AIOps platforms combine big data and machine learning functionality to support all primary IT operations functions through the scalable ingestion and analysis of the ever-increasing volume, variety, and velocity of data generated by IT. The platform enables the concurrent use of multiple data sources, data collection methods, and analytical and presentation technologies" & 
    Performance Analysis, Anomaly Detection, Event Correlation, IT Service Management, and Automation \\ \midrule
    \cite{dang2019aiops} & "AIOps is about empowering software and service engineers (e.g., developers, program managers, support engineers, site reliability engineers) to efficiently and effectively build and operate online services and applications at scale with artificial intelligence (AI) and machine learning (ML) techniques" & 
    High Service Intelligence, High Customer Satisfaction, High Engineering productivity \\ \midrule
     \cite{rijal2022aiops} & "AIOps is a methodology that is on the frontier of enterprise IT operations. AIOps automates various aspects of IT and utilizes the power of artificial intelligence to create self-learning programs that help revolutionize IT services" & 
    Improving human-AI collaboration, Monitoring and Proactive IT work, Efficient time saving, Faster Mean Time To Repair \\ \midrule
    \cite{bogatinovski2021artificial} & "AIOps is an emerging interdisciplinary field arising in the intersection between the research areas of machine learning, big data, streaming analytics, and the management of IT operations" & 
    Efficient Resource Management and Scheduling, Complex failure management \\ \bottomrule
    \end{tabular}%
    }
\end{table}

\subsection{Focus of this Review: AIOps for Incident Management}

Building upon the work of~\cite{shen2020evolving, prasad2018market}, we propose that a prototypical AIOps system comprises six fundamental abilities that give rise to various tasks within the incident management procedure.

\noindent \textbf{Perception.} This capability centers on the ability to gather heterogeneous data sources, including log and event data, key performance metrics, network traffic data, and more, from a multitude of sources, such as networks, infrastructure, and applications. It is essential that the ingestion process accommodates both real-time streaming and historical data analysis. Additionally, powerful data visualization, querying, and indexing mechanisms are also necessary elements.
    
\noindent \textbf{Prevention.} This process entails actively identifying potential failures and forecasting high-severity outages in the system. Preventing incidents is crucial for maintaining a healthy and robust system. Therefore, implementing an automated system that continuously monitors system health and promptly alerts administrators about potential problems is essential.
    
\noindent \textbf{Detection.} If errors occur, it is imperative for the system to detect the associated anomalies or symptoms. This is achieved by analyzing vast amounts of perceptual and historical data to identify abnormal content in either the time domain or spatial domain, or both. This process includes discovering abnormal patterns in data and detecting flexible abnormal conditions that exceed static thresholds, while minimizing noise in data, such as false alarms or redundant events.

\noindent \textbf{Location.} The objective of this process is to identify and analyze potential root causes and faulty actions responsible for the underlying incidents by conducting a causality and correlation study. This study must be contextualized within a unified topology to ensure its accuracy. Without the context and constraint of topology, the detected patterns, while valid, may be unhelpful and distracting. By deriving patterns from data within a topology, the number of recurrent and redundant patterns can be reduced, exceptionalities in data can be highlighted, and hidden dependencies can be identified.

\noindent \textbf{Action.} This includes conducting reactive triage on problems and prioritizing incidents once detected or predicted, as well as implementing a series of corrective actions based on the current scenario and past solutions that have already been provided. However, it is important to note that automatic healing actions need to be executed safely.

\noindent \textbf{Interaction.} It is referred to as human-computer intelligent interaction. This involves bidirectional interactive analysis between the intelligent models and the expertise of users. For instance, the system can integrate human expertise to enhance its models or similarly leverage model insights to enrich and update the user background knowledge. Furthermore, this includes facilitating communication and collaboration between different maintenance teams and with customers, promoting efficient information sharing and effective issue escalation.

Drawing of these capabilities, several companies have started dispensing AIOps tools as commodities within the last few years, while a number of technology giants have adopted an AIOps algorithmic viewpoint to maintain their on-premises or cloud computing infrastructures and manage incidents~\citep{dang2019aiops,li2020predicting, levin2019aiops,li2019aiops,qu2017next,lin2018hardware,chen2020towards}, thereby inducing the academic field to evolve and deliver more ingenious and innovative solutions. In actuality, the notion of utilizing AI to refine IT and maintenance operations, despite its recent emergence as a research field, is not entirely novel~\citep{notaro2021survey,bogatinovski2021artificial}. Beginning in the mid-1990s, some research work explored software defects in source code by employing statistical models founded on source code metrics~\citep{khoshgoftaar1995neural,chidamber1994metrics,briand1999unified}. Since the start of the new decade, various techniques have been proposed to tackle online software~\citep{zhang2016automated,pitakrat2018hora,zhao2020real} and hardware~\citep{xu2016health,li2016being,zheng2017long} failure prediction and anomaly detection~\citep{chow2014mystery,wang2022identifying,nandi2016anomaly}. Multiple other domains of AIOps, such as event correlation~\citep{wu2010detecting,luo2014correlating,lin2016idice}, bug triage~\citep{xuan2010automatic,xi2019bug,xuan2014towards,chen2019continuous}, and root cause analysis~\citep{kim2013root,liu2019bugs,jeyakumar2019explainit,li2022causal}, have also witnessed significant contributions over the last two decades. In fact, the reliability and maintainability of hardware and software systems have always been a prominent research focus. However, we have recently witnessed an increased interest in this field. This phenomenon is driven by two main factors: firstly, the remarkable advances achieved in the field of artificial intelligence, and secondly, the shift of numerous IT organizations from product delivery to service release, coupled with the transition from traditional to dynamic infrastructures.

Despite the promising benefits that AIOps offers, it remains federated and unstructured as a research and practical topic~\citep{notaro2021survey,bogatinovski2021artificial,rijal2022aiops}. It involves a diverse array of contributions stemming from various specialized disciplines involving both industry and academia. Given its novelty and cross-disciplinary nature, AIOps contributions are widely dispersed, lacking standardized taxonomic conventions for data management, targeted areas, technical implementation details, and requirements. As such, discovering and comparing these contributions has proven to be challenging~\citep{notaro2021survey}. The lack of a unified terminology results in the absence of guidelines and a clear roadmap for addressing the gaps in the state-of-the-art within AIOps. In fact, while various data-driven approaches may be attributed to the AIOps research area, findings from disparate domains, such as machine learning, may not necessarily apply to software analytics domains like AIOps~\citep{lyu2021towards}. Therefore, it is important to determine within the purview of AIOps, the optimal taxonomy that must be driven by an industrial need necessitating domain expertise in both IT operations and AI. It is also highly important to outline the requirements (desiderata) to construct effective AIOps models, including interpretability, scalability and robustness among others. Additionally, one must also inquire about the metrics that should be employed to compare AIOps methods that belong to the same category, such as anomaly detection or root cause analysis. Metrics based on machine learning, such as contingency metrics, do not always reflect the real accuracy of models when deployed in actual scenarios and hence require contextual and/or temporal adaptation. Multiple other factors and particularities should be taken into account (e.g., human involvement in the loop.).

\subsection{Outline and Contributions}

Our work focuses on providing a holistic framework to the knowledge base of AIOps including technical and research facets.  This framework is explicitly designed to address the  challenge of adeptly managing incidents within IT environments. In pursuit of this objective, our contributions can be outlined as follows:

\begin{itemize}[itemsep=3pt,leftmargin=8mm]
    \item[\ding{111}] We have established a unified and agnostic terminology to define the most relevant terms and key concepts that have been variably adopted in the existing research work within the field of incident management (e.g., hardware/software failure management, anomaly detection, bug triage, fault localization, etc.). Furthermore, we have expounded upon various dimensions of this process, including maintenance protocols levels. 

    \item[\ding{111}] This effort has led us to reveal existing challenges and pain points and define the fundamental building blocks required to achieve a systematic AIOps approach for an intelligent and effective incident management procedure. This includes providing technical specifications for data management such as data collection, storage, and visualization, establishing clearly defined incident management tasks, and emphasizing crucial requirements to be considered when adopting this approach.   

    \item[\ding{111}] Following this, based on the provided terminology, we establish a comprehensive taxonomy to categorize the notable research contributions from prominent conferences and journals in machine learning, data mining, and software engineering. This taxonomy is proposed considering clearly delineated data sources, specific research areas, properties of models and evaluation metrics.

    \item[\ding{111}] The proposed taxonomy also sets itself apart from previous work by closely aligning with the distinct requirements of both industry and research. It facilitates the discovery, implementation, and comparison of various methods and tools, while also addressing existing gaps in the field and hence highlighting potential areas for improvement.

    \item[\ding{111}] We also provide publicly available datasets across various incident tasks and application areas. This represents a valuable contribution, given that identifying the most appropriate datasets for a specific task can be a non-trivial endeavor. Moreover, it facilitates the replication and comparison of techniques within the same research area. It is noteworthy that none of the existing surveys provide such extensive coverage of datasets.

\end{itemize}

\noindent\textbf{Roadmap.} The paper is structured as follows. In Section~\ref{sec:incidentProcedure}, we begin by providing an overview of the incident management procedure, which includes clear definitions, terminology, existing protocols, and targeted maintenance layers. Next, in Section~\ref{sec:towardsAutomated}, we delve into the usage of AIOps to standardize the incident management process, emphasizing important considerations in designing and implementing AIOps solutions. Moving on, Section~\ref{sec:taxonomy} introduces the taxonomy, outlining its components such as data sources and evaluation metrics. Following that, in Section~\ref{sec:review}, we review the most relevant works in the field of incident management based on the proposed taxonomy. Additionally, Section~\ref{sec:reviewDatasets} presents publicly available AIOps datasets and benchmarks used for evaluating AIOps approaches. Finally, the paper concludes with a discussion that includes concluding remarks, open challenges, and areas for improvement in AIOps research.
\section{Streamlining Incident Management Procedure}
\label{sec:incidentProcedure}

\subsection{Terminology and Definitions}
In the following, we aim to provide an easy-to-understand explanation of various terms that are commonly and interchangeably  used in incident management and AIOps. Taking inspiration from the work of~\citet{salfner2010survey}, we build upon their terminology by offering a formal definition that helps clarify the meaning and relevance of these terms in the field.

\begin{figure}[t]
	\centering
	\includegraphics[width=1.0\linewidth]{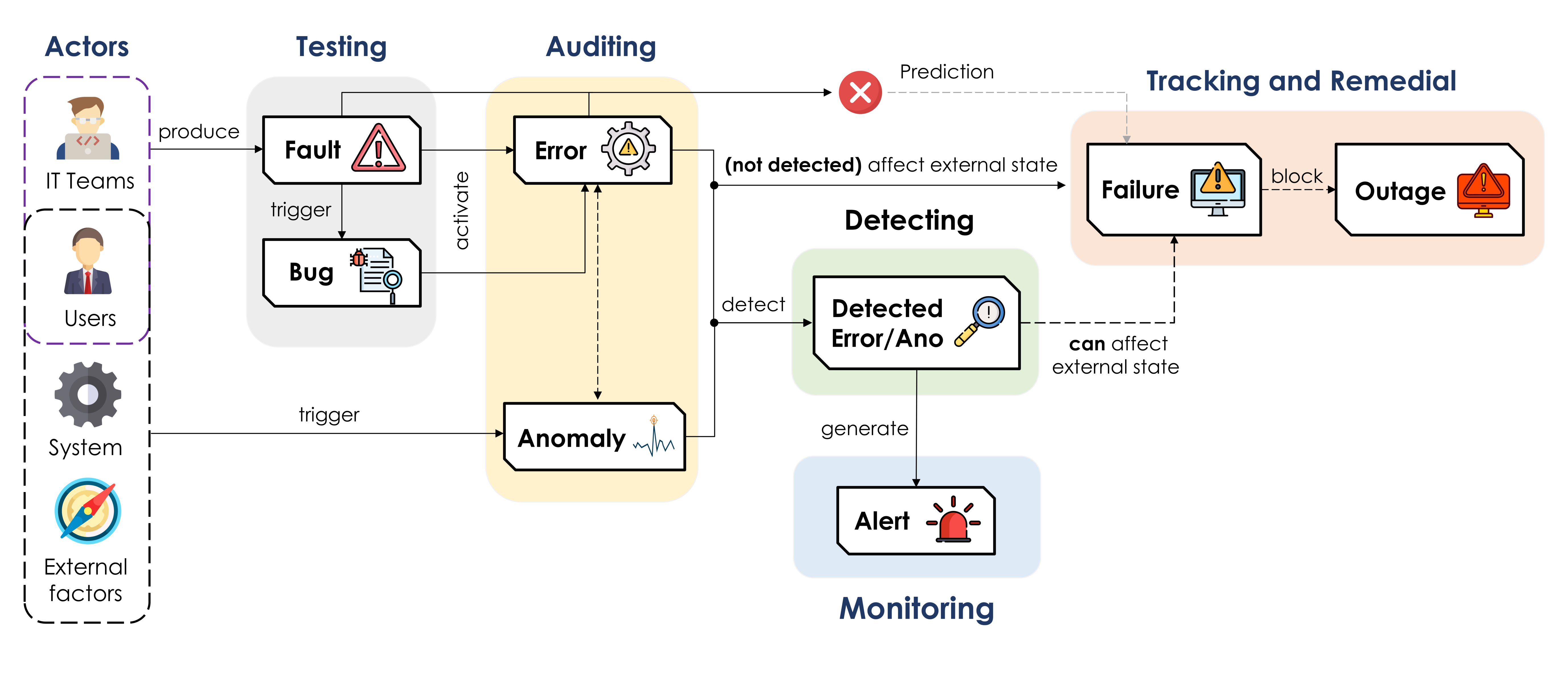}
    \captionsetup{font=footnotesize}
	\caption{Comprehensive chronological schema highlighting the distinctions and key connections among Faults, Bugs, Errors, Anomalies, Failures, and Outages.}%
	\label{fig:sota_terminology}
\end{figure}

Various terms related to incidents, such as fault, error, bug, failure, outage, and anomaly, have been widely used in the field, often without a thorough examination of their precise meanings. For instance, the most commonly used term in the literature is \textit{failure} to indicate system downtime, hardware and software crashes, service disruptions, network traffic disturbances, etc.~\citep{zhang2016automated,pitakrat2018hora,li2016being,zheng2017long}. On the other hand, some other methods utilize the term \textit{outage} to refer to the most severe cases that can significantly reduce system availability and impact user experience~\citep{chen2019outage,zhao2020real}. However, there are also studies that focus on \textit{anomaly} detection, which deals with identifying abnormal behaviors that are already present in the data, often in system metrics~\citep{chow2014mystery,wang2022identifying,nandi2016anomaly}. Regarding the analysis of root causes, the majority of research falls under the category of \textit{fault} localization, which generally identifies faulty components in the source code~\citep{DBLP:conf/issta/WuZCK14,renieres2003fault,abreu2009spectrum}. It may also extend to faulty actions~\citep{li2022actionable}, but it is agreed that this can be the initial point that ultimately leads to failures. The distinctions between these terms based on faulty behavior, underlying causes, and resulting consequences have not received sufficient attention. Although some attempts have been proposed to categorize and differentiate these terms, such as the work of~\citep{salfner2010survey,avizienis2004basic}, they do not provide a comprehensive framework. In order to establish a precise terminology that clarifies the meaning of each interrelated term, we propose a coherent lexicon that covers a broader range of concepts compared to the framework proposed by~\citep{salfner2010survey}. Moreover, we present a chronological schema in Figure~\ref{fig:sota_terminology} illustrating the key relationships among these terms. We also identify the actors involved in initiating faulty behaviors within a system. Subsequently, we provide formal definitions for each term from our perspective.

\noindent \textbf{Failures.} A failure refers to an event that occurs when a system, a component, or a service is unable to fulfill its intended primary function and deviates from it. Failures are often evident and can be observed by either the user or the system, typically stemming from errors or anomalies that give rise to observable issues or malfunctions. Note that within systems, various issues may arise, but unless they result in an undesired output, they do not qualify as failures. Users typically report failures, which can prompt the need for troubleshooting or repairs to prevent outages.

\noindent \textbf{Outages.} An outage refers to a period in which a system, a service, or a network becomes entirely inaccessible or unavailable to users. Outages can stem from failures such as hardware malfunctions or software glitches indicating complete interruptions and unavailability of the required service. These situations often necessitate immediate palliative measures to restore normal operations before delving into the underlying problem for a curative maintenance. 

\noindent \textbf{Errors.} Errors signify instances where the system deviates from its correct and normal state, indicating the presence of an actual issue. These errors may not always be immediately apparent to the user and can remain implicit until they manifest as failures, especially if they are not accurately detected at the opportune moment. Alternatively, errors can be identified through the use of specialized tools or specific algorithms designed for detection purposes.

\noindent \textbf{Anomalies.} Anomalies are defined as unexpected or abnormal behaviors of patterns that deviate from the expected state. They represent irregularities or unusual occurrences that may or may not indicate an error. Unlike errors, anomalies can serve as early indications of underlying issues, but they can also be harmless or temporary deviations that do not directly result in failures. Various factors, such as unusual data patterns or external influences like cyber attacks, can contribute to the emergence of anomalies. These anomalies can be identified and detected through the monitoring and analysis of system metrics. 

\noindent \textbf{Faults and Bugs.}  A fault pertains to an abnormality or defect discovered in a hardware or software component that exhibits incorrect behavior, potentially leading to errors and failures if not promptly detected. These faults generally arise from inherent problems or weaknesses within the system's components. They can be caused by various factors, including human interventions by end-users or administrators, design flaws, system settings, or improper handling. In software development, faults manifest as bugs, which stem from coding mistakes. Identifying faults that lead to bugs often takes place during the testing phase. Conversely, in the case of hardware or setup issues, faults directly result in errors during system operation.

\noindent \textbf{Alerts.} In addition to leading to failures, both undetected and detected errors and anomalies can cause system to deviate from normal behavior as a side effect. This condition is commonly referred to as symptoms~\citep{grottke2008fundamentals}. These symptoms typically manifest as alerting reports, indicating a specific event or condition that demands attention or action. Alerts are usually triggered based on predefined rules or thresholds associated with the symptoms, particularly in the case of anomalies.

Figure~\ref{fig:sota_terminology} illustrates the progression of a fault or anomaly, stemming from internal or external factors, towards a failure and potentially an outage. To illustrate this, let's consider a scenario of a fault-tolerant system with a memory leak problem. The fault in this system is a missing \texttt{free} statement in the source code, which prevents the proper deallocation of memory. As long as the specific part of the software responsible for memory deallocation is never executed, the fault remains dormant and does not affect the system's operation. However, when the piece of code that should free memory is executed, the software enters an incorrect state, turning into an error. In this case, memory is consumed but never freed, even though it is no longer needed. Initially, if the amount of unnecessarily allocated memory is small, the system may still deliver its intended service without any observable failures from the outside. As the code with the memory leak is executed repeatedly, the amount of free memory gradually decreases over time. This out-of-norm behavior, where the system's parameter \texttt{free-memory} deviates from the expected state, can be considered as a symptom of the error. It serves as an indication that something is amiss within the system. At some point, when the memory leak has consumed a significant amount of available memory, there may not be enough resources left for certain memory allocations. This leads to the detection of the error, as the system encounters a situation where it cannot allocate memory when required. In a fault-tolerant system, even if a failed memory allocation occurs, it does not necessarily result in a service failure. The system may employ mechanisms such as spare units to complete the operation and maintain service delivery. Therefore, a single failed memory allocation, by itself, may not cause a service failure or outage. However, if the entire system becomes incapable of delivering its service correctly due to a series of errors or significant resource depletion, a failure occurs. This failure indicates that the system is no longer able to fulfill its intended function, impacting its users and potentially leading to an outage. During an outage, the system becomes completely unavailable, and its services cannot be accessed or utilized by users. In the context of the given example, an outage could happen if the memory leak issue is not addressed in a timely manner, leading to severe resource exhaustion that renders the system inoperable. In summary, the presence of a fault and subsequent error can be indicated by symptoms like memory consumption or depletion. Anomaly detection and monitoring can help identify deviations from expected system behavior. Alerts can be generated to notify system administrators or developers about these anomalies, allowing them to take corrective actions. If the errors and issues persist and prevent the system from delivering its services correctly, a failure occurs, potentially resulting in an outage.

Aiming to provide unified terminology and avoid confusion, in the following, we will refer to all these terms as \textit{incidents}. This term comprises a broader scope and universally applies to any unplanned event or occurrence that disrupts the normal state, behavior, or output of a system, a network, or a service.

\subsection{Existing Maintenance Protocols in Incident Management}

The incident management process should adhere to standardized maintenance protocols and strategies that are universally accepted by IT organizations. These protocols dictate how incidents should be handled based on their occurrence time and the available physical and human resources. They also measure the impact of incidents based on key factors such as availability, performance, and quality. These protocols serve as a framework for assessing the impact of various incident management tasks depicted in Figure~\ref{fig:taxo_aiops}. Unlike some previous works that classified different incident management methods based on these protocols~\cite{davari2021survey,fink2020data}, we present them as abstractions of how incident management tasks help achieve optimal and effective strategies. We do not consider these protocols as standalone phases in the incident management process since our taxonomy primarily focuses on using data-driven approaches to handle reported, detected, or predicted incidents in a phased manner, from reporting to mitigation, regardless of the chosen protocol. The different maintenance strategies can be categorized into two main approaches, Reactive and Proactive maintenance

\begin{figure}
	\centering
	\includegraphics[width=1.0\linewidth]{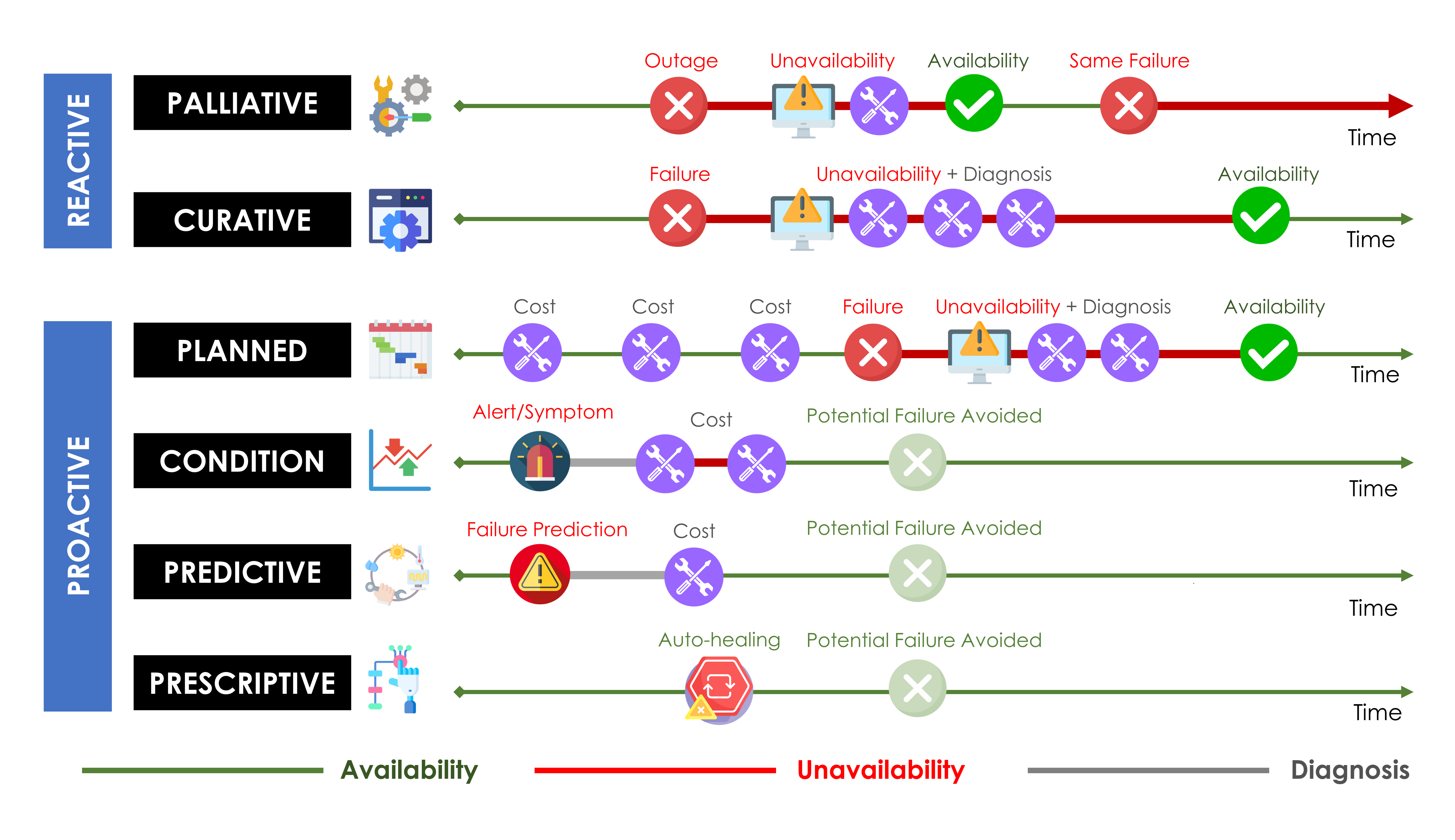}
    \captionsetup{font=footnotesize}
	\caption[Behavioral scheme of the different maintenance protocols.]{Behavioral scheme of the different maintenance protocols. Adapted and improved from~\citep{fink2020data}.}%
	\label{fig:intro_maint_protocols}
\end{figure}

\textbf{Reactive} maintenance is performed in response to incidents that are detected or reported by end users or internal maintenance staff (see detecting and tracking zones in Figure~\ref{fig:sota_terminology}). On the other hand, \textbf{Proactive} maintenance aims to prevent potential problems from occurring and intervenes proactively to rectify them, typically through auditing and testing (as illustrated in Figure~\ref{fig:sota_terminology}). Figure~\ref{fig:intro_maint_protocols} provides a visual representation of the distinct patterns of maintenance strategies analyzed in our study. In reactive maintenance, there is often a time constraint, leading to palliative measures that partially or completely resolve the issue, allowing the affected activity to proceed (especially in the case of outages). For instance, a temporary technical configuration can be implemented. However, curative maintenance must be conducted to implement a stable solution and prevent the issue from recurring for the same or other customers (in the case of failures and/or outages). Proactive maintenance, on the other hand, uses a range of measures, including scheduled maintenance routines and conditional maintenance protocols. These protocols assess system functionality, identify anomalies and errors, and mitigate potential performance degradation and malfunctions (see detecting, auditing, and monitoring zones). This approach heavily relies on AIOps and leverages advanced machine learning algorithms and big data mining to forecast potential system malfunctions by analyzing historical data patterns. Prescriptive maintenance surpasses predictive maintenance by taking a proactive approach to intelligently schedule and plan asset maintenance. Unlike predictive maintenance, which relies solely on historical data, prescriptive maintenance incorporates current equipment conditions to provide precise instructions for repairs or replacements. Moreover, prescriptive maintenance has the capability to recommend optimal palliative or curative actions.
 
\subsection{Target Maintenance Strata}\label{subsec:maint_layers}

The approach to both proactive maintenance and reactive diagnosis involves a comprehensive examination of its components across multiple layers. 
\begin{itemize}[itemsep=3pt,leftmargin=8mm]
    \item[\ding{111}] \textbf{Technical or Physical layer.} This layer focuses on the machines and their various components. For instance, the technical layer would involve monitoring the application server machine, the database, and elements such as RAM, SWAP, processor and disk usage, network identities, and connectors. It also involves monitoring and analyzing the hardware and infrastructure aspects of the system (e.g., remaining useful lifetime of physical components). Examples of checks within this layer include analyzing CPU utilization to ensure it stays within acceptable limits, monitoring disk space to prevent potential storage issues, and checking network connectivity to ensure smooth communication between components.

    \item[\ding{111}] \textbf{Application Layer.} This layer centers around the key components of the software application. It includes the application server, the database server, and user workstations.  For example, the application layer examination would involve analyzing factors such as heap utilization, cache code performance, resource consumption, launch configuration controls, and dump files. This can help to identify incident such as memory leaks, inefficient cache utilization, or misconfigured launch settings that may impact the application performance.

    \item[\ding{111}] \textbf{Functional Layer.} This layer focuses on evaluating the processes executed by the software application to ensure efficient smooth data exchanges, optimal latency, accurate statistics execution, and more. For instance, the maintenance team might analyze the response time of data retrieval or the accuracy of statistical calculations performed by the application. 

    \item[\ding{111}] \textbf{Business Layer.} This layer assesses the control of critical business parameters. It involve evaluating key metrics and business-related aspects of the system to ensure that it aligns with the overall goals and objectives such as the number of transactions processed, the success rate of data transfers, or the adherence to predefined service-level agreements.
\end{itemize}

\section{Towards an Automated AIOps Solution for Incident Management}
\label{sec:towardsAutomated}

The application of AIOps to revolutionize incident management is a complex procedure, requiring a shift from traditional and conventional practices to a fully automated process. Designing, implementing, and deploying machine learning models in real-world software systems scenarios is not a simple task and requires careful consideration. To unlock the full potential of an integrally intelligent solution for incident management and transfer knowledge from AI models to the operating systems domain, a thorough evaluation of the current landscape is necessary. This evaluation should anticipate potential roadblocks stemming from human, resource, or knowledge factors that may arise during implementation. It is also essential to identify areas for improvement and determine the cost associated with transitioning from old practices to a new methodology while considering feedback, reviews, and the trust of practitioners. In the following, we will outline a comprehensive list of pain points and challenges that need to be addressed before building AIOps solutions to manage incidents.

\subsection{Pain points and Challenges}

Within the software industry, there is ongoing development and adoption of AIOps solutions, but it is still in its early stages~\cite{dang2019aiops}. Building and implementing AIOps solutions in real-world scenarios continues to pose challenges from both technical and non-technical standpoints~\cite{chen2020aiops,reiter2021aiops,rijal2022aiops,bogatinovski2021artificial,chen2020towards,lou2013software}. To elaborate further, we outline the significant challenges involved in constructing AIOps solutions.

\smallbreak
\noindent \textbf{Novelty of AIOps.} AIOps remains a relatively new and unstructured area of research that lacks a clear and comprehensive definition~\cite{notaro2021survey,dang2019aiops}. It involves various research domains such as system and pattern design, software engineering, big data, machine learning, distributed computing, and information visualization, among others. Due to its novelty and interdisciplinary nature, AIOps contributions and methods are widely scattered, lacking standardized conventions for data management, target areas, implementation details, and requirements. Consequently, selecting the most suitable techniques from these diverse contributions to achieve specific objectives has proven to be challenging. Hence, it is essential to identify the appropriate methods and tools based on defined goals, and establish an objective procedure to compare between them. Additionally, these methods should align with the IT organization's policies, such as interpretability and trustability, while also considering the available resources and scalability issues.
According to~\citet{{dang2019aiops}}, the domain of AIOps presents several unique challenges that require a comprehensive understanding of the overall problem space, including considerations of business value, data, models, as well as system and process integration.

\smallbreak
\noindent \textbf{Data Management.} Efficient data management is the most critical component for the successful implementation of an AIOps framework. To achieve high observability and real-time analysis, it is essential to integrate and process large volumes of data from diverse sources. However, this integration presents several challenges that need to be addressed. The foremost challenge is to ensure that real-time or near-real-time data collection does not adversely affect the performance of the monitored applications. It is crucial to prevent overwhelming the memory or network resources of these applications. Therefore, the data management system employed must exhibit optimal performance in terms of data storage, ingestion, compression, and querying capabilities across different data types.

\smallbreak
\noindent \textbf{Data Normalization.} AIOps models have specific data requirements that differ from those typically used in general machine learning models. Despite the vast amount of telemetry data collected by major cloud services, which can reach terabytes or even petabytes on a daily or monthly basis, the quality and quantity of available data still fall short of meeting the needs of AIOps solutions. Studies by~\citet{dang2019aiops,levin2019aiops} and~\citet{chen2020aiops} highlight the challenges of working with diverse data from various sources, which often come in disparate formats and structures, making normalization and cleaning complex. This data can be unstructured or semi-structured, including logs, execution traces, source code, hierarchical and graph data, and network traffic, requiring specific preprocessing techniques. Moreover, AIOps models that heavily rely on supervised machine learning algorithms require labeled data for training. However, data often contains noise or missing values, and obtaining labeled data can be challenging. This makes it difficult to build accurate and robust AIOps models. In many AIOps scenarios, constructing supervised machine learning models poses challenges due to data quality issues, such as the absence of clear ground truth labels, the need for manual efforts to obtain high-quality labels, imbalanced datasets, and high levels of noise. Overcoming data quality limitations and managing noise and imbalanced datasets are key areas to focus on when building AIOps models.

\smallbreak
\noindent \textbf{Human Interaction with AIOps.} One of the main challenges in this context is the difficulty of shifting the mindset of IT practitioners towards abandoning old maintenance routines and adopting entirely new approaches~\cite{dang2019aiops}. AIOps-oriented engineering is still at a very early stage, and the establishment of recognized best practices and design patterns for AIOps in the industry is far from complete. Experienced practitioners struggle to let go of their manual activities that are based on adaptation and auditing tasks. Meanwhile, the fundamental methodology of AIOps solutions revolves around learning from historical data to predict the future. The traditional engineering mindset on the other hand, which involves investigating individual cases and reproducing incident steps based on log data, is inefficient or even impractical in large-scale service scenarios.

In fact, there are two distinct perspectives and opinions among practitioners when it comes to AIOps. On one hand, there is a belief that AI can address all challenges, but this expectation is not grounded in reality~\cite{dang2019aiops}. On the other hand, some express doubts about the efficiency of machine learning models in the industry~\cite{rijal2022aiops}. This skepticism stems from the fact that AIOps solutions primarily rely on learning from past experiences to predict future trends using large datasets. However, experienced IT professionals question the effectiveness of these models, even after recognizing the need for digital transformation. They may argue that past experiences do not always provide sufficient coverage to accurately forecast system states, necessitating always human interventions and checking. They might also cite instances where similar past experiences yielded different outcomes. Consequently, businesses require additional time to build confidence in the reliability and dependability of AIOps recommendations. Therefore, it is crucial to invest significant effort in both directions to simplify this transition. Ensuring trust, involving humans in the decision-making process, and providing interpretability and explainability of AIOps solutions are essential to instill confidence in these approaches~\cite{lyu2021towards}. Additionally, setting realistic expectations and clearly defining the scope and goals of AIOps are also crucial considerations.

\smallbreak
\noindent \textbf{Implementation and Integration of AI models.} Building machine learning models for AIOps applications presents unique challenges that are not commonly encountered in other ML/AI scenarios~\cite{dang2019aiops}. For instance, natural language processing models, often used in machine learning, tend to generate inaccurate results when applied to software engineering-related data according to~\citet{menzies2019five} and~\citet{ray2016naturalness}. On the other hand, constructing a supervised machine learning model faces challenges related to data quality and availability, as mentioned earlier. These challenges include imbalanced datasets and the absence of clear ground truth labels. To address these issues, unsupervised or semi-supervised machine learning models can be explored. However, acquiring sufficient labels to understand the "what is abnormal" pattern proves difficult because of the dynamic nature of system behavior and shifting customer requirements and infrastructure. Furthermore, developing high-quality unsupervised models is difficult due to the intricate dependencies and relationships among components and services~\cite{chen2020incidental,lou2013software,bogatinovski2021artificial}. ~\citet{lou2013software} argue that the diagnosis of service incidents cannot solely rely on learning models, but rather requires substantial knowledge about the service system. However, in practice, this type of knowledge is often poorly organized or inadequately documented. Moreover, the need for frequent model updates and online learning poses challenges to DevOps/MLOps practices, when it comes to complex feature engineering efforts.

Another challenge is to ensure that the behavior of the model during the training phase is consistent with its performance in the testing phase. Traditional metrics used to assess models are susceptible to the contamination zone phenomenon~\cite{fourure2021anomaly}, which may lead to erroneous assessments. Indeed,~\citet{fourure2021anomaly}, highlight that by parameterizing the proportion of data between training and testing sets, the F1-score of anomaly detection models can be artificially increased.

Finally, the effectiveness of machine learning models is often proportional to their complexity. Highly accurate models, known as black box models, lack transparency and fail to provide explanations of their decision-making process~\cite{DBLP:journals/csur/GuidottiMRTGP19,molnar2019}. This lack of transparency significantly hampers their adoption by industry practitioners who require a clear understanding of maintenance processes and tool behavior. While leveraging robust models for cost optimization and task automation is valuable, it comes at the expense of transparency. Recent studies in the field of AIOps suggest that interpretable models, even with slightly lower performance, are preferred over high-performing yet non-interpretable models~\cite{lyu2021towards}. Thus, successfully automating incident management processes necessitates establishing practitioners' trust by providing explanations for model decisions, aligning with the concept of eXplainable Artificial Intelligence (XAI)~\cite{DBLP:journals/csur/GuidottiMRTGP19}.

\subsection{AIOps Framework for Data and Incident Management Procedure}

Implementing intelligent solutions for incident management procedures is a complex task that does not solely rely on implementing and training machine learning models on extensive data sets to uncover actionable patterns. One of the significant challenges in adopting AIOps is transitioning from an existing architectural platform that relies on separate modules and executing scripts~\cite{dang2019aiops,shen2020evolving,reiter2021aiops}. This transition involves the adoption and integration of various digital technologies to fundamentally transform the delivery of maintenance and incident management services to end customers or internal maintenance staff. To accomplish this transformation, a reliable, scalable, and secure infrastructure architecture needs to be designed. Such an architecture should be capable of effectively collecting, ingesting, and managing large volumes of data from diverse sources with varying formats. Then, it leverages artificial intelligence approaches to extract meaningful insights from this data, following an iterative and standardized procedure that will be discussed in Section~\ref{subsec:incident_tasks}. Furthermore, this architecture needs to facilitate user interaction, catering to data scientists and engineers, MLOps engineers, and end users. The purpose is to aid in incident management by providing corrective or preemptive actions, predictive alerts, and other valuable insights. Hence, successful implementation of intelligent incident management solutions requires addressing numerous factors, including architectural transformations, digital technology integration, robust data management, and user-friendly interfaces, to fully leverage the power of AIOps in managing incidents, enhancing operational efficiency, and driving better business outcomes.

\begin{figure}
	\centering
	\includegraphics[width=1.0\linewidth]{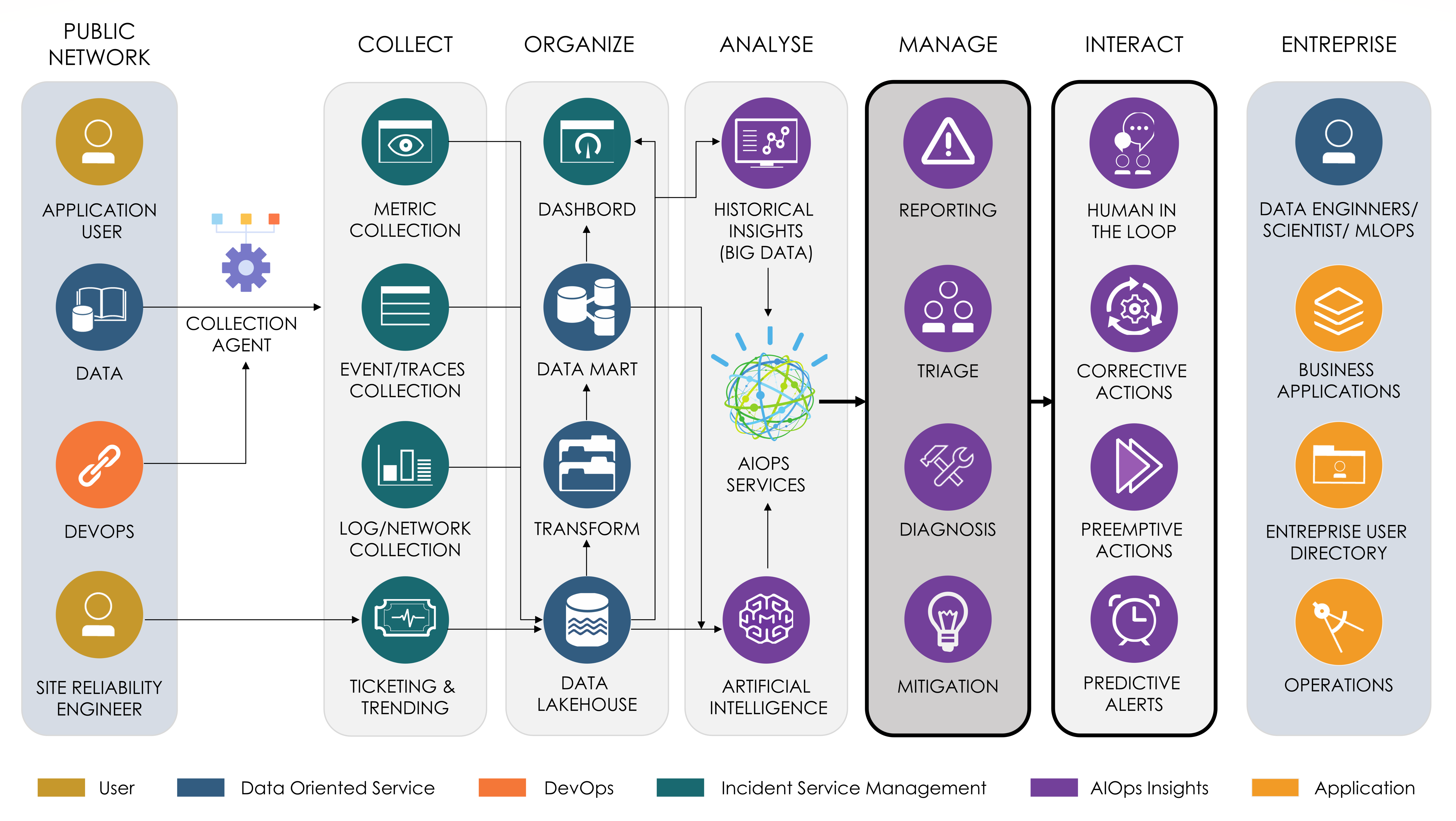}
    \captionsetup{font=footnotesize}
	\caption{Comprehensive AIOps reference architecture for Incident Management Procedure~\cite{prasad2018market,ibmAIOpsArchi}}%
	\label{fig:aiops_archi}
\end{figure}

Figure~\ref{fig:aiops_archi} presents an overview of a layered AIOps reference architecture designed to support intelligent incident management procedures and seamlessly integrate key tools and modules. This architecture is suitable for both traditional in-house and cloud-based application environments, providing a flexible framework for implementing AIOps solutions. In the following, we provide a brief overview of the essential modules that are part of this framework, while also covering potential available and existing open-source tools that can be used to perform the underlying tasks.

\smallbreak
\noindent\textbf{Data Collection and Ingestion.} Dynamic infrastructures generate vast amounts of operational and performance data including infrastructure health, application performance, activity logs, event notifications, network traffic flows, user interactions, and more. Understanding the diverse formats, protocols, and interfaces employed by various data sources is crucial in this context. Extracting monitoring data typically involves installing and configuring a collection agent on the monitored servers. Numerous components within the infrastructures can be effectively monitored, including hardware components, application servers, associated databases, operating systems, virtual machines, and business trends. These components align with the maintenance layers previously discussed in Section~\ref{subsec:maint_layers}.

Numerous open-source collection agents have been introduced to facilitate monitoring tasks. For example, InfluxData offers Telegraf~\cite{telegraf}, a versatile tool capable of connecting with over 300 popular applications. The ELK stack provides the Beats suite~\cite{beats}, which enables the collection and transmission of metrics to ElasticSearch. Another notable agent is Fluentd~\cite{fluentd}, offering the capability to collect data events and send them to various destinations such as files or DBMS. These open-source tools come with several advantages, including their popularity and ease of deployment and configuration. They have gained significant traction in the industry as they are widely accessible and customizable. However, in some cases, it may be beneficial to develop custom collection agents. This allows organizations to collect specific use-case data, leverage pre-built data manipulation libraries, and utilize existing BI tools within their infrastructure~\cite{BendimeradAIOpsInfra23}. Furthermore, custom agents provide the flexibility to tailor the configuration according to specific requirements.

To handle data from different sources with varying formats, it is essential to implement a data ingester module. This module ensures efficient and reliable capture and aggregation of received data into a unified format that can be easily understood by the system. The process involves normalizing data structures, resolving data inconsistencies, and ensuring seamless interoperability. In some cases, appropriate transformations are performed to ensure consistency and compatibility across different data sources before persisting them. This may include data parsing, normalization, filtering, and enrichment. Data ingestion can be achieved through various tools and techniques, including ETL (extract, transform, load) processes, API integrations, or specialized data ingestion platforms. A range of tools and technologies are available to assist with data ingestion in AIOps systems. Open-source platforms like Kafka~\cite{kafka}, Apache NiFi~\cite{nifi}, RabbitMQ~\cite{rabbitmq}, or RocketMQ~\cite{rocketmq} offer capabilities for data streaming, message queuing, data routing, and reliable data ingestion from diverse sources. Additionally, there are proprietary/managed service tools such as AWS Kinesis~\cite{kinesis}, Azure Event Hubs~\cite{azurehubs}, and Google Cloud Pub/Sub~\cite{googlecloudpub}, among others, which provide similar functionalities for data ingestion.

\smallbreak
\noindent\textbf{Data Storage and Organization.} Data collected in the context of AIOps can be categorized into structured, semi-structured, and unstructured data. This encompasses a range of data types, such as time series data that represent system metrics or performance indicators, event logs, sequential templates, network traffic that can be visualized as graphs, and incident tickets reported by end-users in text form. Due to the diversity in data characteristics, it becomes challenging to store all this data in a single warehouse using a "one size fits all" approach. To address this challenge, organizations need to understand and organize the data before feeding it into AI models. Depending on specific requirements and use cases, different approaches to store and query the data are available. Historically, there were two primary options for data persistence: Database Management Systems (DBMS) for structured data and data lakes for unprocessed data. DBMS solutions, like relational databases, are designed for structured data with predefined schemes. They offer predefined structures, enforce data integrity through tables, columns, and relationships, and provide ACID properties (Atomicity, Consistency, Isolation, Durability), ensuring transactional consistency. On the other hand, data lakes serve as storage repositories for vast amounts of raw and unprocessed data in its native format, accommodating semi-structured and unstructured data. The choice between data lakes and DBMS depends on factors such as data volume, variety, velocity, query requirements, data governance needs, and the organization's specific use cases and goals. 

\begin{figure}
	\centering
	\includegraphics[width=1.0\linewidth]{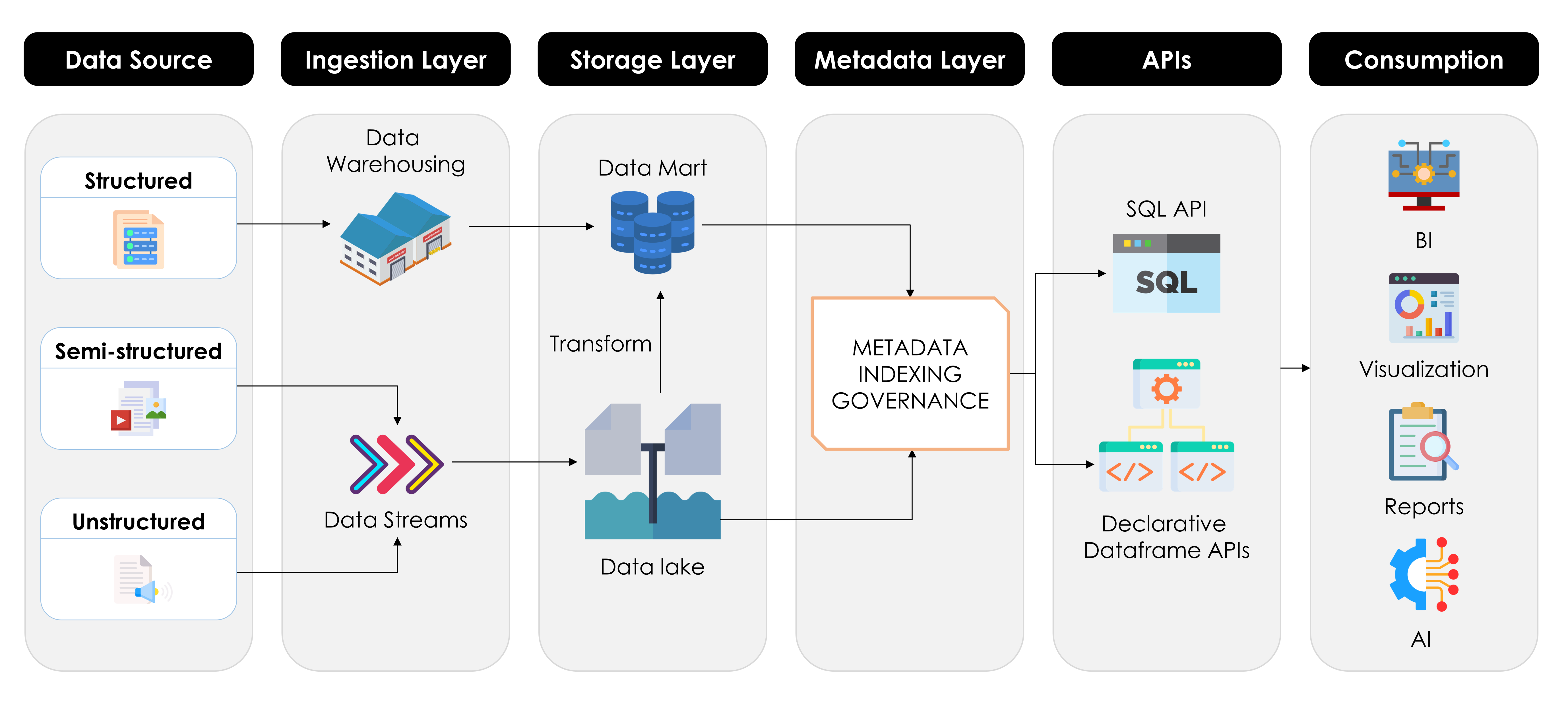}
    \captionsetup{font=footnotesize}
	\caption{Data lakehouse reference architecture}%
	\label{fig:datalakehouse}
\end{figure}

In the context of AIOps as shown in Figure~\ref{fig:aiops_archi}, to support business intelligence, machine learning, and big data mining techniques, it is often beneficial to maintain both data structures simultaneously and link the systems together. Adopting a hybrid approach that combines a data lake and a data warehouse is preferable, which is referred to as a data lakehouse architecture. This approach allows organizations to leverage the strengths of each method for different aspects of their data management requirements. Figure~\ref{fig:datalakehouse} illustrates the data lakehouse architecture and highlights its key components and workflow. A data lakehouse is a centralized, powerful, and flexible big-data storage architecture that integrates various data types from multiple sources, assuming different formats. It combines the best features of data warehouses and data lakes. Initially, data lakehouses start as data lakes containing raw data in its original format, with no constraints on account size or file. This raw data is later processed, transformed, and structured as needed for analysis and querying, creating a data mart repository. To achieve this, data lakehouses leverage technologies like Delta Lake (built on Apache Parquet and Apache Avro)~\cite{delta} to provide capabilities such as schema evolution, ACID transactions, and efficient data indexing. Additionally, they maintain a data catalog and metadata repository, which holds information about available datasets, their schemes, and other relevant details, ensuring data governance and assisting users in discovering and understanding the data. Data-oriented personnel, along with other users, can access the data through SQL or other querying languages. The data lakehouse enables fast and efficient data retrieval and analysis by optimizing data for querying while preserving the raw data for exploration.

Various tools have been proposed to efficiently store and organize data, catering to different needs and use cases. For instance, Elasticsearch~\cite{elasticsearch} stands out as a distributed and open-source search full-text search engine. With an HTTP web interface and schema-free JSON documents, Elasticsearch offers powerful search and indexing capabilities. It excels in managing log and event data in AIOps systems. Similarly, InfluxDB~\cite{influxDB} serves as a time series database designed to handle high write and query loads of time-stamped data. A popular choice for storing and analyzing metrics and sensor data in AIOps systems. These databases come equipped with their own domain-specific languages (DSL). InfluxDB employs the language flux, while Elasticsearch uses ESQL. On the other hand, Clickhouse~\cite{clickhouse} has recently emerged as an open-source columnar database management system. It stands apart with its specific optimization for online analytical processing (OLAP) workloads, making it highly suitable for real-time analytics and reporting. Clickhouse distinguishes itself with its high performance, scalability, and low-latency query capabilities. Furthermore, its SQL support which is familiar to many data analysts and engineers, making it easier to integrate with existing data warehouse repositories like PostgreSQL and Oracle. In a recent study~\cite{BendimeradAIOpsInfra23}, the authors provided specific criteria to choose the most suitable solution for a given context. These criteria include license, notoriety and popularity, adaptability, and performance.

\smallbreak
\noindent\textbf{Data Visualization and Monitoring.} In data management, whether data is organized in a data mart repository or kept within a data lake, the potential for deriving valuable insights is required through data visualization dashboards. These tools play a pivotal role in facilitating a comprehensive understanding of data, enabling in-depth inspections, and supporting data-driven decision-making processes. These visualization tools lies in their ability to offer a user-friendly interface, coupled with a range of visualization panels including versatile options like time series graphs, gauges, dynamic topologies, and more, catering to the diverse needs of IT professionals engaged in monitoring activities. By utilizing such tools, we gain the ability to continuously monitor the infrastructure, diagnose potential problems, adeptly identify intricate relationships between various components, navigate to the root causes of incidents, leading to quicker resolution times. Moreover, the tools excel in anomaly detection by visually representing abnormal behavior, thus allowing for the early identification of potential issues and threats. Among the plethora of available data visualization tools, some noteworthy open-source options include Grafana~\cite{grafana}, Kibana~\cite{kibana}, Metabase~\cite{metabase} and Apache Superset~\cite{superset}.

\smallbreak
\noindent\textbf{Intelligent Incident Management Procedure.} AIOps insights are founded on intelligent algorithms that operate on both historical and real-time data, aiming to provide actionable patterns and prescriptive recommendations. These insights are designed to optimize operational performance, enhance situational awareness, and improve incident management procedures. More specifically, the incident management procedure can be viewed as a systematic approach that encompasses a predefined sequence of steps and processes to be consistently followed, whether incidents occur unexpectedly or are foreseen. Its primary objective is to ensure a consistent and well-coordinated response, regardless of the source of the incident report. This procedure involves the use of robust artificial intelligence algorithms to aid in reporting, classification, prioritization, assignment, diagnosis, and resolution of incidents. Throughout the entire process, it is essential to establish clear communication channels to keep stakeholders informed about the incident's status, progress in resolution, and any pertinent actions taken. Time sensitivity is also crucial, and every effort is naturally directed towards minimizing various time-related metrics, which include time to detect (TTD)\footnote{Also known as time to report, accounting for both manual reporting and predicted incidents}, time to engage (TTE), time to acknowledge or diagnose (TTA), and time to repair or mitigate (TTR). Furthermore, upon the successful resolution of incidents, an essential post-incident review is initiated to record the corrective or preventive actions taken for a specific incident. This phase also plays a pivotal role in identifying valuable lessons learned and proactively preventing similar incidents in the future.

\subsection{Intelligent Incident Management Procedure Tasks}
\label{subsec:incident_tasks}

\begin{figure}[t]
	\centering
	\includegraphics[width=1.0\linewidth]{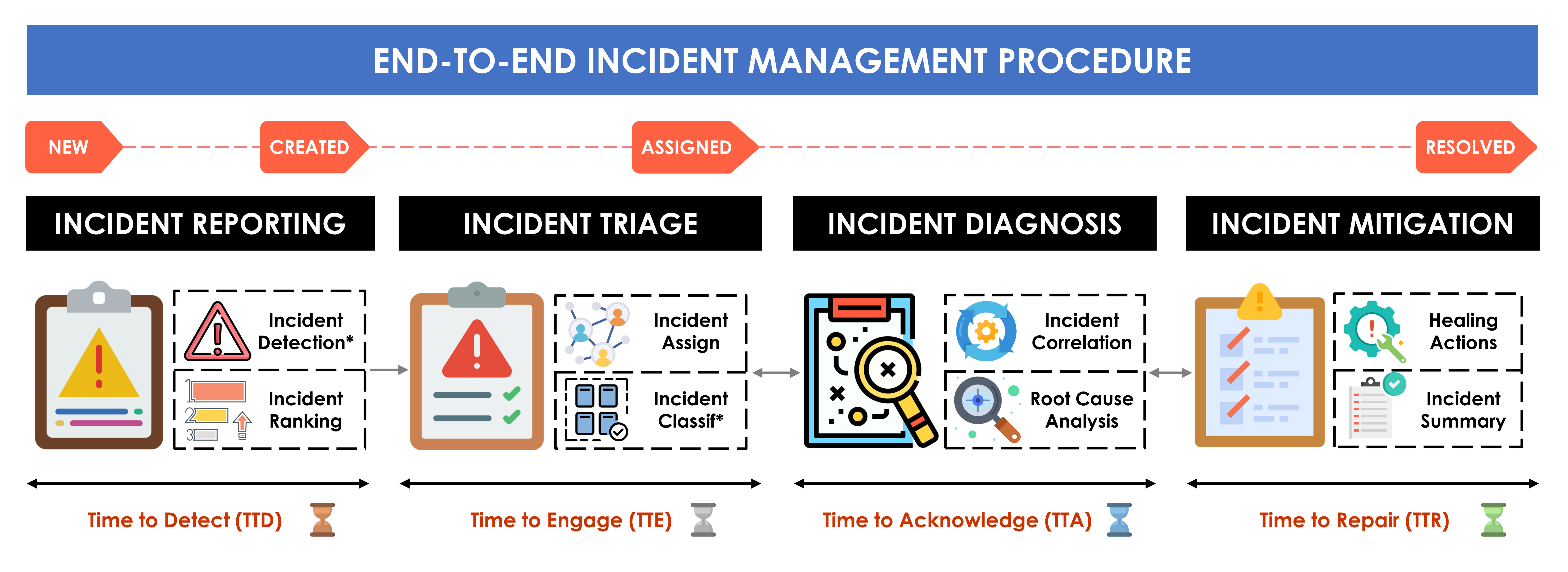}
    \captionsetup{font=footnotesize}
	\caption{Standardized end-to-end procedure proposed for Incident Management in the context of AIOps.}%
	\label{fig:aiops_incident_management_procedure}
\end{figure}

Drawing from the capabilities of AIOps for incident management, our proposal aims to organize maintenance routines within a standardized approach to conduct incidents from creation through diagnosis to resolution. This redesigned process involves a sequential workflow, guiding incidents through four well-defined phases, starting from initial reporting and resulting in mitigation and postmortem analysis. Indeed, the objective is to automate as many maintenance tasks as possible, with a view to optimize the reporting time, diagnosis and triage time, and resolution time. Such a protocol should not only facilitate the resolution of incidents but also serve to document the maintenance context. Figure~\ref{fig:aiops_incident_management_procedure} provides a clear visual representation of this categorization\footnote{For clarity,  detection involves both detection and prediction, while classification also includes deduplication.}. It should be noted that an incident typically goes through many sub-phases in the illustrated workflow, but there may be instances where certain phases may not be necessary. In such cases, a phase may be skipped if it is considered unnecessary or doesn't contribute to resolving the reported issue. For example, if an incident is categorized in a manner that has already undergone prior investigation, there might be no need for a root cause analysis phase.

In the following, we provide from our perspective an extensive review of the essential subcategories within the incident management procedure. Our goal is to examine the most relevant approaches proposed for each of these phases. It is worth mentioning that some previous work has already defined certain terms related to incident management. For instance,~\citet{chen2020towards} defined the incident management procedure as a three-step process involving incident reporting, triage, and mitigation. However, their definition remains generic and does not delve into the subcategories of these phases, such as addressing the problem of incident classification and correlation. On the other hand,~\citet{notaro2021survey} focused on studying failures and developed a taxonomy based on proactive and reactive approaches, with significant emphasis on the reporting phase. However, their research appears to neglect other important phases. Additionally,~\citet{zhang2015survey} provided a formal definition that includes detailed phases like assignment, prioritization, fault localization, and mitigation. However, this survey solely concentrates on software bugs and does not generalize to other specific areas. In this work, our aim is to cover all these use cases under a unified taxonomy, regardless of the terminology used (failures, bugs, anomalies, etc.) or the specific industry focus.

\smallbreak
\noindent \textbf{Incident Detection.} Incident detection refers to the process of identifying and recognizing deviations from normal operations indicating the presence of abnormal behavior or faulty events that may indicate the occurrence of an incident (e.g., error or anomaly). This process involves monitoring and analyzing various data sources (e.g., KPI metrics, system logs, and user reports). For instance, the \textit{anomaly detection} research domain falls within this subcategory.

\begin{figure}[t]
	\centering
	\includegraphics[width=0.75\linewidth]{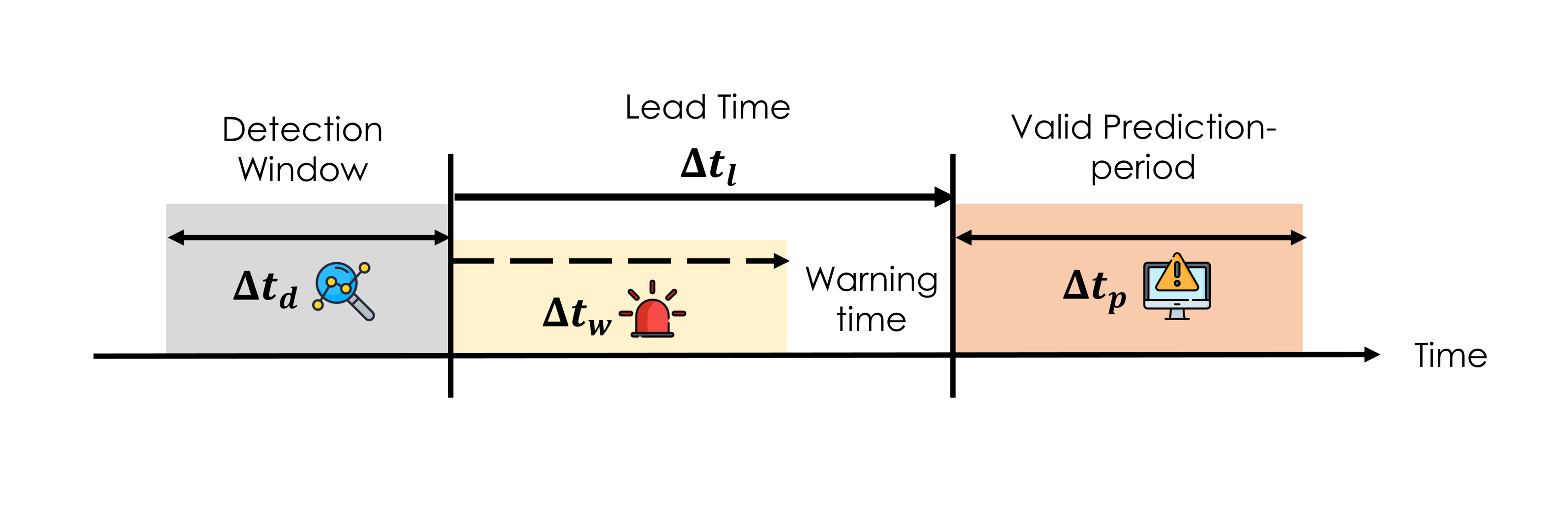}
    \captionsetup{font=footnotesize}
	\caption{Time relations in online failure prediction.}%
	\label{fig:sota_failure_timeline}
\end{figure}

\smallbreak
\noindent \textbf{Incident Prediction.} Incident prediction refers to the process of forecasting, anticipating, or estimating the likelihood of potential incidents, primarily failures, before they occur. It involves leveraging available historical data, along with other relevant factors, to proactively identify and assess potential risks and vulnerabilities. By analyzing patterns, trends, abnormal events, and anomalies in historical and current data using advanced analytics techniques, incident prediction aims to take preventive actions and minimize the impact of future incidents. Incident prediction can be categorized into offline and online methods. The offline category includes Software Defect Prediction (SDP)~\cite{dejaeger2012toward,li2017software}, which entails assessing failure risks by executing specific functional units or analyzing portions of the source code. Another technique in this category is fault injection~\cite{siami2008sufficient,natella2012fault}, where deliberate faults are introduced into a functioning system through stress-testing to evaluate its fault tolerance level. Conversely, online prediction occurs during system runtime. It involves techniques like software rejuvenation~\cite{alonso2010adaptive,vaidyanathan1999measurement}, which addresses resource exhaustion and prevents unexpected system failures caused by aging or accumulated faults in software systems. Additionally, online prediction involves estimating the remaining useful lifetime of the system. This category also involves real-time predictions of hardware and software failures, taking into account time constraints, as depicted in Figure~\ref{fig:sota_failure_timeline}. For a prediction to be considered valid, it must be made with a lead-time $(\Delta t_l)$ greater than the minimal warning time $(\Delta t_w)$. Moreover, the prediction is only considered valid if the failure occurs within a specific time period called the prediction period $(\Delta t_p)$. To make these predictions, data up to a certain time horizon $(\Delta t_d)$, referred to as the data window size, is used.  

\smallbreak
\noindent \textbf{Incident Prioritization.} Incident prioritization is the process of categorizing and ranking incidents based on their urgency, impact, and business priorities. It involves evaluating the severity of the incident, considering factors such as the affected systems, services, and the potential business impact. Incident prioritization ensures as well that resources are allocated appropriately, with relatively critical incidents receiving immediate attention and resources.

\smallbreak
\noindent \textbf{Incident Assignment.} Incident assignment entails the allocation of incidents to the relevant individuals or teams responsible for investigating and resolving them. This process involves analyzing the information contained in incident reports, considering factors such as the nature and complexity of the incident, as well as the skills and availability of the assigned personnel. This process is commonly referred to as incident triage in several work~\cite{chen2019continuous,zhang2016literature}. However, triage in a general sense, does not refer only to assignment but also to other tasks such as classification and identification of duplicate incidents.

\smallbreak
\noindent \textbf{Incident Classification.} Incident classification involves the systematic grouping and/or categorization of incidents based on their distinct characteristics, symptoms, or impact. This classification process establishes a structured framework that enhances the understanding and effective management of incidents. This process can be seen as a refinement of the assignment procedure. In fact, it is necessary to categorize incidents and assign them to specific teams based on their respective topics, whenever possible. For instance, the technical service team within an IT organization may be tasked with resolving incidents related to resource saturation, processes that put a significant demand on the CPU and SWAP, security vulnerabilities, etc. To efficiently manage these issues, the technical team should assign knowledgeable staff to each topic. As such, it would be optimal, once an incident is assigned to the responsible team, to promptly identify the appropriate topic as the initial step in the procedure. Surprisingly, despite its crucial role in optimizing incident management time response, incident classification has not received sufficient coverage or extensive attention. While some studies, such as~\cite{pingclasai2013classifying,antoniol2008bug}, have approached this issue by treating it as a content optimization problem, others have included it as part of prioritization~\cite{lamkanfi2011comparing,tian2012information}. Some researchers have even considered it within the scope of duplicate detection~\cite{banerjee2012automated,jalbert2008automated}. However, we believe that there are inherent differences between these categories. Incident classification primarily focuses on associating the incident with a specific topic or category, regardless of the assigned personnel, incident priority, presence of similar incidents, or whether it is a new incident. 

\smallbreak
\noindent \textbf{Incident Deduplication.} Near-duplicate incident detection is the process that efficiently identifies incidents that are closely related or exhibit significant similarities, grouping them into specific buckets that correspond to distinct problems. This process involves real-time analysis of incoming incidents to determine their similarities, overlaps, and commonalities with historical incidents that pertain to the same topic. By identifying duplicates, incident deduplication reduces redundancy and incident management efforts and prevents unnecessary resource allocation. In fact,~\citet{anvik2005coping} conducted a comprehensive empirical study using bug repositories from Eclipse and Firefox, which revealed that a significant portion (20\%-40\%) of bug reports are flagged as duplicates by developers. This study provides concrete evidence of the necessity to detect duplicate incidents. This process can also be considered as a further refinement of the incident classification problem.

\begin{figure}[t]
	\centering
	\includegraphics[width=0.75\linewidth]{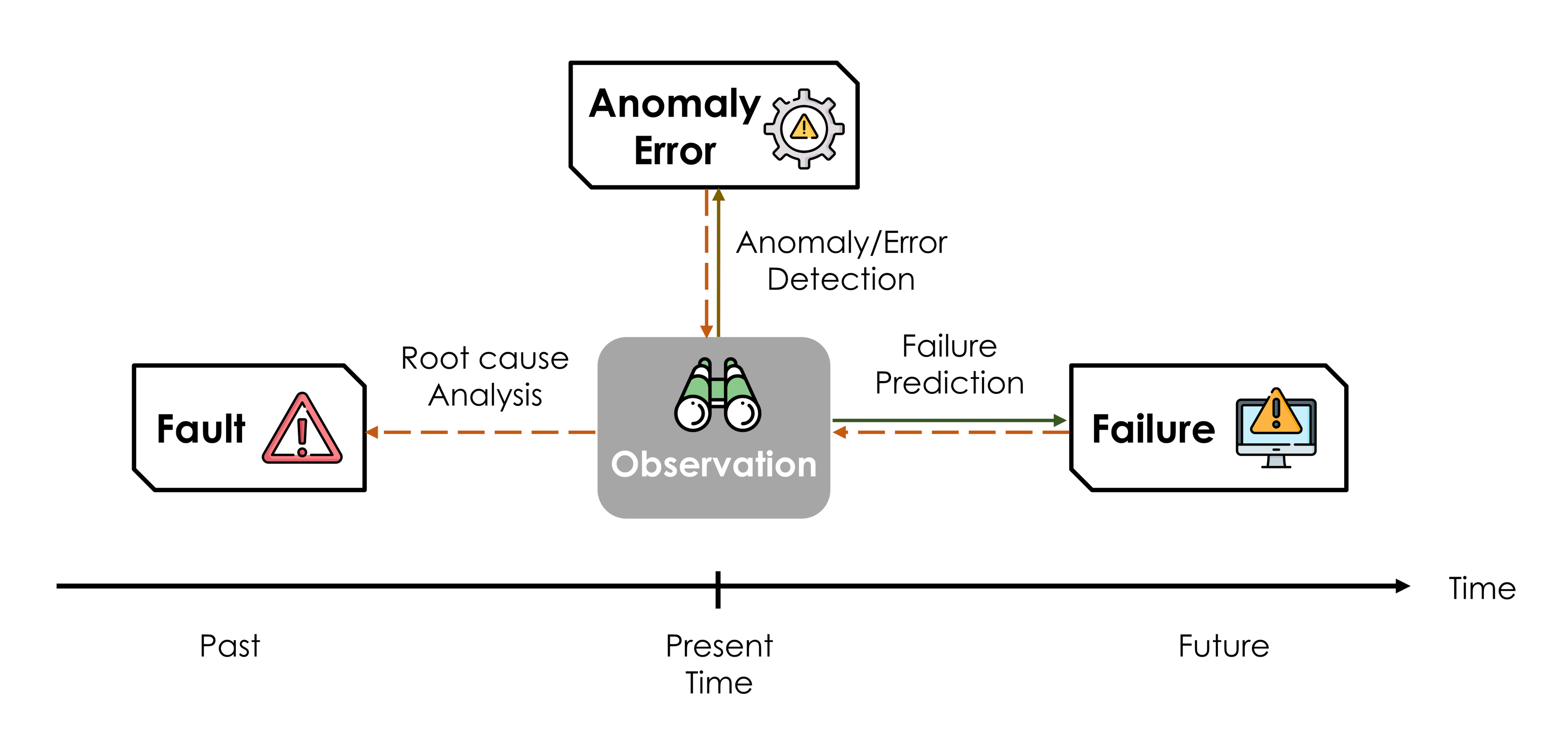}
    \captionsetup{font=footnotesize}
	\caption{Distinction between Root Cause Analysis, Anomaly Detection, and Failure Prediction.}%
	\label{fig:failure_fault}
\end{figure}

\smallbreak
\noindent \textbf{Root Cause Analysis.} Root cause analysis (RCA), also known as root cause diagnosis, plays a pivotal role in the incident management procedure. It is a systematic process that aims to investigate and identify the underlying causes and contributing factors of incidents, which we commonly refer to as faults. As depicted in Figure~\ref{fig:failure_fault}, RCA delves into the fundamental fault behind the occurrence of an incident, recognizing that it may not necessarily stem from a faulty manipulation or defective source code. Its objective goes beyond addressing the symptoms. Instead, it seeks to uncover and understand the root cause, even if it originates from external factors. In related studies, this process is often referred to as Fault Localization~\cite{wong2016survey}. These studies focus on pinpointing the specific set of components (such as devices, network links, hosts, software modules, etc.) that are associated with a fault that has caused or may cause a particular failure. However, root cause analysis goes beyond component localization and extends its investigation to faulty actions or risky external factors that may lead to abnormal behaviors within the system.

\smallbreak
\noindent \textbf{Incident Correlation.} Incident correlation study involves the analysis of multiple incidents or events with the aim of identifying relationships, dependencies, or shared characteristics among them. Through this process, a holistic perspective of the incident landscape can be achieved, allowing to uncover potential hidden patterns or trends. Incident correlation works alongside root cause analysis to assess how the underlying causes and faulty components or behaviors can affect other incidents. This helps in facilitating more efficient incident resolution. The task is considered challenging due to the inherent complexity and interdependence among components in software systems~\cite{attariyan2012x}.

\smallbreak
\noindent \textbf{Incident Mitigation.} Incident mitigation also known as remediation, refers to the process of minimizing the impact and severity of an incident. It involves taking proactive and automatic measures to contain, resolve, or reduce the effects of the incident. Incident mitigation can include implementing temporary workarounds, applying fixes or patches, activating backup systems, or engaging specialized resources or teams to restore normal operations. As mentioned in~\cite{notaro2021survey}, contributions related to remediation actions have been relatively fewer compared to incident prediction, detection, triage, and diagnosis tasks. This could be attributed to the fact that once the underlying problem is identified through diagnosis, the necessary recovery steps become readily identifiable and achievable. In many cases, historical incidents with similar resolutions can be referenced, eliminating the need for using complex models.

\subsection{Desiderata for Effective Intelligent Incident Management}
\label{subsec:desiderata}

Creating intelligent, data-driven strategies for incident management is an intricate endeavor that extends beyond the design of effective machine learning techniques. Simply relying on high-performing machine learning or big data mining models is insufficient for successfully adopting AIOps solutions. To ensure the effectiveness of such solutions, they must adhere to a set of established criteria, which we refer to as desiderata. Drawing from numerous reviewed studies, including ~\citep{dang2019aiops, lyu2021towards,lou2013software,zhao2021empirical,li2020predicting}, we have compiled a comprehensive list of requirements that should be considered, either fully or partially, when constructing AIOps solutions. These requirements are as follows:

\begin{itemize}[itemsep=3pt,leftmargin=8mm]
    \item[\ding{111}]\textbf{Trustablity and Human in the loop.} The literature claims that the requirements for employees skills and mindsets change with the introduction of AIOps~\citep{reiter2021aiops}. Manual activities tend to shift towards adaptation and auditing tasks while dealing with AI requires a different approach focused on recognizing patterns from raw data, deviating from the traditional developer mindset. This transition raises questions about trust in AI capabilities and what it can offer. Consequently, adopted AIOps approaches should incorporate years of field-tested engineer-trusted domain expertise iteratively and interactively into the learning, updating, and explanation phases of sophisticated machine learning models built on raw data. IT professionals possess valuable domain knowledge and insights acquired through years of experience in managing and troubleshooting IT systems. While not all of their best practices may scale with the AIOps trends, their expertise often extends beyond raw data analysis. They have a deep understanding of the underlying technology, infrastructure, applications, and business requirements. Hence, it is crucial to fully leverage and model this expertise into AIOps solutions. This can be achieved by providing mechanisms that incorporate the human in the loop, allowing for interaction, updates, and corrections to the models when necessary. Active Learning~\cite{DBLP:journals/csur/RenXCHLGCW22} can be specifically beneficial in this context. 
 
    \item[\ding{111}] \textbf{Interpretability.} AIOps solutions should prioritize interpretability, even if it comes at the expense of model performance. In the context of AIOps, interpretable models are preferred when high-performing models lack interpretability. Model transparency enables users to fully understand, interact with, and reason about the recommendations made by the model, which can help gain support from upper management in following those recommendations. However, interpreting AIOps models comes with certain constraints and requirements. In a study by~\citep{lyu2021towards}, different factors influencing AIOps model interpretation are investigated across three key dimensions. (1) \textit{Internal Consistency} which assesses the similarity between interpretations derived from AIOps models trained under the same setup. It examines whether the interpretations obtained from an AIOps model are reproducible when the model is trained with the same data and implementation across multiple executions. (2) \textit{External Consistency} which focuses on the similarity between interpretations derived from similar-performing AIOps models on a given dataset. Intuitively, interpretations derived from a low-performing interpretable model could be trustworthy only if the interpretable model has the same interpretation as other machine learning models on a given dataset. (3) \textit{Time Consistency} which captures the similarity between interpretations derived from an AIOps model across different time periods. AIOps models should not only reflect the trends observed in the most recent training data but also capture and reflect trends observed over a longer period. It is important to note that some previous work, such as~\citep{bangash2020time} in defect prediction, has shown that models trained on one time period may not generalize well when tested on a different time period. Additionally, the size of the training data can impact the derived interpretations of the models.

    \item[\ding{111}] \textbf{Scalability.} AIOps solutions must efficiently handle large-scale data in complex IT environments where significant amounts of monitoring and log data are expected. These environments can encompass thousands to millions of components, including servers, network devices, and applications. To go beyond effective modeling and accurate results, it is crucial for AIOps solutions to be implemented within robust architectures that excel at ingesting, storing, and processing big data efficiently. Scalable architectures and data processing frameworks play a key role in distributing the workload and effectively handling the high volume of data. Additionally, when considering the adopted approaches, AIOps solutions should leverage scalable computing techniques, such as distributed and federated learning, as discussed in studies like~\citep{diaz2023orfeon,becker2020towards,nedelkoski2020multi}, to enable parallel processing and distributed data analysis. Scalability also involves optimizing the utilization of computational resources. AIOps solutions should possess the capability to dynamically allocate and distribute resources based on the data volume and processing requirements.
    
    \item[\ding{111}] \textbf{Maintainability and Adaptability.} The concepts of maintainability and adaptability are crucial in AIOps, as they aim to minimize the need for ongoing maintenance and repetitive fine-tuning. This consideration is essential because DevOps engineers, who are responsible for managing and maintaining these solutions, often have a multitude of responsibilities and may not possess extensive expertise in machine learning. Therefore, AIOps solutions should strive for a high degree of automation and self-management of routine tasks such as data preprocessing and regular model training to reduce the reliance on continuous manual interventions. To achieve this, self-adjusting algorithms and automated pipelines can be employed~\citep{BendimeradAIOpsInfra23}. In addition, leveraging advanced machine learning techniques such as transfer learning~\citep{pan2010survey} and one-shot learning~\citep{vinyals2016matching} can greatly benefit AIOps solutions. Instead of training models from scratch, pre-trained models that have been developed and fine-tuned by machine learning experts can be utilized to handle new data patterns.
    
    \item[\ding{111}] \textbf{Robustness.} AIOps solutions need to be built upon robust and stable machine learning models that can handle a wide range of scenarios and exhibit resilience to variations in data patterns. These models should be designed to be less sensitive to the noisy and incomplete data commonly encountered in real-world IT environments~\citep{dang2019aiops}. To ensure the reliability of the modeling process, robust preprocessing techniques, such as systematic data cleaning and effective imputation methods, can be employed. In addition, AIOps solutions must be capable of detecting and adapting to concept drift, which refers to the shifts in underlying data distributions that occur in dynamic IT environments~\citep{lyu2021empirical}. Robust algorithms and models, such as those based on online learning, can be leveraged to handle concept drift and maintain up-to-date insights in the face of evolving data patterns~\citep{saul2004overview,bukhsh2023maintenance}. Furthermore, AIOps solutions should generalize well across different IT environments. To achieve this, they should be trained on diverse and representative data that captures the underlying patterns and relationships applicable across various scenarios.
 
    \item[\ding{111}] \textbf{In-context Evaluation.}  Unlike conventional machine learning evaluation scenarios, such as cross-validation, which are often not directly applicable to AIOps due to the unique characteristics of real-world IT environments, the concept of in-context evaluation emphasizes the need to assess the solution's performance in a context that closely resembles its actual production usage. Traditional evaluation methods typically assume that the data used for evaluation is identically distributed, which may not hold true in IT environments. Real-world data often exhibits temporal dependencies, concept drift, and dynamic patterns, which require specialized evaluation techniques that consider these factors. To conduct in-context evaluation, it is important to create evaluation frameworks that capture the particularities of the production environment. This involves using datasets with a broad range of scenarios, including normal operations, various types of incidents, and different environmental and temporal conditions. In addition to dataset selection, evaluating AIOps solutions in context also requires defining appropriate evaluation metrics and benchmarks that align with the desired outcomes and objectives. (See Section~\ref{subsec:eval_metrics} for more details)

\end{itemize}

\section{Proposed Taxonomy}
\label{sec:taxonomy}

\begin{table}[t]
\caption{Proposed taxonomy to categorize AIOps approaches for Incident Management.}
\label{tab:taxonomy}
\centering
\resizebox{\textwidth}{!}{%
\begin{tabular}{>{\centering\arraybackslash}m{2cm} >{\centering\arraybackslash}m{2cm} >{\raggedright\arraybackslash}m{12cm}}
\toprule
\multirow{6}{*}{\textbf{Context}}         & \textbf{Incident Task}           & Refers to the specific research area in which the proposed approach fits within the incident management procedure. More precisely, it addresses one of the distinct phases when handling the incident, as identified in our categorization, including reporting, triage, diagnosis, or mitigation.                                                                                                                                                                     \\ \cmidrule(l){2-3} 
                                          & \textbf{Focus Area}       & Refers to the specific application area. Some methods may be exclusively dedicated to a particular application domain. Different applications have varying requirements and constraints, and the reviewed method may need to be tailored accordingly.                                                                                                                                                                                                                                            \\ \cmidrule(l){2-3} 
                                          & \textbf{Maintenance Layer}    & Refers to one of the different layers of the system as highlighted in Section~\ref{subsec:maint_layers} that are targeted by the reviewed method.                                                                                                                                                                                                                                                                                                                                                                                                                                                                        \\ \midrule
\multirow{6}{*}{\textbf{Data }} & \textbf{Data Source}          & Refers to the nature of data that the method is built upon. The proposed taxonomy includes various types of data sources, such as log metrics, source code, key performance indicators, topology (environmental characteristics), alerting signals, execution traces, network traffic, etc.                                                                                                                                                                                                                                                                          \\ \cmidrule(l){2-3} 
                                          & \textbf{Data Type}            & Consideration should be attributed to how data is represented. Even when using the same data source, different representations are possible and then imply different methodologies (e.g., source code can be represented as an Abstract Syntax Tree (AST), a sequence of predefined frames, or plain natural language text). To facilitate this understanding, a taxonomy of data types has been adopted. This taxonomy includes structured data, sequential data, graph data, time series (both univariate and multivariate), textual data, hierarchical data, etc.                                                                                                                                                                                                                                                                                                                                                                                                                    \\ \midrule
\multirow{7}{*}{\textbf{Model }}    & \textbf{Approach}             & Represents the principal approach employed to address the problem at hand, taking into consideration the context and the data utilized. It elucidates the way in which the authors formalized and tackled the problem (e.g., clustering, nearest neighbor search, dimensionality reduction, deep learning, transformers, etc).                                                                                                                                                                                                                                       \\ \cmidrule(l){2-3} 
                                          & \textbf{Paradigm}             & Provides an abstract demonstration of how the method was approached to attain its objective. For instance, in the case of training predictive models, it may involve discerning whether the approach was supervised, semi-supervised, or unsupervised, or whether it entailed one-shot, multitasking, reinforcement, or transfer learning, etc.                                                                                                                                                                                                                      \\ \cmidrule(l){2-3} 
                                          & \textbf{Evaluation Metrics}   & Relates to the evaluation metrics used to assess how well the method performs compared to other techniques currently used in the field.                                                                                                                                                                                                                                                                                                                                                                                                                              \\ \cmidrule(l){2-3} 
                                          & \textbf{Package Availability} & This factor is highly important to reveal the accessibility of both data and model packages for reproducibility purposes.                                                                                                                                                                                                                                                                                                                                                                                                                                            \\ \midrule
\multicolumn{2}{c}{\textbf{Particularities}}                              & The final category of our taxonomy concerns essential factors that highlight specific attributes and desired outcomes associated with AIOps which aligns with the requirements detailed in Section~\ref{subsec:desiderata}.                                                                                                                                                                                                                                                                                                                     \\ \bottomrule

\end{tabular}}
\end{table}

In this paper, we present a comprehensive taxonomy that categorizes research work related to AIOps for incident management. Our taxonomy encompasses primary groups, covering diverse factors that are essential to consider when evaluating the necessity, design, implementation, and reproducibility of the reviewed methods. Carefully choosing these categories enables us to analyze many dimensions that significantly impact AIOps for incident management. These dimensions include the context, representing environmental factors that drive and surround the proposed approach. Additionally, we examine factors pertaining to the characteristics of the data and their representation. Moreover, we explore the requisite measures necessary to render the data actionable for efficient analysis and interpretation. Of utmost importance is the detailed exploration of the design and implementation of the approach. We clarify the methodology adopted, the learning paradigm or research area employed, and how the approach was evaluated. Furthermore, we highlight particularities exhibited and interesting requirements validated by the proposed approach. Each of these categories is explained in detail in Table~\ref{tab:taxonomy} to ensure inclusiveness and brings together all relevant aspects in a structured manner. Concerning the specific attributes highlighted for each method, which correspond to the requirements outlined for constructing AIOps models, our focus is solely on instances where a method explicitly addresses these criteria in the paper (e.g., robustness to noise or incorporating mechanisms for explainability).  To illustrate, the adherence of a method to contextual evaluation criteria is established only if the experimental study explicitly aligns with the temporal requisites outlined in the evaluation protocol (refer to Section~\ref{subsec:eval_metrics} for more details). An example of this would involve ensuring that anomalies within the test set must strictly postdate anomalies observed in both the training and validation sets.

In the following, we will delve into major factors that are exclusively relevant to our studied domain. Our focus is specifically to explore the different data sources and outline the evaluation methods employed for the proposed AI approaches.

\subsection{Data Sources and Types}
\label{subsec:data_sources}

Data plays the most crucial role in incident management, serving as the fundamental building block that guides the design of the approach used to identify predictive or descriptive patterns within it. The primary objective is to use this data to effectively accomplish the desired task at hand. In an industrial setting, data is derived from a multitude of sources, including physical or software components, and can also be generated or edited by humans. One key characteristic of this data is its unstructured nature, lacking a standardized format and displaying non-homogeneity. Consequently, data collected from different sources require a pre-processing stage to perform subsequent analysis. Furthermore, data from the same source can be represented in various ways. We categorize data sources based on the following convention.

\begin{figure}[t]
	\centering
	\includegraphics[width=0.90\linewidth]{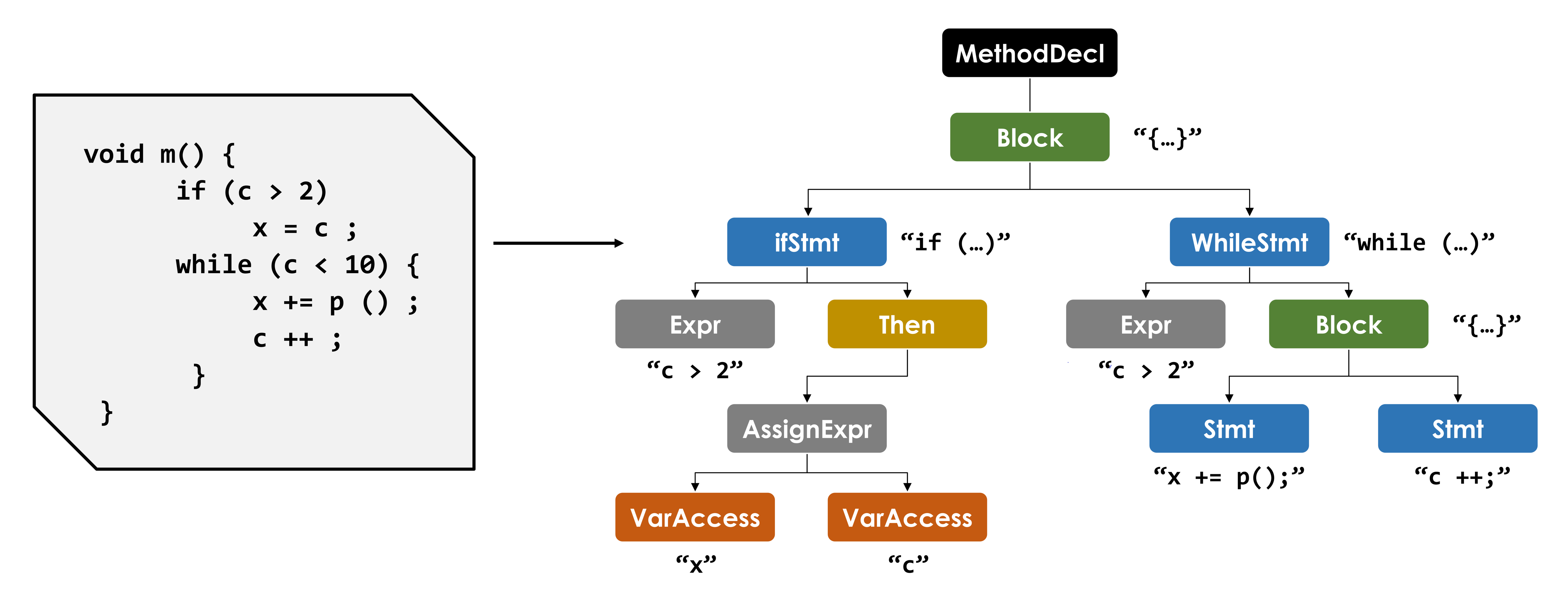}
    \captionsetup{font=footnotesize}
	\caption{Sample Java method and its Abstract Syntax Tree (AST).}%
	\label{fig:sota_example_ast}
\end{figure}

\smallbreak
\noindent \textbf{Source Code.} It represents the fundamental building block of software, including various units such as functions, file codes, classes, and modules. It reflects the design, structure, and logic of different functionalities and services within the software system. In previous works, source code has been represented in diverse ways, including code metrics, which are handcrafted features derived directly from the source code. Code metrics play a critical role in software defect prediction by quantifying different aspects of the codebase that can impact software quality. These metrics cover a wide range, from static module-level metrics~\citep{mccabe1976complexity} for procedural languages, such as Lines of Code (LOC), Coupling Between Objects (CBO), Lack of Cohesion in Methods (LCOM), and Depth of Inheritance Tree (DIT), which provide insights into module complexity and potential defect-proneness. At the class level~\citep{chidamber1994metrics}, metrics like Number of Methods (NOM), Weighted Methods per Class (WMC), and Response for a Class (RFC) offer indications of class complexity and the potential occurrence of defects. Machine learning algorithms are trained using these metrics to identify patterns and relationships between code quality and defects. Another representation of source code is in the form of Abstract Syntax Trees (AST)~\citep{wang2016automatically}, which capture the syntactic structure of the code by breaking it down into constituent elements such as expressions, statements, functions, classes, and variables (Figure~\ref{fig:sota_example_ast}). Source code has been modeled using the program spectrum~\citep{abreu2009spectrum,cellier2008formal}, which provides execution information from specific perspectives, such as conditional branches or loop-free intra-procedural paths and also as a sequence of tokens~\citep{nessa2009fault,you2012statistical} in many research works, particularly in Software Fault Localization.

\smallbreak
\noindent \textbf{Topology (Environment Features).} Topology refers to the structure, either physical or logical, of the IT environment. It includes information about the components, connections, and spatial relationships within the system. This data source offers valuable insights into the overall architecture, providing details about servers, network devices, databases, etc. Additionally, topology data may include configuration settings, software versions, and other relevant features of the system. The presence of topology is important as it establishes the context necessary for identifying actionable and contextualized patterns within the data. Without the constraints and context provided by the topology, the detected patterns, while valid, may be misleading or distracting~\citep{prasad2018market}. Topology has found extensive use in various areas, such as determining causality and localizing faults~\citep{lin2016idice,DBLP:conf/kbse/RemilBMCK21}, ranking incident~\citep{lin2018predicting}, prediction of incident~\citep{li2020predicting}, as well as enhancing models explainability~\citep{DBLP:conf/dsaa/RemilBPRK21}.

\begin{figure}[t]
	\centering
	\includegraphics[width=0.7\linewidth]{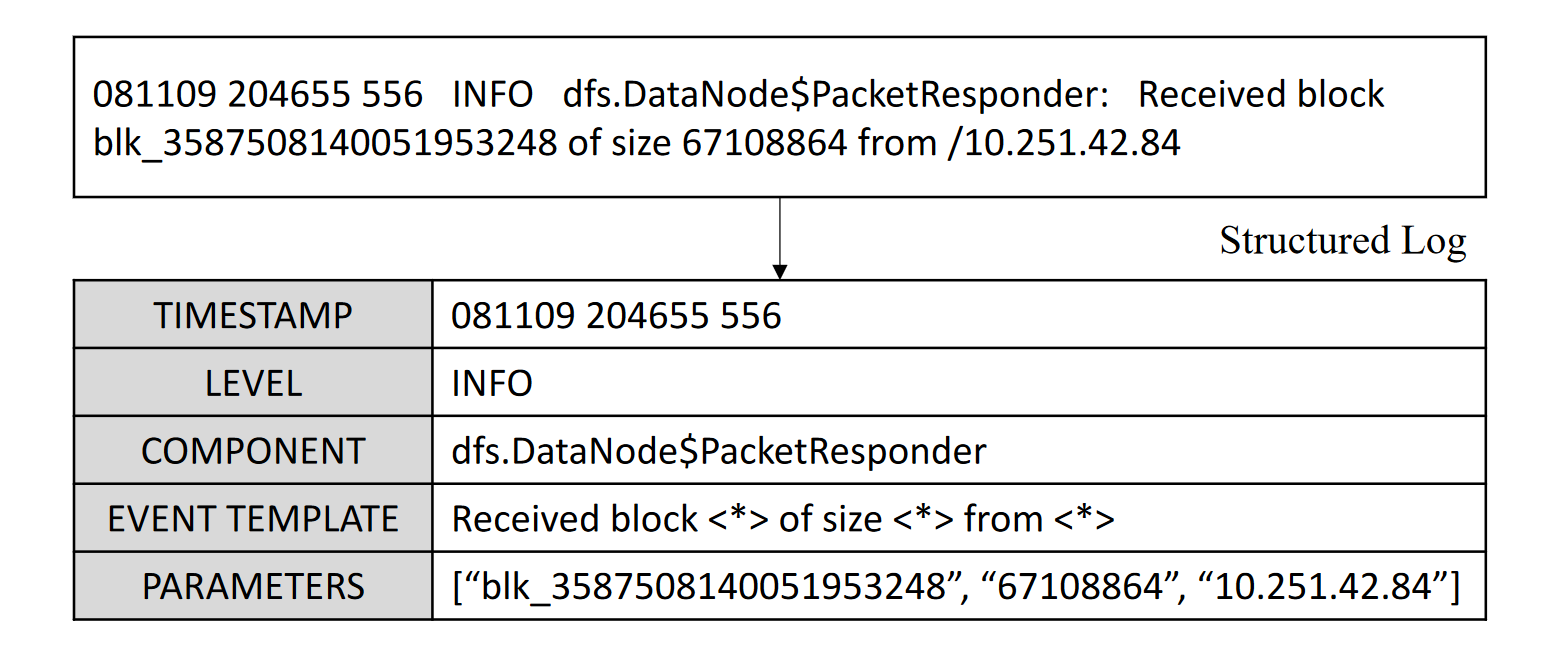}
    \captionsetup{font=footnotesize}
    \caption{Log parsing example~\citep{xu2009detecting}.}%
	\label{fig:sota_example_log2}
\end{figure}

\smallbreak
\noindent \textbf{Event Logs.} Logs consist of human-readable statements generated by software applications, operating systems, or devices to describe events or actions. They serve as valuable records, providing information about system activities, errors, warnings, and other relevant events. Timestamps are typically included in logs, along with details such as the event's source, severity, and description. Analyzing logs is essential for understanding the sequence of events leading to an incident, identifying abnormal behaviors, debugging issues, and ultimately pinpointing the root cause. In general, logs are semi-structured text that is produced by logging statements (e.g., \texttt{printf()}, \texttt{logger.info()}) within the source code. Once logs are collected, they need to be parsed to be utilized in various downstream log mining tasks, such as incident detection. 
    
Parsing log messages is a crucial step in making logs usable for different analytical tasks. This process aims to transform the semi-structured log messages into structured log events by extracting constant parts and variables~\citep{he2021survey}. In the given example depicted in Figure~\ref{fig:sota_example_log2}, a log parsing scenario is presented, showcasing a log message obtained from the Hadoop Distributed File System (HDFS)~\citep{xu2009detecting}. A log message consists of two main components, the message header, and the message content. The message header, which is determined by the logging framework, is relatively straightforward to extract, including details like verbosity levels (e.g., INFO). On the other hand, extracting essential information from the message content proves to be more challenging due to its unstructured nature, primarily consisting of free-form natural language written by developers. Typically, the message content comprises both constants and variables. Constants represent fixed text provided by developers (e.g., the word "Received"), describing a particular system event. On the other hand, variables correspond to the dynamic runtime values of program variables, carrying contextual information. The set of constants forms the event template. 
    
Numerous log parsing techniques have been proposed, including clustering-based approaches (e.g., LKE~\citep{fu2009execution}, LogSig~\citep{tang2011logsig}), heuristic-based methods (e.g., iPLoM~\citep{makanju2011lightweight}, SLCT~\citep{vaarandi2003data}), Evolutionary Algorithms such as MoLFI~\citep{messaoudi2018search}, and Frequent Pattern Mining techniques like Logram~\citep{dai2020logram}. These algorithms are evaluated based on factors such as offline or online parsing mode, coverage, and alignment with domain knowledge. Parsed logs have been used in log mining algorithms with different approaches, including structured features~\citep{xu2009detecting,he2016experience}, log sequences~\citep{du2017deeplog,meng2019loganomaly}, graphs~\citep{nandi2016anomaly} and Finite State Automata (FSA)~\citep{lou2010mining}. For more details, please refer to the work of~\citep{he2016experience,he2021survey}.   

\smallbreak
\noindent \textbf{Key Performance Indicators (KPIs).} These metrics serve as performance indicators for assessing the health status of IT infrastructures and services. They provide quantitative measurements that offer insights into system performance, availability, reliability, and response times. Examples of these metrics include service response time, error rates, disk reads, resource utilization (such as CPU, Swap, and memory consumption), etc. Typically, these metrics are represented as univariate or multivariate time series, which are utilized for various purposes such as detecting abnormal trends~\citep{wang2022identifying,li2018robust}, predicting failures by monitoring certain measures~\citep{li2016being, lu2020making}, estimating the remaining useful lifetime of physical components, or storage capacity~\citep{zheng2017long}, identifying recurrent and unknown performance issues~\citep{lim2014identifying}, and conducting root cause analysis and incident correlations \citep{jeyakumar2019explainit,shan2019diagnosis}.

\smallbreak
\noindent \textbf{Network Traffic.} This type of data involves the analysis of data packet flow within a computer network, including source and destination IP addresses, ports, protocols, and packet sizes. This data source provides valuable insights into communication patterns, network congestion, anomalies, and potential security threats. By analyzing network traffic, it is possible to identify and address network-related issues, pinpoint performance bottlenecks, and detect signs of malicious activities that could lead to significant incidents. Different approaches have been employed to model network traffic data. For instance, in one study~\citep{lakhina2004diagnosing}, SNMP data was utilized to monitor network links and diagnose anomalies in network traffic. The authors treated flow measurements as multivariate time series collected over time, enabling the separation of traffic into normal and anomalous subspaces. In a subsequent work by the same authors~\citep{lakhina2005mining}, static features such as source and target destination addresses or ports were incorporated into the traffic data to detect and diagnose security threats and service outages. Another approach, presented in~\citep{wang2017end}, involved representing network traffic as image-like structures using the IDX file format for encrypted traffic classification. This process involved mapping the network traffic data onto a 2D grid, where each grid cell represented a specific traffic attribute, such as packet size, source IP address, or destination port. The intensity or color of each pixel in the image reflected the value or frequency of the corresponding network traffic attribute at that particular location. Furthermore, network traffic has been represented as graphs by~\citep{bahl2007towards} to localize the sources of performance problems in networks. Probabilistic inference graphs were constructed from the observation of packets exchanged in the network infrastructure. Nodes of the inference graph are divided into root cause nodes (corresponding to internal IP entities), observation nodes (corresponding to clients), and meta-nodes, which model the dependencies between the first two types of nodes. Each node is also associated with a categorical random variable modeling the current state (up, troubled, down), which is influenced by other nodes via the dependency probabilities.

\begin{figure}[t]
	\centering
	\includegraphics[width=0.65\linewidth]{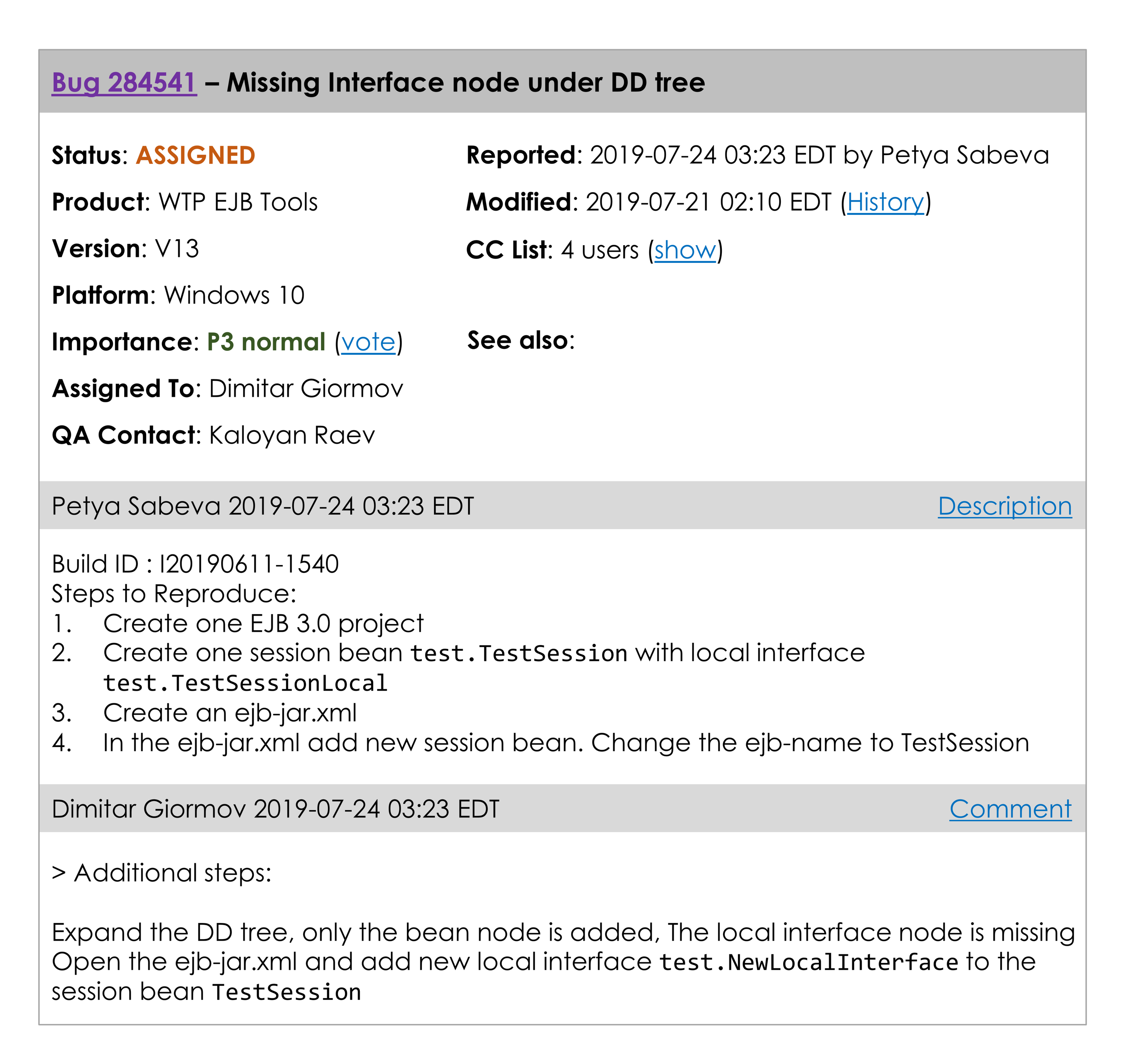}
    \captionsetup{font=footnotesize}
	\caption{Example of an incident report for a bug in Eclipse. This bug is about a missing node of XML files in Product Web Tools Platform~\citep{xuan2014towards}.}%
	\label{fig:sota_example_report}
\end{figure}

\smallbreak
\noindent \textbf{Incident Reports.} Incident reports are valuable sources of information that comprehensively document the details, impact, discussions, and resolution of incidents. These reports are typically initiated by developers, testers, or end-users, capturing crucial information to aid incident management. As illustrated in Figure~\ref{fig:sota_example_report}, incident reports commonly include an identification number and a title, along with contextual features such as the timeline of when the incident was reported and modified, the affected system or service's topology, the severity or importance of the incident, the assigned programmer responsible for resolution, and the incident's resolution status (e.g., new, unconfirmed, resolved). Of utmost importance, incident reports contain a detailed description of the issue, including information on how to reproduce the problem, stack traces (in the case of a bug), and the expected behavior. Additional comments within the report may include discussions about potential solutions, diagnosis and root cause analysis, and actions taken to mitigate the incident. Attachments such as proposed patches, test cases, or screenshots may also be included to provide further context and support. Furthermore, incident reports should offer historical context to leverage past knowledge from similar incidents. Tagging relevant information from previous incidents enables the application of valuable lessons learned in the current resolution processes.

The challenge in processing incident reports lies in the diversity of data types, including structured, semi-structured, and unstructured data. For example, environment characteristics are typically presented in structured or tabular formats \citep{chen2020towards, xuan2014towards, DBLP:conf/kdd/PhamJDOJ20}, while stack traces, problematic SQL queries and user traces fall into the category of semi-structured data \citep{zhang2015survey, DBLP:conf/kdd/PhamJDOJ20}. On the other hand, the description of the problem and the comments section dedicated to the analysis and diagnosis of the problem consist of unstructured natural language text \citep{chen2019continuous, xuan2014towards, DBLP:conf/sigsoft/LeeHLKJ17}, which requires data normalization. Various approaches have been employed to encode these different data types. For instance, both \citet{chen2019continuous} and \citet{DBLP:conf/kdd/PhamJDOJ20} utilized the FastText algorithm \citep{bojanowski2017enriching} for text encoding. Another approach employed in \citep{DBLP:conf/sigsoft/LeeHLKJ17} was the use of Word2Vec \citep{mikolov2013efficient}, which builds pretrained subword vectors based on an external corpus and then fine-tunes them using historical incident data. Contextualization data, on the other hand, was handled using exponential family embeddings in the work of \citet{DBLP:conf/kdd/PhamJDOJ20}. In addition, some researchers explored assignment information in incident reports to create what is referred to as a \textit{Tossing Graph}, aiming to reduce the need for reassigning incidents to other developers \citep{xi2019bug, bhattacharya2010fine, jeong2009improving}.

\begin{figure}[t]
	\centering
	\includegraphics[width=1.0\linewidth]{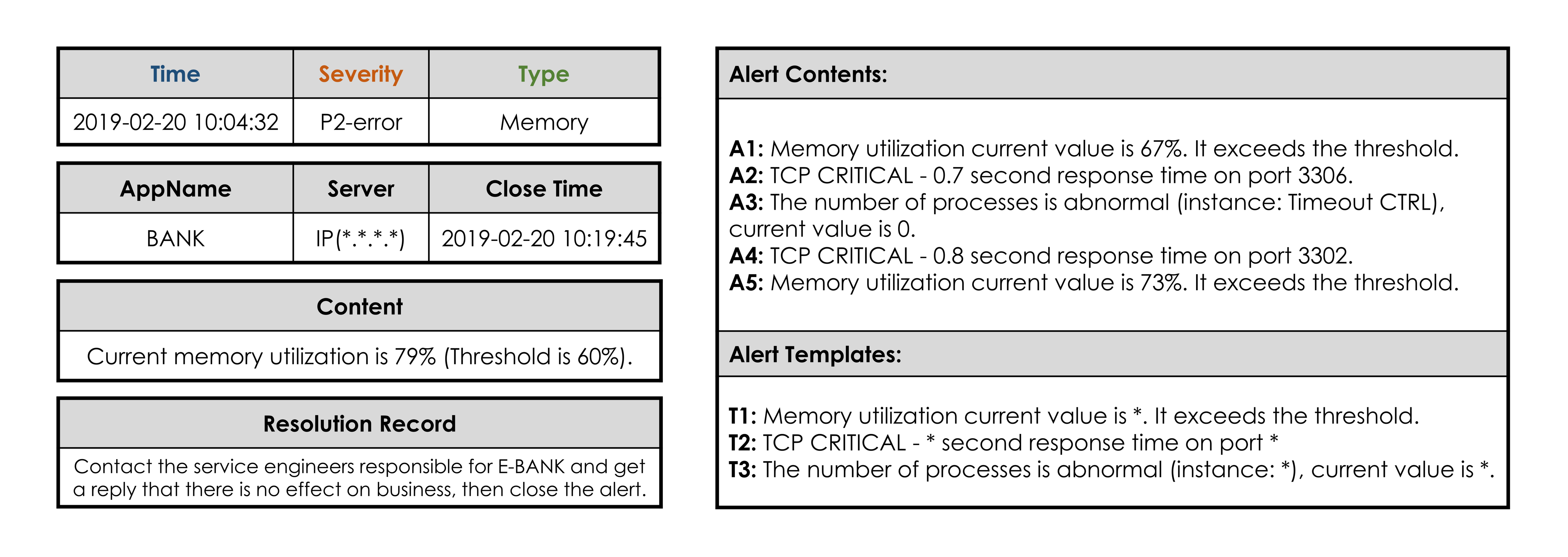}
    \captionsetup{font=footnotesize}
	\caption[Example of an alert on memory consumption and explanation of alert template extraction.]{[Left] Example of an alert on memory consumption from~\citep{zhao2020automatically}. [Right] Explanation of alert template extraction.}%
	\label{fig:sota_example_alert}
\end{figure}

\smallbreak
\noindent \textbf{Alerting Signals.} Alerting signals are notifications generated by monitoring systems or tools when specific thresholds on time series metrics or conditions on event occurrences are violated. These signals serve as indications of abnormal behaviors, potential issues, or breaches of established thresholds. Examples of triggering events include high CPU usage, low disk space, high latency time, or application errors. Alerting signals act as an early warning system, allowing proactive identification and resolution of emerging problems before they escalate into failures. It is important to note that alerting signals are not raw or unprocessed data directly collected from monitored systems, such as KPI metrics or event logs. Instead, they are refined data derived from metrics or events based on a set of predefined rules. For example, Figure~\ref{fig:sota_example_alert} (left) illustrates an example alert generated by the \texttt{AlertRank} Framework \citep{zhao2020automatically}, indicating that the current memory utilization has exceeded the threshold of 79\%, leading to a P2-error severity. This alert is derived from the analysis of memory consumption over time and generated using alert templates within the source code, similar to log events, as shown in Figure~\ref{fig:sota_example_alert} (right). Alerting signals are commonly used to identify and prioritize critical issues within a system \citep{zhao2020automatically}. They can also be utilized for categorizing similar or identical problems into specific categories \citep{zhao2020understanding}, as well as deducing correlations among multiple simultaneous events \citep{mirheidari2013alert}.

\smallbreak
\noindent \textbf{Execution Traces.} In addition to the main data sources mentioned above, which are utilized in the incident management process to develop data-driven approaches for incident detection, diagnosis, triage, and resolution, there are other data sources that are specifically relevant to certain tasks. One such example is code traces, which provide a hierarchical description of the modules and services invoked to fulfill a user request and are employed in the diagnosis task. These traces capture the flow of control, method invocations, input/output data, and interactions with external dependencies. Execution traces are particularly valuable for diagnosing complex or intermittent problems that are challenging to reproduce. Stack traces, for example, are detailed reports that provide information about the executed methods and their associated packages during a crash. They can be obtained through system calls in various programming languages. Stack traces have primarily been utilized in the context of crash deduplication, which involves identifying near-duplicate reports that indicate the same bug or error. Stack traces have been modeled using graphical representations~\citep{DBLP:conf/dsn/KimZN11}., sequence-based approaches~\citep{DBLP:conf/icac/BrodieMLMMWCS05,DBLP:conf/icsm/DhaliwalKZ11}, or vectorization techniques such as n-grams and TF-IDF for information retrieval purposes~\citep{DBLP:conf/csmr/LerchM13,DBLP:conf/qrs/SaborHL17}. Another prominent type of execution trace comprises SQL queries executed to retrieve a service or important data. Parsing SQL queries efficiently is crucial for feeding them into analytical models to extract essential components such as tables, predicates, and projections~\citep{aouiche2006clustering,aligon2014similarity}. This enables their utilization as static features to identify schema issues in data models within databases and improve performance, such as recommending and selecting indexes~\citep{deep2020comprehensive,chaudhuri2002compressing,DBLP:conf/kbse/RemilBMCK21}, detecting anti-patterns~\citep{chen2014detecting}, and identifying insider threats~\citep{kul2016ettu}. Alternatively, some methods treat SQL queries as natural language, employing techniques such as Query2Vec~\citep{jain2018query2vec} to capture their semantic meaning. Heap dumps in Java memory analysis are another type of data that can be considered. They are snapshots of the Java heap memory taken at a specific moment in time. The Java heap is the region of memory where objects are allocated and deallocated during the execution of a Java application. A heap dump captures the complete state of the Java heap, including all objects and their attributes, such as instance variables and references. This provides a detailed view of the memory usage within the Java application. Heap dumps are particularly useful for analyzing memory-related issues, such as memory leaks~\citep{DBLP:conf/icse/JungLRP14,DBLP:journals/tosem/XuR13}. In the related literature, heap dumps are commonly represented as trees~\citep{lee2014detecting}, graphs~\citep{maxwell2010diagnosing}, or hierarchies~\citep{RemilSCA23}.

\subsection{Evaluation Metrics}
\label{subsec:eval_metrics}

\begin{table}[]
    \captionsetup{font=footnotesize}
    \caption{Contingency table.}
    \label{tab:sota_contigency_table}
    \centering
    \scalebox{0.9}{
    \begin{tabular}{c|c|c|c}
    \toprule
                                                                                          & \textbf{True Failure }                                                                   & \textbf{True Non-failure }                                                                   & \textbf{Sum}                                                       \\ \midrule 
    \begin{tabular}[c]{c}\textbf{Prediction: Failure}\\ (Failure Warning)\end{tabular}       & \begin{tabular}[c]{c}True positive (TP)\\ (Correct Warning)\end{tabular}  & \begin{tabular}[c]{c}False positive (FP)\\ (False Warning)\end{tabular}       & \begin{tabular}[c]{c}Positives\\ (POS)\end{tabular} \\ \midrule
    \begin{tabular}[c]{c}\textbf{Prediction: No failure}\\ (No Failure Warning)\end{tabular} & \begin{tabular}[c]{c}False Negative (FN)\\ (Missing Warning)\end{tabular} & \begin{tabular}[c]{c}True Negative (TN)\\ (Correctly no Warning)\end{tabular} & \begin{tabular}[c]{c}Negatives\\ (NEG)\end{tabular} \\ \midrule
    \textbf{Sum}                                                                                   & Failures (F)                                                                    & Non-Failures (NF)                                                                   & Total (N)                                                 \\ \bottomrule
    \end{tabular}}
\end{table}

To comprehensively evaluate the quality and performance of data-driven approaches in incident management tasks, it is crucial to assess them using appropriate metrics, also known as figures of merit. While machine learning metrics are commonly used to evaluate predictive models, it is important to note that relying solely on these metrics, such as contingency metrics, may not accurately reflect the models' performance in real-world scenarios, especially when considering time constraints. For example, in the case of incident prediction, the goal is to predict incidents while minimizing false alarms and maximizing the coverage of actual incidents. However, predicting incidents after the designated prediction period ($\Delta t_p$) is not considered accurate since the incident has already occurred, leading to suboptimal allocation of time and resources. Therefore, we introduce a set of established metrics, focusing primarily on two main tasks: detecting and predicting incidents. It is worth mentioning that these metrics can be adapted for other tasks as well, and several other metrics have been proposed to evaluate triage, diagnosis, and the effectiveness of automated remediation actions, which will be briefly covered. To organize the metrics effectively, we categorize them based on the nature of the model output. 

\begin{figure}[h]
	\centering
	\includegraphics[width=1.05\linewidth]{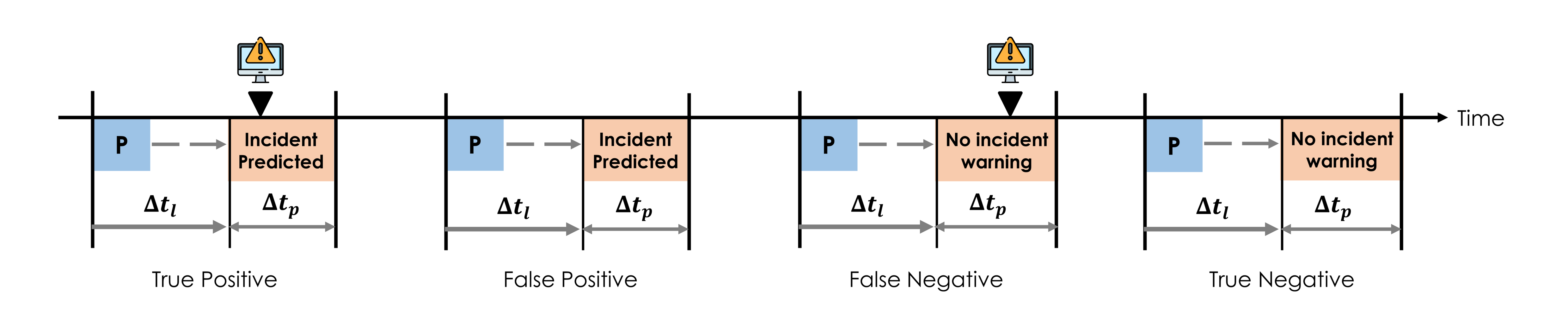}
    \captionsetup{font=footnotesize}
	\caption{A timeline showing true incidents and all four types of predictions TP, FP, FN, TN.}%
	\label{fig:sota_timeline_metrics}
\end{figure}

\begin{table}[]
    \captionsetup{font=footnotesize}
    \caption{Metrics obtained from the contingency table.}
    \label{tab:sota_binary_metrics}
    \centering
    \scalebox{1.0}{
    \begin{tabular}{c|c|c}
    \toprule
    \textbf{Metric}                    & \textbf{Formula}                                                                           & \textbf{Other names}                                                    \\ \midrule 
    Precision                 & $\frac{{\text{TP}}}{{\text{TP} + \text{FP}}}$                                     & Confidence                                                     \\ \midrule
    Recall                    & $\frac{{\text{TP}}}{{\text{TP} + \text{FN}}} $                  & \begin{tabular}[c]{c}Support \\ Sensitivity \\ True positive rate (TPR) \end{tabular} \\ \midrule
    False positive rate (FPR)      & $\frac{{\text{FP}}}{{\text{FP} + \text{TN}}} $                                  & Fall-out                                                       \\ \midrule
    Specificity               & $\frac{{\text{TN}}}{{\text{TN} + \text{FP}}}$                                     & True negative rate (TNR)                                            \\ \midrule
    False negative rate (FNR)      & $\frac{{\text{FN}}}{{\text{FN} + \text{TP}}}$                                     & 1 - recall                                                     \\ \midrule
    Negative predictive value & $ \frac{{\text{TN}}}{{\text{TN} + \text{FN}}}$                                    &                                                                \\ \midrule
    False positive error rate & $ \frac{{\text{FP}}}{{\text{FP} + \text{TP}}}$                                    & 1 - precision                                                  \\ \midrule
    Accuracy                  & $\frac{{\text{TP} + \text{TN}}}{{\text{TP} + \text{TN} + \text{FP} + \text{FN}}}$ &                                                                \\ \midrule
    Odds ratio                & $\frac{{\text{TP} \times \text{TN}}}{{\text{FP} \times \text{FN}}}$               &                                                                \\ \bottomrule
    \end{tabular}}
\end{table}
\smallbreak
\noindent \textbf{Classification Metrics.} Classification metrics are commonly derived from four cases, as shown in Table~\ref{tab:sota_contigency_table}. A prediction is classified as a true positive if an incident occurs within the prediction period and a warning is raised. Conversely, if no incident occurs but a warning is given, the prediction is considered a false positive. If the algorithm fails to predict a true incident, it is categorized as a false negative. Finally, if no true incident occurs and no incident warning is raised, the prediction is labeled as a true negative. To compute metrics like precision and recall, the contingency table is populated with the number of true positives (TP), false positives (FP), false negatives (FN), and true negatives (TN). The prediction algorithm is applied to test data that was not used to determine the parameters of the prediction method. This allows the comparison of prediction outcomes against the actual occurrence of incidents. The four possible cases are illustrated in Figure~\ref{fig:sota_timeline_metrics}. It's worth noting that the prediction period ($\Delta t_p$) is instrumental in determining whether an incident is counted as predicted or not. Therefore, the choice of $\Delta t_p$ also has implications for the contingency table and should align with the requirements of subsequent steps in the incident management process.

The metrics presented in Table~\ref{tab:sota_binary_metrics} are derived from the contingency table (see Table~\ref{tab:sota_contigency_table}). They are commonly used in pairs, such as precision/recall, true positive rate/false positive rate, sensitivity/specificity, and positive predictive value/negative predictive value. Different research areas may use different names for the same metrics, hence, the leftmost column indicates the commonly used terminology, while the rightmost column lists alternative names. It's important to note that improving precision (reducing false positives) often results in a decrease in recall (increasing false negatives) and vice versa. To balance the trade-off between precision and recall, the F-Measure is used as the harmonic mean of the two, assuming equal weighting. One limitation of precision and recall is that they don't consider true negative predictions. Therefore, it is necessary to consider other metrics in combination with precision and recall. The false positive rate is the ratio of incorrectly predicted incidents to the total number of non-incidents. A lower false positive rate is desirable, provided that the other metrics do not deteriorate.

Accuracy appears to be an appropriate metric for incident prediction due to the rarity of incidents. Achieving high accuracy by always classifying the system as non-faulty may be misleading because it fails to capture any incidents, resulting in a recall of zero. However, it is important to consider true negatives when assessing incident prediction techniques. Let's consider an example from~\citep{salfner2010survey}. Two prediction methods perform equally well in terms of true positives (TP), false positives (FP), and false negatives (FN), resulting in the same precision and recall. However, one method makes ten times more predictions than the other because it operates on more frequent measurements. The difference between these methods is reflected only in the number of true negatives (TN), which becomes apparent in metrics that include TN. True negatives are counted by considering predictions made when no incident was imminent and no warning was issued.

It is noteworthy that the quality of predictions does not depend only on algorithms but also on factors such as the data window size ($\Delta t_d$), lead-time ($\Delta t_l$), and prediction period ($\Delta t_p$). Predicting incidents at an exact point in time is highly unlikely, so predictions are typically made within a specific time interval (prediction period). The number of true positives is influenced by ($\Delta t_p$): a longer prediction period captures more incidents, increasing the number of true positives and impacting metrics like recall. Incident predictors often utilize an adjustable decision threshold. When the threshold is set low, incident warnings are raised easily, increasing the chances of capturing true incidents (resulting in high recall). However, a low threshold also leads to many false alarms, resulting in low precision. Conversely, if the threshold is set very high, the situation is reversed. To visualize this trade-off, precision/recall curves are used, plotting precision over recall for various threshold levels. Similarly, the receiver operating characteristic (ROC) curve plots the true positive rate versus the false positive rate (sensitivity/recall versus 1-specificity, respectively). This curve assesses the model's ability to distinguish between incidents and non-incidents. The closer the curve is to the upper-left corner of the ROC space, the more accurate the model is. The Area Under the Curve (AUC) is defined as the area between the ROC curve and the x-axis and measures the probability that a data point from an incident-prone situation receives a higher score than a data point from a non-incident-prone situation. By summarizing the capacity of a prediction algorithm to discriminate between incidents and non-incidents, the AUC converts the ROC curve into a single number. A random predictor has an AUC of 0.5, while a perfect predictor achieves an AUC of 1.

\begin{table}[]
    \captionsetup{font=footnotesize}
    \caption{Metrics used for regression tasks in incident management tasks.}
    \label{tab:sota_regressor_tasks}
    \centering
    \scalebox{0.95}{
    \begin{tabular}{c|c}
    \toprule
    \textbf{Metric}                                & \textbf{Formula}                                                                              \\ \midrule 
    Mean Absolute Error (MAE)             & $\frac{1}{n} \sum_{i=1}^{n} |\hat{y}_i - y_i|$                                     \\ \midrule
    Root Mean Squared Error (RMSE)        & $\sqrt{\frac{1}{n} \sum_{i=1}^{n} (\hat{y}_i - y_i)^2}$                              \\ \midrule
    Mean Absolute Percentage Error (MAPE) & $\frac{1}{n} \sum_{i=1}^{n} \left| \frac{\hat{y}_i - y_i}{y_i} \right| \times 100\%$ \\ \midrule
    R-squared (R2)                        & $ 1 - \frac{\sum_{i=1}^{n} (y_i - \hat{y}_i)^2}{\sum_{i=1}^{n} (y_i - \bar{y})^2}$   \\ \midrule
    Explained Variance Score              & $ 1 - \frac{\text{Var}(y - \hat{y})}{\text{Var}(y)}$                                 \\ \bottomrule
    \end{tabular}}
\end{table}

\smallbreak
\noindent \textbf{Regression Metrics.} Regression metrics are commonly used in tasks such as remaining useful lifetime estimation or anomaly detection, particularly when applied to time series metric data to identify outliers. Unlike classification metrics, the metrics used for regressors remain consistent across conventional machine learning models, as shown in Table~\ref{tab:sota_regressor_tasks}. Mean Absolute Error (MAE) is a metric that represents the average absolute difference between the predicted values $\hat{y}_i$ and the actual values $y_i$. It provides a measure of the average magnitude of errors. Several works have utilized MAE for regression tasks~\citep{ordonez2019hybrid}. Root Mean Squared Error (RMSE) is similar to MAE, but it takes the square root of the average squared differences between the predicted values and the actual values. RMSE penalizes larger errors more significantly. It has been used, for example, in anomaly detection~\citep{lee2017anomaly}. Mean Absolute Percentage Error (MAPE) measures the average percentage difference between the predicted values and the actual values. It is particularly useful when evaluating the accuracy of predictions relative to the scale of the target variable~\citep{li2015prediction}. R-squared (R2) is a metric used to represent the proportion of variance in the target variable that is explained by the predicted values. R2 ranges from 0 to 1, where a value of 1 indicates a perfect fit and a value of 0 indicates no improvement over a naive baseline~\citep{siami2008sufficient}. Finally, the explained variance score, similar to R2, measures the proportion of variance in the target variable that is explained by the predicted values. However, it is based on the variance of the residuals and can be used as an alternative metric for evaluation~\citep{siami2008sufficient}.

\smallbreak
\noindent \textbf{Other Metrics.} In addition to the commonly used regression and classification metrics, there exists a set of specialized metrics that are particularly relevant for evaluating the effectiveness and performance of specific incident management tasks, such as fault localization, incident correlation, and incident deduplication. These metrics are often specifically designed for incident management purposes and may have limited applicability outside this domain. Some of these metrics have been developed in alignment with specific research studies, while others are unique to the field of incident management. For example, \citet{wang2013network} have introduced the metric of "normality" for detecting anomalies within a network traffic flow sequence.

In fault localization research, several specific metrics are used to evaluate the effectiveness of different techniques. One such metric is T-Score, which estimates the percentage of code that a programmer can ignore before identifying the first faulty location in the program~\citep{renieres2003fault,liu2006statistical}. Another metric commonly used is EXAM (Expense metric), which measures the percentage of program statements that need to be examined before encountering the first faulty statement~\citep{jones2005empirical,ju2014hsfal}. In addition to these metrics, other research work have also employed the Wilcoxon signed-rank test \citep{wilcoxon1992individual,wong2013dstar} as a statistical evaluation method. This test serves as an alternative to the paired Student's t-test when the assumption of a normal distribution in the population cannot be made to compare the effectiveness of two techniques, denoted as $\alpha$ and $\beta$. The test examines the one-tailed alternative hypothesis that $\beta$ requires the examination of an equal or greater number of statements compared to $\alpha$.

In software defect prediction, there are cases where it is more useful to evaluate the classes based on their predicted number of defects in a ranking manner. One approach to assess the performance of the prediction model is by calculating Spearman's correlation coefficient \citep{spearman1961proof}, as demonstrated in the study conducted by~\citep{d2010extensive}. Another method used in this context is the cumulative lift chart, which compares the performance of two different models or strategies by plotting the cumulative gain against the number of cases. This approach has been also employed by~\citep{d2012evaluating}.

In intrusion detection systems within practical network settings,~\citet{mirheidari2013alert} conducted a comprehensive comparison of alert correlation algorithms. The study aimed to evaluate the performance of these algorithms using both quantitative and qualitative measures. The quantitative assessment focused on accuracy, while the qualitative evaluation delved into additional aspects such as extendibility and flexibility. These aspects refer to the algorithm's adaptability, localizability, and capacity to adjust to new conditions. Furthermore, the evaluation considered the algorithm's ability to parallelize tasks and the associated memory requirements. This evaluation aligns with the desirable attributes emphasized in Section~\ref{subsec:desiderata}, which outlines the desired characteristics for AIOps solutions in incident management.

In the context of incident triage and deduplication approaches, specific performance metrics have been adopted to improve the evaluation of these processes. These metrics include Mean-reciprocal rank~\citep{Craswell2009,RemilDeepLSH23}, Recall rate of order k~\citep{DBLP:journals/ftir/Sanderson10,DBLP:conf/sigsoft/VasilievKCKLP20,DBLP:conf/msr/KhvorovVCRKP21}, and Average Hit Ratio~\citep{akila2014bug}. It is noteworthy that there are some metrics specifically designed to assess the process, rather than a particular algorithm or data-driven approach. While we previously discussed metrics such as Mean Time To Repair (MTTR), Mean Time To Engage (MTTE), and Mean Time to Detect (MTTD), other research works~\citep{chen2020towards,prasad2018market} have introduced alternative terminology. For example, MTTR may be referred to as Mean Time to Repair, and MTTD as Mean Time to Detect. Furthermore, there are additional valuable measures that have been considered, such as Mean Time Between Failures (MTBF)~\citep{braglia2012data}, which evaluates the average duration between consecutive incidents or failures to assess system reliability.

\section{A Comprehensive Review of AIOps-based Data-Driven Approaches}
\label{sec:review}

In the following sections, we present a review of research efforts centered around AIOps-driven data-centric approaches tailored specifically for incident management. We begin with an insightful overview of related and analogous studies, serving as an invaluable compass for those keenly interested in specific incident phases or application domains. We also review noteworthy surveys that offer targeted insights into particular facets of the field. Subsequently, we delve into an in-depth examination of the pioneering techniques that have found their place within the incident management literature. Each facet of incident management tasks is described based on our taxonomy through summaries and evaluations featured in dedicated tables. These tables serve as efficient reference points, providing readers with a clear and concise overview of each task. Furthermore, we assess the extent to which the reviewed papers meet the requisite criteria.

\subsection{Prior Research Efforts in AIOps and Incident Management}

\begin{table}[h]

\caption{Summary and categorization of Incident Management related papers}
\label{tab:aiops_papers}
\centering
\renewcommand{\arraystretch}{1.3}
\resizebox{\textwidth}{!}{%
\begin{tabular}{ccccc*{9}{c}*{4}{c}*{3}{c}}\toprule
\multirow{6}{*}{\textbf{Ref.}} & \multirow{6}{*}{\textbf{Year}} & \multirow{6}{*}{\textbf{\begin{tabular}[c]{c}Study \\ Approach\end{tabular}}} & \multirow{6}{*}{\textbf{AIOps}} &\multirow{6}{*}{{\textbf{\begin{tabular}[c]{c}Classif.\\ Taxonomy\end{tabular}}}} & \multicolumn{9}{c}{\textbf{Focus Area}}                                                                                                                                                                                                                                                                                                                                         & \multicolumn{4}{c}{\textbf{Findings}}                                                                                                             & \multicolumn{3}{c}{\textbf{Practical Adoption}}                                                                                                                                                                                                                            \\ \cmidrule(l){6-14} \cmidrule(l){15-18} \cmidrule(l){19-21}
      &                     &                     &         &                                                                                    & \multicolumn{1}{c}{\rotatebox[origin=c]{90}{\textbf{Detection}}} & \multicolumn{1}{c}{\rotatebox[origin=c]{90}{\textbf{Prediction}}} & \multicolumn{1}{c}{\rotatebox[origin=c]{90}{\textbf{Ranking}}} & \multicolumn{1}{c}{\rotatebox[origin=c]{90}{\textbf{Assignation}}} & \multicolumn{1}{c}{\rotatebox[origin=c]{90}{\textbf{Classification}}} & \multicolumn{1}{c}{\rotatebox[origin=c]{90}{\textbf{Deduplication}}} & \multicolumn{1}{c}{\rotatebox[origin=c]{90}{\textbf{Root Cause}}} & \multicolumn{1}{c}{\rotatebox[origin=c]{90}{\textbf{Correlation}}} & \rotatebox[origin=c]{90}{\textbf{Mitigation}} & \multicolumn{1}{c}{\rotatebox[origin=c]{90}{\textbf{Trends}}} & \multicolumn{1}{c}{\rotatebox[origin=c]{90}{\textbf{Capabilities}}} & \multicolumn{1}{c}{\rotatebox[origin=c]{90}{\textbf{Challenges}}} & \rotatebox[origin=c]{90}{\textbf{Desiderata}} & \multicolumn{1}{c}{\rotatebox[origin=c]{90}{\textbf{\begin{tabular}[c]{c}Implementation \\ Details\end{tabular}}}} & \multicolumn{1}{c}{\rotatebox[origin=c]{90}{\textbf{\begin{tabular}[c]{c}Specialized\\ Techniques\end{tabular}}}} & \rotatebox[origin=c]{90}{\textbf{\begin{tabular}[c]{c}Research \\ Innovation\end{tabular}}} \\ \midrule

\rowcolor{gray!15} \citep{lou2013software}      &  2013 &      Experience  &                                 &               $\bullet$                                                                              & \multicolumn{1}{c}{$\bullet$}                   & \multicolumn{1}{c}{}                    & \multicolumn{1}{c}{$\bullet$}                 & \multicolumn{1}{c}{$\bullet$}                     & \multicolumn{1}{c}{}                        & \multicolumn{1}{c}{$\bullet$}                       & \multicolumn{1}{c}{$\bullet$}                    & \multicolumn{1}{c}{$\bullet$}                     &       $\bullet$              & \multicolumn{1}{c}{}                & \multicolumn{1}{c}{}                      & \multicolumn{1}{c}{$\bullet$}                    &       $\bullet$              & \multicolumn{1}{c}{$\bullet$}                                                                           & \multicolumn{1}{c}{$\bullet$}                                                                          &                                                                         \\ 
\citep{zhang2015survey}      &  2015 &  Survey    &                                              &                    $\bullet$                                                                         & \multicolumn{1}{c}{}                   & \multicolumn{1}{c}{}                    & \multicolumn{1}{c}{$\bullet$}                 & \multicolumn{1}{c}{$\bullet$}                     & \multicolumn{1}{c}{$\bullet$}                        & \multicolumn{1}{c}{$\bullet$}                       & \multicolumn{1}{c}{$\bullet$}                    & \multicolumn{1}{c}{}                     &         $\bullet$            & \multicolumn{1}{c}{}                & \multicolumn{1}{c}{}                      & \multicolumn{1}{c}{}                    &                     & \multicolumn{1}{c}{}                                                                           & \multicolumn{1}{c}{$\bullet$}                                                                          &   $\bullet$   \\

\rowcolor{gray!15} \citep{zhang2016literature}      & 2016 &    Review   &                                          &                    $\bullet$                                                                         & \multicolumn{1}{c}{}                   & \multicolumn{1}{c}{}                    & \multicolumn{1}{c}{$\bullet$}                 & \multicolumn{1}{c}{$\bullet$}                     & \multicolumn{1}{c}{$\bullet$}                        & \multicolumn{1}{c}{$\bullet$}                       & \multicolumn{1}{c}{$\bullet$}                    & \multicolumn{1}{c}{}                     &         $\bullet$            & \multicolumn{1}{c}{}                & \multicolumn{1}{c}{}                      & \multicolumn{1}{c}{$\bullet$}                    &                     & \multicolumn{1}{c}{}                                                                           & \multicolumn{1}{c}{$\bullet$}                                                                          &   $\bullet$   \\

\citep{li2017data}     &  2017 &   Survey        &    & $\bullet$                                                                                                    & \multicolumn{1}{c}{$\bullet$}                   & \multicolumn{1}{c}{$\bullet$}                    & \multicolumn{1}{c}{}                 & \multicolumn{1}{c}{}                     & \multicolumn{1}{c}{$\bullet$}                        & \multicolumn{1}{c}{}                       & \multicolumn{1}{c}{$\bullet$}                    & \multicolumn{1}{c}{}                     &                    & \multicolumn{1}{c}{}                & \multicolumn{1}{c}{}                      & \multicolumn{1}{c}{}                    &                     & \multicolumn{1}{c}{$\bullet$}                                                                           & \multicolumn{1}{c}{$\bullet$}                                                                          &      \\

\rowcolor{gray!15}\citep{mukwevho2018toward}     &      2018        &                   Review    &           &        $\bullet$                                                                                      & \multicolumn{1}{c}{$\bullet$}                   & \multicolumn{1}{c}{$\bullet$}                    & \multicolumn{1}{c}{}                 & \multicolumn{1}{c}{}                     & \multicolumn{1}{c}{}                        & \multicolumn{1}{c}{}                       & \multicolumn{1}{c}{}                    & \multicolumn{1}{c}{}                     &   $\bullet$                 & \multicolumn{1}{c}{}                & \multicolumn{1}{c}{}                      & \multicolumn{1}{c}{$\bullet$}                    &    $\bullet$                & \multicolumn{1}{c}{$\bullet$}                                                                           & \multicolumn{1}{c}{$\bullet$}                                                                          &   $\bullet$  \\

 \citep{prasad2018market}     &  2018 & Report          &   $\bullet$                                        &     $\bullet$                                                                                        & \multicolumn{1}{c}{$\bullet$ }                   & \multicolumn{1}{c}{$\bullet$ }                    & \multicolumn{1}{c}{}                 & \multicolumn{1}{c}{}                     & \multicolumn{1}{c}{$\bullet$ }                        & \multicolumn{1}{c}{}                       & \multicolumn{1}{c}{$\bullet$ }                    & \multicolumn{1}{c}{$\bullet$ }                     &      $\bullet$               & \multicolumn{1}{c}{$\bullet$ }                & \multicolumn{1}{c}{$\bullet$ }                      & \multicolumn{1}{c}{$\bullet$ }                    &       $\bullet$               & \multicolumn{1}{c}{$\bullet$ }                                                                           & \multicolumn{1}{c}{$\bullet$ }                                                                          &      \\

\rowcolor{gray!15} \citep{dang2019aiops}     &      2019      &             Experience       &   $\bullet$         &                                                                                             & \multicolumn{1}{c}{}                   & \multicolumn{1}{c}{}                    & \multicolumn{1}{c}{}                 & \multicolumn{1}{c}{}                     & \multicolumn{1}{c}{}                        & \multicolumn{1}{c}{}                       & \multicolumn{1}{c}{}                    & \multicolumn{1}{c}{}                     &                    & \multicolumn{1}{c}{}                & \multicolumn{1}{c}{$\bullet$}                      & \multicolumn{1}{c}{$\bullet$}                    &    $\bullet$                 & \multicolumn{1}{c}{}                                                                           & \multicolumn{1}{c}{}                                                                          &   $\bullet$   \\

 \citep{chen2020towards}     &   2020  &  Experience  &   $\bullet$                            &       $\bullet$                                                                                      & \multicolumn{1}{c}{$\bullet$}                   & \multicolumn{1}{c}{}                    & \multicolumn{1}{c}{$\bullet$}                 & \multicolumn{1}{c}{$\bullet$}                     & \multicolumn{1}{c}{}                        & \multicolumn{1}{c}{}                       & \multicolumn{1}{c}{$\bullet$ }                    & \multicolumn{1}{c}{$\bullet$ }                     &       $\bullet$              & \multicolumn{1}{c}{}                & \multicolumn{1}{c}{$\bullet$ }                      & \multicolumn{1}{c}{$\bullet$ }                    &           $\bullet$           & \multicolumn{1}{c}{$\bullet$ }                                                                           & \multicolumn{1}{c}{$\bullet$ }                                                                          &   $\bullet$   \\
     
\rowcolor{gray!15} \citep{shen2020evolving}     &   2020     &  Experience  &   $\bullet$    &                                                                                              & \multicolumn{1}{c}{$\bullet$}                   & \multicolumn{1}{c}{}                    & \multicolumn{1}{c}{}                 & \multicolumn{1}{c}{}                     & \multicolumn{1}{c}{}                        & \multicolumn{1}{c}{}                       & \multicolumn{1}{c}{$\bullet$}                    & \multicolumn{1}{c}{$\bullet$}                     &           $\bullet$         & \multicolumn{1}{c}{}                & \multicolumn{1}{c}{$\bullet$}                      & \multicolumn{1}{c}{}                    &                     & \multicolumn{1}{c}{$\bullet$}                                                                           & \multicolumn{1}{c}{$\bullet$}                                                                          &      \\
     
 \citep{bogatinovski2021artificial}     &     2021   &      Survey  &   $\bullet$    &     $\bullet$                                                                                        & \multicolumn{1}{c}{$\bullet$}                   & \multicolumn{1}{c}{}                    & \multicolumn{1}{c}{}                 & \multicolumn{1}{c}{}                     & \multicolumn{1}{c}{}                        & \multicolumn{1}{c}{}                       & \multicolumn{1}{c}{$\bullet$}                    & \multicolumn{1}{c}{}                     &                    & \multicolumn{1}{c}{}                & \multicolumn{1}{c}{$\bullet$}                      & \multicolumn{1}{c}{$\bullet$}                    &                     & \multicolumn{1}{c}{}                                                                           & \multicolumn{1}{c}{}                                                                          &      \\

\rowcolor{gray!15} \citep{reiter2021aiops}     &      2021      &        Review  &   $\bullet$                    &   $\bullet$                                                                                          & \multicolumn{1}{c}{$\bullet$}                   & \multicolumn{1}{c}{$\bullet$}                    & \multicolumn{1}{c}{}                 & \multicolumn{1}{c}{}                     & \multicolumn{1}{c}{}                        & \multicolumn{1}{c}{}                       & \multicolumn{1}{c}{$\bullet$}                    & \multicolumn{1}{c}{}                     &        $\bullet$            & \multicolumn{1}{c}{$\bullet$}                & \multicolumn{1}{c}{$\bullet$}                      & \multicolumn{1}{c}{$\bullet$}                    &         $\bullet$            & \multicolumn{1}{c}{$\bullet$}                                                                           & \multicolumn{1}{c}{$\bullet$}                                                                          &  $\bullet$    \\

 \citep{notaro2021survey}     &     2021       &          Survey    &   $\bullet$                &    $\bullet$                                                                                         & \multicolumn{1}{c}{$\bullet$ }                   & \multicolumn{1}{c}{$\bullet$ }                    & \multicolumn{1}{c}{}                 & \multicolumn{1}{c}{}                     & \multicolumn{1}{c}{$\bullet$}                        & \multicolumn{1}{c}{}                       & \multicolumn{1}{c}{$\bullet$ }                    & \multicolumn{1}{c}{$\bullet$ }                     &     $\bullet$                & \multicolumn{1}{c}{$\bullet$ }                & \multicolumn{1}{c}{$\bullet$ }                      & \multicolumn{1}{c}{}                    &                     & \multicolumn{1}{c}{}                                                                           & \multicolumn{1}{c}{$\bullet$ }                                                                          &      \\    

\rowcolor{gray!15} \citep{lyu2021towards}     &    2021        &        Case Study    &   $\bullet$              &                                                                                             & \multicolumn{1}{c}{}                   & \multicolumn{1}{c}{}                    & \multicolumn{1}{c}{}                 & \multicolumn{1}{c}{}                     & \multicolumn{1}{c}{}                        & \multicolumn{1}{c}{}                       & \multicolumn{1}{c}{}                    & \multicolumn{1}{c}{}                     &                    & \multicolumn{1}{c}{}                & \multicolumn{1}{c}{}                      & \multicolumn{1}{c}{$\bullet$}                    &       $\bullet$              & \multicolumn{1}{c}{$\bullet$}                                                                           & \multicolumn{1}{c}{$\bullet$}                                                                          &  $\bullet$          \\ 
 \citep{rijal2022aiops}     &       2022     &         Review    &   $\bullet$            &                                                                                             & \multicolumn{1}{c}{}                   & \multicolumn{1}{c}{}                    & \multicolumn{1}{c}{}                 & \multicolumn{1}{c}{}                     & \multicolumn{1}{c}{}                        & \multicolumn{1}{c}{}                       & \multicolumn{1}{c}{}                    & \multicolumn{1}{c}{}                     &                    & \multicolumn{1}{c}{$\bullet$ }                & \multicolumn{1}{c}{$\bullet$}                      & \multicolumn{1}{c}{$\bullet$}                    &                     & \multicolumn{1}{c}{}                                                                           & \multicolumn{1}{c}{}                                                                          &      \\

\rowcolor{gray!15} Our Survey     &       2023     &         Review     &   $\bullet$           &                                                                            $\bullet$                 & \multicolumn{1}{c}{$\bullet$}                   & \multicolumn{1}{c}{$\bullet$}                    & \multicolumn{1}{c}{$\bullet$}                 & \multicolumn{1}{c}{$\bullet$}                     & \multicolumn{1}{c}{$\bullet$}                        & \multicolumn{1}{c}{$\bullet$}                       & \multicolumn{1}{c}{$\bullet$}                    & \multicolumn{1}{c}{$\bullet$}                     &   $\bullet$                 & \multicolumn{1}{c}{$\bullet$}                & \multicolumn{1}{c}{$\bullet$}                      & \multicolumn{1}{c}{$\bullet$}                    &              $\bullet$       & \multicolumn{1}{c}{$\bullet$}                                                                           & \multicolumn{1}{c}{$\bullet$}                                                                          & $\bullet$
\\
\bottomrule
\end{tabular}}
\end{table}
\begin{table}[]
\caption{Summary and categorization of specific data-driven surveys and reviews with a focus on particular Incident Management phases}
\label{tab:related_surveys}
\resizebox{\textwidth}{!}{%
\begin{tabular}{cc>{\centering\arraybackslash}m{2.5cm}ccccccccccccl>{\centering\arraybackslash}m{3cm}}
\toprule
\multirow{6}{*}{\textbf{Ref.}} & \multirow{6}{*}{\textbf{Year}} & \multirow{6}{*}{\textbf{\begin{tabular}[c]{@{}c@{}}Focus\\ and Scoop\\ Area\end{tabular}}}                                                                  & \multicolumn{4}{c}{\textbf{Target Area}}                                            & \multicolumn{8}{c}{\textbf{Data sources}}                                                                                                                             & \multicolumn{1}{c}{\multirow{6}{*}{\textbf{Type of AI Techniques}}}                                                                                                                                                                     & \multirow{6}{*}{\textbf{Evaluation Metrics}}                                           \\ \cmidrule(lr){4-7} \cmidrule(lr){8-15}
                               &        &                                                                                                 & \multicolumn{1}{c}{\rotatebox[origin=c]{90}{\textbf{Technical}}} &  \multicolumn{1}{c}{\rotatebox[origin=c]{90}{\textbf{Application}}} &  \multicolumn{1}{c}{\rotatebox[origin=c]{90}{\textbf{Functional}}} &  \multicolumn{1}{c}{\rotatebox[origin=c]{90}{\textbf{Business}}} & \multicolumn{1}{c}{\rotatebox[origin=c]{90}{\textbf{Source Code}}} &  \multicolumn{1}{c}{\rotatebox[origin=c]{90}{\textbf{Topology}}} &  \multicolumn{1}{c}{\rotatebox[origin=c]{90}{\textbf{Event Logs}}} &  \multicolumn{1}{c}{\rotatebox[origin=c]{90}{\textbf{KPIs/SLOs}}} &  \multicolumn{1}{c}{\rotatebox[origin=c]{90}{\textbf{Traffic Network}}} &  \multicolumn{1}{c}{\rotatebox[origin=c]{90}{\textbf{Reports}}} &  \multicolumn{1}{c}{\rotatebox[origin=c]{90}{\textbf{Alerts}}} & \rotatebox[origin=c]{90}{\textbf{Traces}} & \multicolumn{1}{c}{}                                                                                                                                                                                                                    &                                                                                        \\ \midrule
                              \rowcolor{gray!15} \citep{chandola2009anomaly} & 2009                           & Cyber Intrusion Detection & $\bullet$             &                      &                     &                   &                      & $\bullet$            &                     & $\bullet$             & $\bullet$                   &                  &                 &                 & \begin{tabular}[c]{@{}l@{}}Statistical Models\\ Conventional ML Classifiers\\ Clustering\end{tabular}                                                                                                                                   & N/A            \\
                              
                               \citep{wang2013network} & 2013                           & Network Anomaly Detection                                                                       & $\bullet$             &                      &                     &                   &                      &                   &                     &                    & $\bullet$                   &                  &                 &                 & \begin{tabular}[c]{@{}l@{}}Statistical Models\\ Conventional ML Classifiers\\ Clustering\end{tabular}                                                                                                                                   & Domain-specific metrics                                                                \\
                               \rowcolor{gray!15} \citep{ibidunmoye2015performance} & 2015                           & Bottleneck Identification                                                                       & $\bullet$             & $\bullet$               &                     &                   &                      & $\bullet$            &                     & $\bullet$             & $\bullet$                   &                  &                 &                 & \begin{tabular}[c]{@{}l@{}}Statistical Models\\ Conventional ML Classifiers\\ Bayesian Models\\ Conventional ML Regressors\\ Rule-based Models\\ Clustering\end{tabular}                                                                & Contingency metrics                                                                                                                          \\
                               \citep{ibidunmoye2015performance} & 2019                           & IoT Systems Anomaly Detection                                                                   & $\bullet$             & $\bullet$               &                     &                   &                      &                   &                     & $\bullet$             & $\bullet$                   &                  &                 &                 & \begin{tabular}[c]{@{}l@{}}Statistical Models\\ Conventional ML Classifiers\\ Bayesian Models\\ Conventional ML Regressors\\ Deep Learning Models\\ Auto Encoders\\ Dimensionality Reduction\\ Clustering\\ Pattern Mining\end{tabular} & N/A                                                                                    \\
                               \rowcolor{gray!15} \citep{he2021survey} & 2021                           & Log Anomaly Detection                                                                           & $\bullet$             & $\bullet$               & $\bullet$              &                   &                      &                   & $\bullet$              &                    &                          &                  &                 &                 & \begin{tabular}[c]{@{}l@{}}Statistical Models\\ Conventional ML Classifiers\\ Deep Neural Networks\\ Dimensionality Reduction\\ Clustering\\ Pattern Mining\end{tabular}                                                                & N/A                                                                                    \\
                               \citep{zhao2021empirical} & 2021                           & Log Anomaly Detection                                                                           & $\bullet$             & $\bullet$               & $\bullet$              &                   &                      &                   & $\bullet$              &                    &                          &                  &                 &                 & \begin{tabular}[c]{@{}l@{}}Conventional ML Classifiers\\ Deep Learning Models\\ Dimensionality Reduction\\ Clustering\\ Pattern Mining\end{tabular}                                                                                     &                                                                                        \\ \midrule
                               \rowcolor{gray!15} \citep{salfner2010survey} & 2010                           & Online Failure Prediction                                                                       & $\bullet$             & $\bullet$               & $\bullet$              & $\bullet$            &                      & $\bullet$            & $\bullet$              & $\bullet$             & $\bullet$                   &                  &                 &                 & \begin{tabular}[c]{@{}l@{}}Statistical Models\\ Conventional ML Classifiers\\ Bayesian Models\\ Conventional ML Regressors\\ Dimensionality Reduction\\ Rule-based Models\\ Clustering\\ Pattern Recognition\end{tabular}               &     \begin{tabular}[c]{@{}c@{}}Contingency metrics\\ ROC/AUC\end{tabular}                                                                                  \\
                               \citep{d2012evaluating} & 2012                           & Software Defect Prediction                                                                      &                    &                      & $\bullet$              & $\bullet$            & $\bullet$               &                   &                     &                    &                          &                  &                 &                 & \begin{tabular}[c]{@{}l@{}}Conventional ML Classifiers\\ Ranking-based Models\end{tabular}                                                                                                                                              & \begin{tabular}[c]{@{}c@{}}ROC/AUC\\ Spearman correlation\\ Friedman yest\end{tabular} \\
                               \rowcolor{gray!15} \citep{karim2017software} & 2017                           & Software Defect Prediction                                                                      &                    &                      & $\bullet$              & $\bullet$            & $\bullet$               &                   &                     &                    &                          &                  &                 &                 & \begin{tabular}[c]{@{}l@{}}Statistical Models\\ Conventional ML Classifiers\\ Bayesian Models\end{tabular}                                                                                                                              & Contingency metrics                                                               \\
                               \citep{jauk2019predicting} & 2019                           & Faults Prediction                                                                               & $\bullet$             & $\bullet$               & $\bullet$              & $\bullet$            &                      &                   & $\bullet$              & $\bullet$             & $\bullet$                   &                  &                 &                 & \begin{tabular}[c]{@{}l@{}}Statistical Models\\ Conventional ML Classifiers\\ Bayesian Models\\ Conventional ML Regressors\\ Deep Learning Models\\ Rule-based Models\end{tabular}                                                      & Contingency metrics                                                               \\
\midrule
\end{tabular}}
\end{table}

\begin{table}[]
\resizebox{\textwidth}{!}{%
\begin{tabular}{cc>{\centering\arraybackslash}m{2.5cm}ccccccccccccl>{\centering\arraybackslash}m{3cm}}
\toprule
\multirow{6}{*}{\textbf{Ref.}} & \multirow{6}{*}{\textbf{Year}} & \multirow{6}{*}{\textbf{\begin{tabular}[c]{@{}c@{}}Focus\\ and Scoop\\ Area\end{tabular}}}                                                                  & \multicolumn{4}{c}{\textbf{Target Area}}                                            & \multicolumn{8}{c}{\textbf{Data sources}}                                                                                                                             & \multicolumn{1}{c}{\multirow{6}{*}{\textbf{Type of AI Techniques}}}                                                                                                                                                                     & \multirow{6}{*}{\textbf{Evaluation Metrics}}                                           \\ \cmidrule(lr){4-7} \cmidrule(lr){8-15}
                               &        &                                                                                                 & \multicolumn{1}{c}{\rotatebox[origin=c]{90}{\textbf{Technical}}} &  \multicolumn{1}{c}{\rotatebox[origin=c]{90}{\textbf{Application}}} &  \multicolumn{1}{c}{\rotatebox[origin=c]{90}{\textbf{Functional}}} &  \multicolumn{1}{c}{\rotatebox[origin=c]{90}{\textbf{Business}}} & \multicolumn{1}{c}{\rotatebox[origin=c]{90}{\textbf{Source Code}}} &  \multicolumn{1}{c}{\rotatebox[origin=c]{90}{\textbf{Topology}}} &  \multicolumn{1}{c}{\rotatebox[origin=c]{90}{\textbf{Event Logs}}} &  \multicolumn{1}{c}{\rotatebox[origin=c]{90}{\textbf{KPIs/SLOs}}} &  \multicolumn{1}{c}{\rotatebox[origin=c]{90}{\textbf{Traffic Network}}} &  \multicolumn{1}{c}{\rotatebox[origin=c]{90}{\textbf{Reports}}} &  \multicolumn{1}{c}{\rotatebox[origin=c]{90}{\textbf{Alerts}}} & \rotatebox[origin=c]{90}{\textbf{Traces}} & \multicolumn{1}{c}{}                                                                                                                                                                                                                    &                                                                                        \\ \midrule
                               \rowcolor{gray!15} \citep{davari2021survey} & 2021                           & Railway Failures Prediction                                                                     & $\bullet$             &                      &                     &                   &                      &                   &                     & $\bullet$             &                          &                  & $\bullet$          &                 & \begin{tabular}[c]{@{}l@{}}Conventional ML Classifiers\\ Bayesian Models\\ Conventional ML Regressors\\ Deep Learning Approaches\\ Auto-Encoders\\ Generative Adversarial Model\end{tabular}                                            & \begin{tabular}[c]{@{}c@{}}Contingency metrics\\ Regression metrics\end{tabular} \\
                               \citep{berghout2022systematic} & 2022                           & Remaining Useful Lifetime                                                                       & $\bullet$             &                      &                     &                   &                      &                   &                     & $\bullet$             &                          &                  &                 &                 & \begin{tabular}[c]{@{}l@{}}Statistical Models\\ Conventional ML Regressors\\ Deep Learning Approaches\\ Auto-Encoders\\ Generative Adversarial Models\\ Transfer Learning\\ Reinforcement Learning\end{tabular}                         & Regression metrics                                                                     \\ \midrule
                               \rowcolor{gray!15} \citep{akila2014bug} & 2014                           & Bug Triage                                                                                            &                    &          & $\bullet$              & $\bullet$            & $\bullet$               &                   &                     &                    &                          & $\bullet$           &                 &                 & \begin{tabular}[c]{@{}l@{}}Conventional ML Classifiers\\ Bayesian Models\\ EM models\\ Information Retrieval\end{tabular}                                                                                                               & Contingency metrics                                                               \\ \midrule
                               \citep{DBLP:journals/jss/SorS14} & 2014                           & Memory Leak Diagnosis                                                                           &                    & $\bullet$               &                     &                   & $\bullet$               &                   &                     &                    &                          &                  &                 & $\bullet$          & \begin{tabular}[c]{@{}l@{}}Statistical Models\\ Graph Mining\end{tabular}                                                                                                                                                               & N/A                                                                                    \\
                               \rowcolor{gray!15} \citep{gao2015survey} & 2015                           & Fault Diagnosis                                                                                 & $\bullet$             & $\bullet$               &                     &                   & \multicolumn{8}{c}{N/A}                                                                                                                                               & \begin{tabular}[c]{@{}l@{}}Statistical Models\\ Signal-based Models\end{tabular}                                                                                                                                                        & N/A                                                                                    \\
                               \citep{wong2016survey} & 2016                           & Fault Localization                                                                              &                    &                      &      $\bullet$           & $\bullet$            & $\bullet$               &            &                     &                    &                          &                  &                 &                 & \begin{tabular}[c]{@{}l@{}}Program Spectrum-based Models\\ Statistical Models\\ Conventional ML Classifiers\\ Rule-based approaches\\ Pattern Mining\end{tabular}                                                                       &                                                                                        \\
                               \rowcolor{gray!15} \citep{sole2017survey} & 2017                           & Root Cause Analysis                                                                                            & $\bullet$             & $\bullet$               & $\bullet$              & $\bullet$            & \multicolumn{8}{c}{N/A}                                                                                                                                               & \begin{tabular}[c]{@{}l@{}}Statistical Models\\ Conventional ML Classifiers\\ Bayesian Models\\ Rule-based Models\\ Pattern Mining\end{tabular}                                                                                         & \begin{tabular}[c]{@{}c@{}}Complexity\\ Domain-specific \\ metrics\end{tabular}           \\ \midrule
                               \citep{mirheidari2013alert} & 2013                           & Intrusion Alerts Correlation                                                                    & $\bullet$             &                      &                     &                   &                      & $\bullet$            &                     &                    & $\bullet$                   &                  &                 &                 & \begin{tabular}[c]{@{}l@{}}Statistical Models\\ Conventional ML Classifiers\\ Similarity-based Models\end{tabular}                                                                                                                      & \begin{tabular}[c]{@{}c@{}}Accuracy\\ Domain-specific \\ metrics\end{tabular}             \\ \bottomrule
\end{tabular}}
\end{table}

Over the past two decades, numerous research endeavors have been dedicated to effectively tackling the automated incident management procedure. This becomes particularly important given the rise of expansive IT systems and the continually expanding scale of their infrastructures. These advancements have led to the generation of various incident types spanning multiple target domains, necessitating swift reporting, comprehension, and mitigation. Numerous research initiatives, motivated by industrial imperatives, have been engaged in structuring the incident management procedure. These efforts involve the formulation of taxonomies or comprehensive frameworks, thoughtfully considering an array of requirements and challenges, while including a review of existing methods that can be adapted to address specific aspects of the proposed framework~\citep{notaro2020systematic,zhang2015survey,zhang2016literature,li2017data,mukwevho2018toward,bogatinovski2021artificial,reiter2021aiops}. Others channel their focus into practical insights gained from real-world industrial scenarios through experience papers and use case studies, offering effective implementation details, specialized techniques and research innovations~\citep{dang2019aiops,chen2020towards,lyu2021towards}. 

As AIOps emerges as a novel and cross-disciplinary research frontier involving diverse domains and applications, it's worth noting that earlier works~\citep{lou2013software,zhang2015survey,zhang2016literature,li2017data,mukwevho2018toward}, particularly predating 2018, do not expressly reference the term AIOps, despite aligning with its conceptual framework and scoop. Additionally, some of these works may introduce non-data-driven methodologies, thus falling beyond the boundaries of the AIOps domain. In the following, we present a comparative analysis of related research that is either closely aligned or shares some similarities to our comprehensive examination of AIOps in incident management. The objective is to pinpoint key differentiating criteria, including their focal areas, whether a taxonomy is proposed, notable findings, and the provision of practical insights to facilitate effective implementation. Furthermore, we delve into surveys and reviews that are centered on specific phases within incident management, such as failure prediction or root cause analysis, without offering an exhaustive examination of the entire process~\citep{chandola2009anomaly,wang2013network,ibidunmoye2015performance,ibidunmoye2015performance,he2021survey,zhao2021empirical,salfner2010survey,d2012evaluating,karim2017software,jauk2019predicting,davari2021survey,berghout2022systematic,akila2014bug,DBLP:journals/jss/SorS14,gao2015survey,wong2016survey,sole2017survey,mirheidari2013alert}. Additionally, we explore works that target specific data sources, such as anomaly detection in log event data.

Table~\ref{tab:aiops_papers} compiles an overview of the most significant AIOps and Incident Management research papers. Our objective is to assist readers in locating pertinent works aligned with their specific inquiries and research subjects. Moreover, we aim to pinpoint research gaps and derive innovative ideas from existing literature. We classify the selected research into distinct factors. This classification assists readers in various approaches, including survey papers that provide a holistic snapshot of the current state-of-the-art, literature reviews that delve into analyzed literature and conduct technical comparisons based on objective criteria, and studies that offer practical insights such as industrial experience papers and use cases. Notably, since 2018, pioneering studies in incident management have increasingly incorporated the AIOps acronym. For each of these papers, we assess the extent to which they cover all phases of incident management as defined in Section~\ref{subsec:incident_tasks}. While a majority of these works concentrate on the reporting phase (detection and prediction), along with root cause analysis and mitigation, a clear void exists in the coverage of aspects such as ranking, assignment, and deduplication. It is also noteworthy to highlight that some research papers have a limited scope concerning the broad concept of incidents. For example,~\citet{zhang2015survey} address solely software bug management, thereby overlooking other areas like remaining useful lifetime estimation. Our analysis also extends to insights into the implementation of AI models in real-world scenarios. We report whether these papers provide a comprehensive depiction of related challenges, requirements, and articulate gaps for potential future research. The study conducted by~\citet{notaro2021survey} aligns closely with our work. Nevertheless, it exhibits gaps in addressing certain aspects, such as assignment and deduplication. Furthermore, it lacks an in-depth technical exploration of the challenges and requirements for establishing AIOps in production scenarios.

In Table~\ref{tab:related_surveys}, we present an overview of the most relevant surveys and reviews that focus on specific phases within the incident management procedure, rather than offering a comprehensive analysis of the entire process. These surveys delve into particular aspects, such as anomaly detection~\citep{wang2013network,chandola2009anomaly}, online failure prediction~\citep{salfner2010survey}, and more. We organize this research into categories based on their primary focal points and scopes. For instance, certain surveys narrow their scope to specific industries, such as the work by~\cite{davari2021survey} concentrating solely on predicting failures within the railway sector. Meanwhile,~\citet{mirheidari2013alert} address the issue of correlating alerts for cyber intrusion detection systems. Importantly, this does not exclude the potential applicability of the reviewed approaches in other practical domains. Furthermore, we outline the targeted maintenance levels, available data sources, and data-driven methodologies employed in these surveys. Some of these studies also emphasize the evaluation metrics used to assess the reviewed methods. In specific cases, like the research conducted by~\citep{sole2017survey} focusing on root cause analysis, domain-specific metrics are employed, as discussed in Section~\ref{subsec:eval_metrics} (e.g., these particular metrics assess aspects such as the ability to pinpoint multiple faults simultaneously, in identifying unknown faults, etc.).

\subsection{Incident Detection Methods}
\label{sec:sota_review_detection}

\begin{table}[]
\caption{Summary of reviewed AIOps Incident Detection methods.}
\label{tab:particularities_detection}
\renewcommand{\arraystretch}{1.2}

\resizebox{\textwidth}{!}{%
\begin{tabular}{cc>{\centering}m{1.5cm}cccccccccccc>{\centering}m{2cm}>{\centering}m{1.5cm}>{\centering}m{2.5cm}cc}
\toprule
\multirow{6}{*}{\textbf{Ref.}} & \multirow{6}{*}{\textbf{Year}} & \multirow{6}{*}{\textbf{Focus}}   & \multicolumn{4}{c}{\textbf{Target Area}}  & \multicolumn{8}{c}{\textbf{Data sources}}                                                                                  & \multicolumn{1}{c}{\multirow{6}{*}{\textbf{Approach}}}    & \multirow{6}{*}{\textbf{Paradigm}}    & \multirow{6}{*}{{\textbf{\begin{tabular}[c]{c}Evaluation\\ Metrics\end{tabular}}}} 
                               & \multirow{6}{*}{\rotatebox[origin=c]{90}{\textbf{Code}}} & \multirow{6}{*}{\rotatebox[origin=c]{90}{\textbf{Dataset}}}
                                \\ \cmidrule(lr){4-7} \cmidrule(lr){8-15}
                               &        &                                                                                                 & \multicolumn{1}{c}{\rotatebox[origin=c]{90}{\textbf{Technical}}} &  \multicolumn{1}{c}{\rotatebox[origin=c]{90}{\textbf{Application}}} &  \multicolumn{1}{c}{\rotatebox[origin=c]{90}{\textbf{Functional}}} &  \multicolumn{1}{c}{\rotatebox[origin=c]{90}{\textbf{Business}}} & \multicolumn{1}{c}{\rotatebox[origin=c]{90}{\textbf{Source Code}}} &  \multicolumn{1}{c}{\rotatebox[origin=c]{90}{\textbf{Topology}}} &  \multicolumn{1}{c}{\rotatebox[origin=c]{90}{\textbf{Event Logs}}} &  \multicolumn{1}{c}{\rotatebox[origin=c]{90}{\textbf{KPIs/Metrics}}} &  \multicolumn{1}{c}{\rotatebox[origin=c]{90}{\textbf{Traffic Network}}} &  \multicolumn{1}{c}{\rotatebox[origin=c]{90}{\textbf{Reports}}} &  \multicolumn{1}{c}{\rotatebox[origin=c]{90}{\textbf{Alerts}}} & \rotatebox[origin=c]{90}{\textbf{Traces}} & \multicolumn{1}{c}{}  & \multicolumn{1}{c}{} & \multicolumn{1}{c}{} & \multicolumn{1}{c}{}                                                                                                                  &                                                                                        \\ \midrule
\rowcolor{gray!15}   \citep{lakhina2004diagnosing}    &  2004    &  Network          &  $\bullet$            &          &        &      &             &          &          &    &  $\bullet$            &        &      &       &  PCA         &  UNSUP         &  {ACC, FPR}        &       &     $\bullet$     \\

   \citep{lakhina2005mining}    &  2005     &  Network         &  $\bullet$            &          &          &      &             &   $\bullet$         &          &      &  $\bullet$            &         &       &       &  FD         &  UNSUP         &  {ACC, FPR}        &      &     $\bullet$   \\

\rowcolor{gray!15}   \citep{pascoal2012robust}    &  2012     &  Network         &  $\bullet$            &          &          &      &             &           &          &      &  $\bullet$            &         &       &       &  PCA         &  UNSUP         &  {PREC, REC, FPR}        &      &     \\

    \citep{karami2015fuzzy}    &  2015     &  Network         &  $\bullet$            &          &          &      &             &           &          &      &  $\bullet$            &         &       &       &  K-Means, PSO         &  UNSUP         &  {F1, PREC, REC, FPR}        &      &  $\bullet$   \\
\rowcolor{gray!15}     \citep{pena2017anomaly} & 2017 & Network  &  $\bullet$            &          &          &      &             &           &          &   $\bullet$   &  $\bullet$    &         &       &       &  ARIMA, PL, ACODS  &  UNSUP         &  {REC, FPR, ROC}        &      &  $\bullet$ \\ 

\citep{bang2017anomaly} & 2017  & Network  &  $\bullet$            &          &          &      &             &           &          &   &  $\bullet$    &         &       &       &  HsMM  &  UNSUP   &  {FPR, FNR, TNR}        &      &   \\ 

\rowcolor{gray!15}  \citep{xie2016distributed} & 2017  & Network  &  $\bullet$            &          &          &      &             &           &          & $\bullet$  &  $\bullet$    &         &       &       &  KL-Div  &  UNSUP   &  {FPR, ACC, ROC}        &      &  $\bullet$ \\ 

\citep{deng2021graph} & 2021  & Network  &  $\bullet$            &    $\bullet$    &          &      &             &           &          &  &  $\bullet$    &         &       &       &  GNN &  UNSUP   &  {F1, PREC, REC}        &   $\bullet$   &  $\bullet$ \\

\rowcolor{gray!15}  \citep{yu2018netwalk} & 2018  & Edge Strm  &  $\bullet$            &      &          &      &             &           &          &  &  $\bullet$  &         &       &       &  AE, K-Means &  UNSUP   &  {AUC}        &   $\bullet$   &  $\bullet$ \\ 

\citep{yoon2019fast} & 2019  & Edge Strm  & $\bullet$ 
        &      &          &      &             &           &          &  &  $\bullet$  &         &       &       &  Anomalousness Score &  UNSUP   &  {PREC}     &   $\bullet$   &  $\bullet$ \\

\rowcolor{gray!15}  \citep{chang2021f} & 2021 & Edge Strm & $\bullet$ 
        &      &          &      &             &           &          &  &  $\bullet$  &         &       &       &  Frequency Factorization &  UNSUP   &  {AUC}     &   $\bullet$   &  $\bullet$ \\ \midrule

\citep{sharma2013cloudpd} & 2013 & Cloud &  $\bullet$
        &   $\bullet$    &          &      &             &           &       & $\bullet$ &   &         &       &       &  KNN, HMMs, K-Means &  UNSUP   &  {PREC, REC, ACC, FPR}  &     &  $\bullet$ \\

\rowcolor{gray!15}  \citep{vallis2014novel} & 2014 &  Cloud &  $\bullet$
        &   $\bullet$    &  $\bullet$   &      &             &           &       & $\bullet$ &   &         &       &       &  ESD Test &  UNSUP   &  {F1, PREC, REC}   &  $\bullet$   &   \\

\citep{liu2015opprentice} & 2015 &  Cloud &  
        &   $\bullet$    &     &      &             &           &       & $\bullet$ &   &         &       &       &  RF &  SUP   &  {AUC}   &     &   \\
\rowcolor{gray!15}  \citep{siffer2017anomaly} & 2017 & N/A & $\bullet$ &   $\bullet$   &     &      &             &           &       & $\bullet$ &   &         &       &       &  EV Theory &  UNSUP   &  {ROC}   &   $\bullet$   &  $\bullet$  \\

\citep{xu2018unsupervised} & 2018 & Web Apps &  &   $\bullet$   &     &      &             &           &       & $\bullet$ &   &         &       &       &  VAE, KDE &  UNSUP   &  {F1, AUC, Alert Delay}   &   $\bullet$   &  $\bullet$  \\

\rowcolor{gray!15}  \citep{li2018robust} & 2018 & Web Apps  &   &   $\bullet$   &     &      &             &           &       & $\bullet$ &   &         &       &       &  DBSCAN &  UNSUP   &  {NMI, F1}   &     &  $\bullet$  \\

\citep{ren2019time} & 2019 & Cloud &  
        &   $\bullet$   &   &      &             &           &       & $\bullet$ &   &         &   $\bullet$   &       &  SR-CNN &  UNSUP   &  {F1, PREC, REC}   &  $\bullet$   &  $\bullet$  \\ \midrule

\rowcolor{gray!15}  \citep{zhao2021identifying} & 2021 & Software Changes &  
        &   $\bullet$   &  $\bullet$  &   $\bullet$   &             &           &    $\bullet$   & $\bullet$ &   &         &     &       &  Multimodel LSTM &  UNSUP   &  {F1, PREC, REC, MTTD}   &  $\bullet$   &  $\bullet$  \\
  
\citep{wang2022identifying}  & 2022 & Software Changes &  
        &   $\bullet$   &  $\bullet$  &     &             &           &    & $\bullet$ &   &         &     &       &  LSTM &  SELFSUP   &  {F1, PREC, REC}   &  $\bullet$   &  $\bullet$  \\ \midrule

\rowcolor{gray!15}  \citep{su2019robust}  & 2019 & N/A &  $\bullet$
        &   $\bullet$   &   &     &             &           &    & $\bullet$ &   &         &     &       &  Stochastic RNN &  UNSUP   &  {F1, PREC, REC}   &  $\bullet$   &  $\bullet$  \\         

\citep{zhang2019deep}  & 2019 & N/A &  $\bullet$
        &   $\bullet$   &   &   &   &           &    & $\bullet$ &   &         &     &       &  CNN-LSTM &  UNSUP   &  {F1, PREC, REC}   &  $\bullet$   &  $\bullet$  \\         

\rowcolor{gray!15}  \citep{audibert2020usad}  & 2020 & N/A &  $\bullet$
        &   $\bullet$   &   &     &             &           &    & $\bullet$ &   &         &     &       &  GAN &  Adversial    &  {F1, PREC, REC}   &  $\bullet$   &  $\bullet$  \\

\citep{li2021multivariate}  & 2021 & N/A &  $\bullet$
        &   $\bullet$   &   &     &             &           &    & $\bullet$ &   &     &     &       &  Hierarchical VAE, MCMC &  UNSUP   &  {F1, Interpretation Score}   &  $\bullet$   &  $\bullet$  \\ \midrule

\rowcolor{gray!15}  \citep{fu2009execution}  & 2009 & Log Anom &  & $\bullet$
        &   $\bullet$   &        &             &           &  $\bullet$  &  &   &     &     &       &  FSM &  UNSUP   &  {FPR}   &  $\bullet$   &  $\bullet$  \\   
\citep{xu2009detecting} & 2009 &   Log Anom &  & $\bullet$
        &   $\bullet$   &        &       $\bullet$       &           &  $\bullet$  &  &   &     &     &       &  PCA &  UNSUP   &  {ACC}   &  $\bullet$   &  $\bullet$  \\

\rowcolor{gray!15}  \citep{lou2010mining} & 2010 &   Log Anom &  & $\bullet$
        &   $\bullet$   &        &           &           &  $\bullet$  &  &   &     &     &       &  Mining Invariants &  UNSUP   &  {ACC, FPR}   &  $\bullet$   &  $\bullet$  \\  

\citep{he2018identifying} & 2018 &  Log Anom &  & $\bullet$
        &   $\bullet$   &      &           &           &  $\bullet$  &  &   &     &     &       & Hierarchical Clustering  &  UNSUP   &  {F1, PREC, REC}  &  $\bullet$   &   \\

\rowcolor{gray!15}  \citep{du2017deeplog} & 2017 &  Log Anom &  & $\bullet$
        &   $\bullet$   &      &           &           &  $\bullet$  &  &   &     &     &       & LSTM &  UNSUP   &  {F1, PREC, REC, FPR, FNR}  &  $\bullet$   &  $\bullet$  \\

\citep{meng2019loganomaly} & 2019 &  Log Anom &  & $\bullet$
        &   $\bullet$   &      &           &           &  $\bullet$  &  &   &     &     &       & Template2VEC, LSTM &  UNSUP   &  {F1, PREC, REC}  &  $\bullet$   &  $\bullet$  \\

\rowcolor{gray!15} \citep{zhang2019robust} & 2019 &  Log Anom &  & $\bullet$
        &   $\bullet$   &      &           &           &  $\bullet$  &  &   &     &     &       & Word2VEC, biLSTM &  UNSUP   &  {F1, PREC, REC}  &  $\bullet$   &  $\bullet$  \\

\citep{xia2021loggan} & 2021 &  Log Anom &  & $\bullet$
        &   $\bullet$   &      &           &           &  $\bullet$  &  &   &     &     &       & LSTM-based GAN &  UNSUP, Adversial   &  {F1, PREC, REC}  &   &  $\bullet$  \\

\rowcolor{gray!15} \citep{guo2021logbert} & 2021 &  Log Anom &  & $\bullet$
        &   $\bullet$   &      &           &           &  $\bullet$  &  &   &     &     &       & BERT &  SELFSUP   &  {F1, PREC, REC}  & $\bullet$  &  $\bullet$  
  
  \\ \bottomrule
  
\end{tabular}}

\end{table}
\begin{table}[]
\caption{Study of particularities of AIOps Incident Detection methods.}
\label{tab:particularities_detection}
\renewcommand{\arraystretch}{1.0}
\resizebox{\textwidth}{!}{%
\begin{tabular}{cccccc}
\toprule
\textbf{Ref.} & \textbf{Interpretability} & \textbf{Scalability} & \textbf{Robustness} & \textbf{Temporal Evaluation}  & \textbf{Human-in-the-loop}                                           \\ \midrule
\rowcolor{gray!15} \citep{lakhina2004diagnosing,bang2017anomaly} & $\bullet$ & $\bullet$ & & & \\
\citep{lakhina2005mining} & $\bullet$ &  & & & \\
\rowcolor{gray!15} \citep{pascoal2012robust} & $\bullet$ & & $\bullet$ & $\bullet$ & \\

\citep{karami2015fuzzy,pena2017anomaly} & $\bullet$ & $\bullet$ & $\bullet$ & & \\

\rowcolor{gray!15} \citep{xie2016distributed} & $\bullet$ & $\bullet$ & $\bullet$ & $\bullet$ & \\

\citep{deng2021graph} & $\bullet$ & & $\bullet$ & & \\

\rowcolor{gray!15}\citep{yu2018netwalk} &  & $\bullet$  & & $\bullet$ &  \\
\citep{yoon2019fast} & $\bullet$ & $\bullet$  & & $\bullet$ & \\

\rowcolor{gray!15}\citep{chang2021f} & $\bullet$ & $\bullet$ & $\bullet$ & $\bullet$ & \\ \midrule

\citep{sharma2013cloudpd, vallis2014novel, siffer2017anomaly, li2018robust} & $\bullet$  & $\bullet$ & $\bullet$ & $\bullet$ & \\

\rowcolor{gray!15}\citep{liu2015opprentice, ren2019time} & $\bullet$ & $\bullet$ & $\bullet$ & $\bullet$ & $\bullet$ \\
\citep{xu2018unsupervised} & $\bullet$ & & $\bullet$ &  $\bullet$ & \\
 \midrule

\rowcolor{gray!15} \citep{zhao2021identifying} & $\bullet$ & $\bullet$ &  & & $\bullet$ \\

\citep{wang2022identifying} & & $\bullet$ & $\bullet$ & \\ \midrule

\rowcolor{gray!15} \citep{su2019robust, zhang2019deep}  & $\bullet$ & & $\bullet$ & &  \\

 \citep{audibert2020usad}  & & $\bullet$ & $\bullet$  &  $\bullet$ & \\

\rowcolor{gray!15}\citep{li2021multivariate}  & $\bullet$ & & $\bullet$ & $\bullet$ & \\ \midrule

 \citep{fu2009execution}  & $\bullet$ & & & $\bullet$ &  \\
\rowcolor{gray!15} \citep{xu2009detecting} & $\bullet$ & $\bullet$ & $\bullet$ & $\bullet$ & $\bullet$ \\
 \citep{lou2010mining, he2018identifying} & $\bullet$ & $\bullet$ & $\bullet$ & & \\
\rowcolor{gray!15} \citep{du2017deeplog} & $\bullet$ & & & $\bullet$ & $\bullet$ \\
\citep{meng2019loganomaly} & & & & $\bullet$ & $\bullet$ \\
\rowcolor{gray!15} \citep{zhang2019robust} & & $\bullet$ & $\bullet$ & $\bullet$ & \\ 
\citep{xia2021loggan} & & $\bullet$ & & & 
\\ \bottomrule
\end{tabular}}
\end{table}

Incident detection approaches are reactive methods of incident management that aim to track and identify abnormal states or behaviors in a system. Their purpose is to either anticipate failures before they occur or mitigate the consequences of failures after they have happened. This is driven by the understanding that, despite employing advanced prediction techniques, it is impossible to completely eliminate the occurrence of failures. These methods also aid in comprehending the causal relationships, as well as understanding the temporal characteristics that lead to incidents. Incident detection methods often leverage unsupervised learning approaches, primarily because acquiring high-quality, sufficient, and balanced data labels poses significant challenges. Among the notable techniques employed in this context, we find clustering methods, dimensionality reduction techniques, and auto-encoders. Additionally, other approaches, such as graph mining, and statistical models, have been employed in this context. 

At the network level, \citet{lakhina2004diagnosing} propose an anomaly detection method for network traffic analysis using SNMP data. The authors apply Principal Component Analysis (PCA) to link flow measurements collected over time to separate traffic into normal and anomalous subspaces. Anomalies are identified by reconstructing new observations using abnormal components. If the reconstruction error exceeds a predefined threshold based on explained variance, the data point is considered anomalous. In a subsequent work~\citep{lakhina2005mining}, the same authors argue that by analyzing the distributions of packet features (IP addresses and ports) in flow traces, it is possible to reveal the presence and structure of various anomalies in network traffic. They use entropy as a summarization tool to quantify the randomness or irregularity in these feature distributions. The method has demonstrated its ability to adapt to new kinds of anomalies. However,~\citet{ringberg2007sensitivity} have raised concerns about the application of PCA in~\citep{lakhina2004diagnosing} by identifying its susceptibility of false positive rates to small normal subspace noises. Thus,~\citet{pascoal2012robust} introduce another approach that incorporated a resilient PCA detector combined with a robust feature selection technique in order to obtain adaptability to distinct network contexts and circumstances. Moreover, this approach eliminates the need for flawless ground-truth data during training, addressing a limitation highlighted in the conventional PCA approach present in~\citep{lakhina2004diagnosing}.~\citet{karami2015fuzzy} introduce a clustering approach for identifying attacks and anomalies, ranging from Denial of Service (DoS) to privacy breaches, within content-centric networks. This approach operates through two distinct phases. Firstly, a unique hybridization of the Particle Swarm Optimization (PSO) and K-means algorithm is employed in the training phase to accurately determine the optimal number of clusters and avoid being trapped in a locally optimal solution. In the subsequent detection phase, a fuzzy approach is employed by combining two distance-based methods, to effectively identify anomalies within newly monitored data in which false positive rates are reduced with reliable detection of intrusive activities.

Statistical approaches have also been leveraged within this context. For instance,~\citet{pena2017anomaly} propose a Correlational Paraconsistent Machine (CPM) that employs non-classical paraconsistent logic (PL) in conjunction with two unsupervised traffic characterization techniques. These techniques involve Ant Colony Optimization for Digital Signature (ACODS) and Auto-Regressive Integrated Moving Average (ARIMA). The purpose of these methods is to analyze historical network traffic data and generate two distinct network profiles, known as Digital Signatures of Network Segment using Flow Analysis (DSNSF), which describe normal traffic behavior. The detection of anomalies is linked to the degrees of certainties and contradictions produced by paraconsistent logic when correlating two prediction profiles with real traffic measurements. In a different study,~\citet{bang2017anomaly} propose an Intrusion Detection System (IDS) utilizing a Hidden Semi-Markov Model (HsMM) tailored specifically for detecting advanced LTE signaling attacks on Wireless Sensor Networks (WSNs). The authors argue that traditional Hidden Markov Models (HMM) are limited in representing various potential transition behaviors, while HsMM overcomes this limitation due to its arbitrary state sojourn time, making it more suitable for analyzing time-series behavior. HsMM is effectively employed to model the spatial-temporal characteristics of the wake-up packet generation process. The detector then compares observed spatiotemporal features of a server's wake-up packet generation with the normal criteria established by the HsMM. Subsequently, an alert is triggered whenever a significant divergence occurs.~\citet{xie2016distributed} present an algorithm to detect long-term anomalies in WSNs, where a group of neighboring nodes is consistently affected by such anomalies over a substantial time frame, particularly in the presence of DoS and sinkhole attacks. To achieve this, the authors employ the Kullback-Leibler (KL) divergence to measure the differences between the global Probability Density Functions (PDFs) across consecutive time intervals. The resulting time series derived from this function was subject to analysis based on an adaptive threshold to identify any noteworthy deviations from the norm. 

More recently,~\citet{deng2021graph} address the challenge of detecting anomalous events in the context of cybersecurity attacks by analyzing high-dimensional time series data, specifically sensor data. The authors propose an approach that combines structure learning with an attention-based graph neural network, which learns a graph of the dependence relationships between sensors and identifies and explains deviations from these relationships. In order to provide useful explanations for anomalies, a forecasting-based approach is used to predict the expected behavior of each sensor at each time based on the past. This allows the user to easily identify the sensors which deviate greatly from their expected behavior. Moreover, the user can compare the expected and observed behavior of each sensor, to understand why the model regards a sensor as anomalous. The approach is also highly versatile, offering extendibility for application across a wide array of use cases beyond just cybersecurity attacks.

Another concern lies in detecting anomalies in edge streams, which exhibit some specificities since they are commonly used to capture interactions in dynamic and evolutionary networks.~\citet{yu2018netwalk} introduce \texttt{NetWalk}, a novel approach for real-time detection of structural anomalies in dynamic networks prevalent in fields like social media, security, and public health. Unlike traditional methods designed for static networks, \texttt{NetWalk} focuses on dynamic environments where the network representations need constant updating. The proposed approach employs a new embedding approach to encode vertices into vector representations. This process minimizes the pairwise distance between vertex representations derived from dynamic network walks and utilizes deep autoencoder reconstruction error as global regularization. The vector representations are computed efficiently using a sampling technique that requires constant memory space. Based on the learned low-dimensional vertex representations, \texttt{NetWalk} employs a clustering-based technique to detect network anomalies incrementally and in real-time as the network evolves. This approach can be applied to various types of networks, making it flexible for different application domains. In the same context,~\citet{yoon2019fast} introduced \texttt{AnomRank} to specifically address the challenge of detecting sudden anomalous patterns, like link spam and follower boosting in dynamic graph streams by employing a unique two-pronged approach, introducing novel metrics for measuring anomalousness. These metrics track derivatives of node scores or importance functions, enabling the detection of abrupt changes in node significance. \texttt{AnomRank} demonstrates through the experiments its scalability, processing millions of edges in mere seconds, as well as theoretical soundness by providing guarantees for its two-pronged approach. \texttt{F-Fade}~\citep{chang2021f} is another approach to detect anomalies in edge streams which proposes a frequency-factorization technique to model the time-evolving distributions of interaction frequencies between node pairs. Anomalies are then detected based on the probability of the observed frequency of each incoming interaction. This approach has also proven to effectively operate in an online streaming setting while requiring constant memory.

Numerous alternative methods involve the analysis of Key Performance Indicators and metric observations to effectively identify anomalies occurring across both technical, application and functional levels. Typically, these methods employ unsupervised models that are trained on either univariate or multivariate time series data. For instance, \texttt{CloudPD}~\citep{sharma2013cloudpd} is a fault management framework for clouds that introduced conventional machine learning techniques for anomaly detection in cloud environments, addressing both the physical and application layers. The approach involved utilizing various measures at the virtual machine (VM) and application machine levels, including operating system variables and application performance metrics. The paper proposed to combine three unsupervised machine learning methods: k-nearest neighbors(k-NN), HMMs, and K-means clustering. Similarly,~\citet{vallis2014novel} addressed the challenge of detecting long-term anomalies in cloud computing environments on Twitter. The authors proposed a new statistical learning approach using the generalized Extreme Studentized Deviate test (ESD) and incorporated time series decomposition and robust statistics to identify anomalies such as a piecewise approximation of the underlying long-term trend reducing the number of false positive while taking into account both intra-day and weekly seasonalities to further minimize false positives. The proposed technique is applicable to various types of time series data, including application metrics like Tweets Per Second and system metrics like CPU utilization. On the other hand,~\citet{liu2015opprentice} introduced a novel supervised approach called \texttt{Opprentice} (Operators' apprentice). In this approach, operators need to periodically label anomalies in performance data using a user-friendly tool. Then, multiple existing detectors are applied to the data in parallel to extract anomaly features. These features, along with the labeled data, are used to train a random forest classifier. This classifier automates the selection of suitable detector-parameter combinations and thresholds. In~\citep{siffer2017anomaly}, the authors propose a novel approach that employs Extreme Value Theory for identifying outliers in streaming univariate KPIs. The proposed approach eliminates the need for manual threshold setting and does not assume any specific data distribution. The key parameter in this method is the risk level, which controls the rate of false positives. Apart from outlier detection, the approach can also automatically set thresholds, making it applicable to a wide range of scenarios including intrusion detection.   

\texttt{Donut}~\cite{xu2018unsupervised} addresses the challenge of detecting anomalies in seasonal Key Performance Indicators (KPIs) within Web applications of large Internet companies. The authors proposed an unsupervised approach based on Variational Autoencoders (VAE) and window sampling. Importantly, the paper also introduces a novel interpretation of Kernel Density Estimation (KDE) for reconstructing anomalies in Donut, providing a solid theoretical foundation and making it the first VAE-based anomaly detection method with a well-defined theoretical explanation. In~\citep{li2018robust}, the authors propose a KPI clustering-based strategy wherein millions of KPIs are grouped into a smaller number of clusters, allowing models to be selected and trained on a per-cluster basis. This approach addresses various challenges due to the unique nature of KPIs, such as their extended length and susceptibility to various distortions like noise, anomalies, phase shifts, and amplitude differences. Therefore, the paper presents a framework named \texttt{ROCKA}, which involves time series preprocessing, baseline extraction, density-based spatial clustering of applications with noise (DBSCAN) with Dynamic Time Warping (DTW) distance, and assignment steps.~\citet{ren2019time} introduced a sophisticated time-series anomaly detection service developed at Microsoft, emphasizing its role in continuous monitoring and incident alerting. The paper offers a detailed insight into the service's underlying pipeline and algorithm, highlighting three key modules: data ingestion, an experimentation platform, and online computation. Notably, the paper introduces a novel algorithm that combines Spectral Residual (SR) and Convolutional Neural Network (CNN) techniques on univariate KPIs.

In a slightly similar context,~\citet{zhao2021identifying} conducted a study utilizing real-world data from a prominent commercial bank to quantify the substantial impact of software changes on incidents, including code defects, configuration errors, resource conflicts, and software version discrepancies. This quantitative analysis is complemented by qualitative insights that expose limitations in current detection practices, particularly when dealing with heterogeneous multi-source data inherent in software modifications. Building upon these insights, the paper introduces an approach dubbed SCWarn that leverages multimodal learning using LSTM model to identify detrimental changes and generate interpretable alerts. The authors also quantitatively demonstrate the efficiency of their approach in reducing the mean time required to detect adverse changes.~\citet{wang2022identifying} addressed a similar problem by introducing Kontrast, a self-supervised and adaptive approach that employs contrastive learning. This approach focuses on comparing KPI time series before and after a software change to ensure the system's stability post-change. To facilitate contrastive learning, a unique data augmentation technique inspired by self-supervised learning is proposed, generating data with pseudo labels. Through the experiments conducted in the paper, the proposed model has shown impressive processing speeds and cross-dataset adaptability. 

More recently, a plethora of approaches have emerged to address the challenge of anomaly detection in multivariate time series data. Two benefits of conducting a multi-dimensional analysis include the capacity to depict the interconnections among diverse metrics and the generation of understandable findings for root-cause investigation~\citep{notaro2021survey}. \texttt{OmniAnomaly}~\citep{su2019robust} focuses on capturing normal patterns in multivariate time series using a stochastic recurrent neural network framework, by learning robust representations through techniques like stochastic variable connection and planar normalizing flow. The approach involves reconstructing input data based on these representations and using reconstruction probabilities to identify anomalies. Additionally, \texttt{OmniAnomaly} provides insightful interpretations for detected anomalies by analyzing the reconstruction probabilities of constituent univariate time series. Multi-Scale Convolutional Recurrent Encoder-Decoder (\texttt{MSCRED})~\citep{zhang2019deep} constructs multi-scale signature matrices to represent different system statuses at various time steps. It utilizes a convolutional encoder to capture inter-sensor correlations and an attention-based Convolutional Long-Short Term Memory (ConvLSTM) network to identify abnormal time steps by capturing temporal patterns. The utilization of residual signature matrices aids in anomaly detection and diagnosis. \texttt{USAD}~\citep{audibert2020usad} employs adversarial training in the auto-encoder architecture to effectively isolate anomalies while maintaining fast training speeds. Experimental results focus also on the scalability and robustness of the approach.~\citet{li2021multivariate} introduced \texttt{InterFusion} that employs a hierarchical variational auto-encoder with two stochastic latent variables to capture normal patterns, learning low-dimensional inter-metric and temporal embeddings. The paper also presents a Markov chain Monte Carlo (MCMC) method to obtain meaningful reconstructions for interpreting anomalies.

Log-based approaches have surfaced as efficient means for detecting anomalies in application, functional, and business services.~\citet{fu2009execution} proposed a clustering-based technique in which log entries are mapped to their corresponding template versions, enabling the identification of line templates. Subsequently, a Finite State Machine (FSM) was learned to model program workflows based on the log evidence. This FSM-based model facilitated the verification of correct program execution and the detection of software problems. In another study by~\citep{xu2009detecting}, text analysis and information retrieval techniques were applied to console logs and source code for anomaly detection in large-scale data centers. State variables and object identifiers were automatically extracted from parsed logs, and their frequency across different documents was analyzed using PCA and term inverse-document frequency (TF-IDF). Anomalies were detected using a threshold-based rule on the reconstruction error.~\citet{lou2010mining} were the first to explore mining invariants from log messages for system anomaly detection. They identified two types of invariants that capture relationships among different log messages: invariants in textual logs describing equivalence relations, and invariants expressed as linear equations representing linear independence relations. \texttt{Log3C}~\citep{he2018identifying} presents a cascading clustering algorithm designed to efficiently detect significant system issues by utilizing both log sequences and system KPIs. This method involves iterative sampling, clustering, and matching of log sequences. Subsequently, the correlation between clusters of log sequences and system KPIs is employed to pinpoint and prioritize impactful problems through the application of a multivariate linear regression model.

Significant efforts have been directed towards exploring sequential Recurrent Neural Networks (RNNs) and their variations. \texttt{DeepLog}~\citep{du2017deeplog} introduced the use of LSTM networks to learn patterns from logs and predict the probability distribution of the next log key based on the observation of previous log keys. The identification of an anomalous log key occurred when it failed to appear among the top-k keys ranked by probability. Furthermore, the research introduced an online learning approach based on user feedback and tried to provide explanations using a finite-state machine. \texttt{LogAnomaly}~\citep{meng2019loganomaly} is a semantic-aware representation framework, that employed the language model method and used template embeddings (template2vec) to automatically combine similar log keys. This approach eliminated the requirement for human input. The issues of log-specific word embedding and out-of-vocabulary are both addressed. \citet{zhang2019robust} discovered that log data frequently include log events or sequences that have not been encountered before, indicating log unpredictability and noise in log processing. To address this issue, they introduced \texttt{LogRobust}, a method that extracts the meaning of log events using readily available word vectors. Subsequently, they employed a bidirectional LSTM model and attention mechanisms to compute an anomaly score directly, rather than predicting the next probable log keys.~\citet{xia2021loggan} proposed \texttt{logGAN}, an LSTM-based generative adversarial network framework, incorporating permutation event modeling for log-level anomaly detection. This approach addresses both sequential characteristics and out-of-order issues of logs, while the generative adversarial network component counteracts class imbalance, thereby enhancing anomaly detection performance. More recently, a novel self-supervised framework known as \texttt{LogBERT}~\citep{guo2021logbert} has been introduced, harnessing the power of Bidirectional Encoder Representations from Transformers (BERT). This approach introduces two innovative self-supervised training tasks: masked log message prediction and volume of hypersphere minimization. These tasks enable \texttt{LogBERT} to capture patterns inherent in normal log sequences. As a result, \texttt{LogBERT} identifies anomalies by detecting deviations from these established patterns.

\subsection{Incident Prediction Methods}
\label{sec:sota_review_prediction}

\begin{table}[]
\caption{Summary of reviewed AIOps Incident Prediction methods.}
\label{tab:tab_prediction}
\renewcommand{\arraystretch}{1.2}

\resizebox{\textwidth}{!}{%
\begin{tabular}{cc>{\centering}m{1.5cm}cccccccccccc>{\centering}m{2cm}>{\centering}m{1.5cm}>{\centering}m{2.5cm}cc}
\toprule
\multirow{6}{*}{\textbf{Ref.}} & \multirow{6}{*}{\textbf{Year}} & \multirow{6}{*}{\textbf{Focus}}   & \multicolumn{4}{c}{\textbf{Target Area}}  & \multicolumn{8}{c}{\textbf{Data sources}}                                                                                  & \multicolumn{1}{c}{\multirow{6}{*}{\textbf{Approach}}}    & \multirow{6}{*}{\textbf{Paradigm}}    & \multirow{6}{*}{{\textbf{\begin{tabular}[c]{c}Evaluation\\ Metrics\end{tabular}}}} 
                               & \multirow{6}{*}{\rotatebox[origin=c]{90}{\textbf{Code}}} & \multirow{6}{*}{\rotatebox[origin=c]{90}{\textbf{Dataset}}}
                                \\ \cmidrule(lr){4-7} \cmidrule(lr){8-15}
                               &        &                                                                                                 & \multicolumn{1}{c}{\rotatebox[origin=c]{90}{\textbf{Technical}}} &  \multicolumn{1}{c}{\rotatebox[origin=c]{90}{\textbf{Application}}} &  \multicolumn{1}{c}{\rotatebox[origin=c]{90}{\textbf{Functional}}} &  \multicolumn{1}{c}{\rotatebox[origin=c]{90}{\textbf{Business}}} & \multicolumn{1}{c}{\rotatebox[origin=c]{90}{\textbf{Source Code}}} &  \multicolumn{1}{c}{\rotatebox[origin=c]{90}{\textbf{Topology}}} &  \multicolumn{1}{c}{\rotatebox[origin=c]{90}{\textbf{Event Logs}}} &  \multicolumn{1}{c}{\rotatebox[origin=c]{90}{\textbf{KPIs/Metrics}}} &  \multicolumn{1}{c}{\rotatebox[origin=c]{90}{\textbf{Traffic Network}}} &  \multicolumn{1}{c}{\rotatebox[origin=c]{90}{\textbf{Reports}}} &  \multicolumn{1}{c}{\rotatebox[origin=c]{90}{\textbf{Alerts}}} & \rotatebox[origin=c]{90}{\textbf{Traces}} & \multicolumn{1}{c}{}  & \multicolumn{1}{c}{} & \multicolumn{1}{c}{} & \multicolumn{1}{c}{}                                                                                                                  &                                                                                        \\ \midrule
  \rowcolor{gray!15} \citep{nagappan2006mining}    &  2006    &  SDP          &              &          &     $\bullet$   &  $\bullet$   &    $\bullet$      &         &          &    &            &        &      &       &  PCA, Lin. Reg        &  SUP      &  {Spearmann, Pearson, $\text{R}^2$}        &       &         \\

    \citep{menzies2006data}    &  2007    &  SDP          &              &          &     $\bullet$   &  $\bullet$   &    $\bullet$      &         &          &    &            &        &      &       &  Naive Bayes         &  SUP      &  {ROC}        &       &  $\bullet$     \\

   \rowcolor{gray!15} \citep{elish2008predicting}    &  2008    &  SDP          &              &          &     $\bullet$   &  $\bullet$   &    $\bullet$      &         &          &    &            &        &      &       &  SVM         &  SUP      &  {F1, ACC, PREC, REC}        &       &  $\bullet$     \\

    \citep{dejaeger2012toward}    &  2013    &  SDP          &              &          &     $\bullet$   &  $\bullet$   &    $\bullet$      &         &          &    &            &        &      &       &  Bayesian Nets    &  SUP      &  {AUC}        &       &  $\bullet$     \\

   \rowcolor{gray!15} \citep{ostrand2005predicting}  &  2005    &  SDP          &              &          &     $\bullet$   &  $\bullet$   &    $\bullet$      &         &          &    &            &        &      &       &  Poisson GLM    &  SUP      &  {ACC}        &       &      \\

    \citep{moser2008comparative}  &  2008    &  SDP          &              &          &     $\bullet$   &  $\bullet$   &    $\bullet$      &         &          &    &            &        &      &       &  Ensemble Model    &  SUP      &  {PREC, REC, FPR, ROC}        &       &  $\bullet$     \\
   \rowcolor{gray!15} \citep{nam2013transfer} & 2013 &  SDP          &              &          &     $\bullet$   &  $\bullet$   &    $\bullet$      &         &          &    &            &        &      &       &  Log. Reg    &  Transfert      &  {F1, PREC, REC}        &  &  $\bullet$     \\

    \citep{wang2016automatically}  &  2016  &  SDP          &              &          &     $\bullet$   &  $\bullet$   &    $\bullet$      &         &          &    &            &        &      &       &  DBN  &  SUP      &  {F1, PREC, REC}        &  &  $\bullet$     \\

   \rowcolor{gray!15} \citep{li2017software}  &  2017  &  SDP          &              &          &     $\bullet$   &  $\bullet$   &    $\bullet$      &         &          &    &            &        &      &       &  CNN   &  SUP      &  {F1, PREC, REC}        &  &  $\bullet$     \\

    \citep{majd2020sldeep}  &  2020  &  SDP          &              &          &     $\bullet$   &  $\bullet$   &    $\bullet$      &         &          &    &            &        &      &       &  LSTM   &  SUP      &  {F1, PREC, REC, ACC}        & $\bullet$  &  $\bullet$     \\ 

   \rowcolor{gray!15} \citep{xu2020defect}  &  2021  &  SDP          &              &          &     $\bullet$   &  $\bullet$   &    $\bullet$      &         &          &    &            &        &      &       &  LDA, GNN   &  SUP      &  {F1, PREC, REC, ACC, AUC}        &   &  $\bullet$     \\  

    \citep{uddin2022software}  &  2022  &  SDP          &              &          &     $\bullet$   &  $\bullet$   &    $\bullet$      &         &          &    &            &        &      &       &  biLSTM, BERT   &  SUP      &  {F1, PREC, REC}        &   &  $\bullet$     \\  \midrule

    \rowcolor{gray!15} \citep{garg1998methodology}  &  1998  &  Software Aging    &              &     $\bullet$      &       &    &         &         &          &   $\bullet$ &            &        &      &       &  Kendall-test, Lin. Reg   &  Forecast   &  {Autocorrelation Plots}        &   &       \\ 

    \citep{vaidyanathan1999measurement}  &  1999  &  Software Aging    &              &     $\bullet$      &       &    &         &         &          &   $\bullet$ &            &        &      &       &  SMM, Lin. Reg, K-Means   &  Forecast  &  {time to exhaustion}   &   &      \\ 

   \rowcolor{gray!15} \citep{alonso2010adaptive}  &  2010  &  Software Aging    & $\bullet$   &     $\bullet$      &       &    &         &         &          &   $\bullet$ &            &        &      &       &  Lin. Reg Ensemble    &  Forecast  &  {MAE}        &   &  $\bullet$    \\

    \citep{sudhakar2014software}  &  2014  &  Software Rejuv.     & $\bullet$   &     $\bullet$      &       &    &         &         &          &   $\bullet$ &            &        &      &       &  Neural Network    &  SUP   &  {MAPE}   &   &    \\ 

    \rowcolor{gray!15} \citep{araujo2014software}  &  2014  &  Software Rejuv.     & $\bullet$   &     $\bullet$      &       &    &         &         &          &   $\bullet$ &            &        &      &       &  MLP    &  SUP   &  {MAPE}   &   &    \\ \midrule

    \citep{zhao2010predicting}  &  2010  &  Disk Fail.  & $\bullet$   &      &       &    &         &         &          &   $\bullet$ &            &        &      &       &  HMM, HSMM    &  SUP   &  {ROC}   &   &  $\bullet$  \\  

    \rowcolor{gray!15} \citep{wang2013online}  &  2013  &  Disk Fail.  & $\bullet$   &      &       &    &         &         &          &   $\bullet$ &            &        &      &       &  mRMR   &  SUP   &  {ROC}   &   &  $\bullet$  \\  

    \citep{xu2016health}  &  2016  &  Disk Fail.  & $\bullet$   &      &       &    &         &         &          &   $\bullet$ &            &        &      &       &  RNN   &  SUP   &  {REC}   &   &  $\bullet$  \\ 

    \rowcolor{gray!15} \citep{xiao2018disk}  &  2018  &  Disk Fail.  & $\bullet$   &      &       &    &         &         &          &   $\bullet$ &            &        &      &       &  Online RF   &  SUP   &  {REC, FPR}   &   &  $\bullet$  \\ \midrule
    
    \citep{davis2017failuresim}  &  2017  &  Datacenter Networks  & $\bullet$   &      &       &    &         &         &          &   $\bullet$ &            &        &      &       &  MLP, RNN   &  SUP   &  {ACC}   &   &  $\bullet$  \\ 

    \rowcolor{gray!15}  \citep{zhang2017syslog}  &  2017  &  Switch Fail.  & $\bullet$   &      &       &    &         &         &     $\bullet$     &   &            &        &      &       &  FT-Tree, HsMM  &  SUP   &  {F1, PREC, REC}   &   &    \\ 

   \citep{sun2019system}  &  2019  &  DRAM Fail.  & $\bullet$   &      &       &    &         &         &     $\bullet$     &  $\bullet$  &            &        &      &       &  TCNN  &  SUP   &  {F1, PREC, REC}   &   &   $\bullet$ \\ 
    \rowcolor{gray!15}  \citep{khalil2020machine}  &  2020  &  Circuit Fail.  & $\bullet$   &      &       &    &         &         &     $\bullet$     &  $\bullet$  &            &        &      &       &  FFT, PCA, CNN  &  SUP   &  {F1, PREC, REC, ACC}   &   &    \\ \midrule
    
    \citep{zheng2017long}  &  2017  &  RUL  & $\bullet$   &      &       &    &         &         &          &  $\bullet$  &            &        &      &       &  LSTM  &  Forecast  &  {RMSE}   &   & $\bullet$   \\ 

    \rowcolor{gray!15}  \citep{wu2018remaining}  &  2018  &  RUL  & $\bullet$   &      &       &    &         &         &          &  $\bullet$  &            &        &      &       &  Vanilla-LSTM  &  Forecast  &  {MSE}   &   & $\bullet$   \\ 

    \citep{ma2020deep}  &  2020  &  RUL  & $\bullet$   &      &       &    &         &         &          &  $\bullet$  &            &        &      &       &  CNN-LSTM  &  Forecast  &  {RMSE, MAE}   &   & $\bullet$   \\ 

    \rowcolor{gray!15} \citep{bellani2019towards}  &  2019  &  RUL  & $\bullet$   &      &       &    &         &         &          &  $\bullet$  &            &        &      &       &  Sequential Decision  &  Transfer, Reinforc.   &  {Reward}   &   &   \\ 

     \citep{li2020data}  &  2020  &  RUL  & $\bullet$   &      &       &    &         &         &          &  $\bullet$  &            &        &      &       &  GAN  &  Adversial   &  {RMSE, MAE, MAPE}   &   & $\bullet$  
    
  \\ \bottomrule
  
\end{tabular}}

\end{table}

\begin{table}[]
\resizebox{\textwidth}{!}{%
\begin{tabular}{cc>{\centering}m{1.5cm}cccccccccccc>{\centering}m{2cm}>{\centering}m{1.5cm}>{\centering}m{2.5cm}cc}
\toprule
\multirow{6}{*}{\textbf{Ref.}} & \multirow{6}{*}{\textbf{Year}} & \multirow{6}{*}{\textbf{Focus}}   & \multicolumn{4}{c}{\textbf{Target Area}}  & \multicolumn{8}{c}{\textbf{Data sources}}                                                                                  & \multicolumn{1}{c}{\multirow{6}{*}{\textbf{Approach}}}    & \multirow{6}{*}{\textbf{Paradigm}}    & \multirow{6}{*}{{\textbf{\begin{tabular}[c]{c}Evaluation\\ Metrics\end{tabular}}}} 
                               & \multirow{6}{*}{\rotatebox[origin=c]{90}{\textbf{Code}}} & \multirow{6}{*}{\rotatebox[origin=c]{90}{\textbf{Dataset}}}
                                \\ \cmidrule(lr){4-7} \cmidrule(lr){8-15}
                               &        &                                                                                                 & \multicolumn{1}{c}{\rotatebox[origin=c]{90}{\textbf{Technical}}} &  \multicolumn{1}{c}{\rotatebox[origin=c]{90}{\textbf{Application}}} &  \multicolumn{1}{c}{\rotatebox[origin=c]{90}{\textbf{Functional}}} &  \multicolumn{1}{c}{\rotatebox[origin=c]{90}{\textbf{Business}}} & \multicolumn{1}{c}{\rotatebox[origin=c]{90}{\textbf{Source Code}}} &  \multicolumn{1}{c}{\rotatebox[origin=c]{90}{\textbf{Topology}}} &  \multicolumn{1}{c}{\rotatebox[origin=c]{90}{\textbf{Event Logs}}} &  \multicolumn{1}{c}{\rotatebox[origin=c]{90}{\textbf{KPIs/Metrics}}} &  \multicolumn{1}{c}{\rotatebox[origin=c]{90}{\textbf{Traffic Network}}} &  \multicolumn{1}{c}{\rotatebox[origin=c]{90}{\textbf{Reports}}} &  \multicolumn{1}{c}{\rotatebox[origin=c]{90}{\textbf{Alerts}}} & \rotatebox[origin=c]{90}{\textbf{Traces}} & \multicolumn{1}{c}{}  & \multicolumn{1}{c}{} & \multicolumn{1}{c}{} & \multicolumn{1}{c}{}                                                                                                                  &                                                                                        \\ \midrule
    \rowcolor{gray!15} \citep{cohen2004correlating}    &  2004    &  Web Apps       &              &     $\bullet$      &     &     &    &         &          &  $\bullet$  &            &        &      &       &  TAN         &  SUP      &  {ACC, FPR, TPR}        &       &         \\

    \citep{chalermarrewong2012failure}    &  2012    &  Datacenters       &              &     $\bullet$      &     &     &    &         &          &  $\bullet$  &            &        &      &       &  ARMA         &  UNSUP      &  {RMSE, PREC, REC, FPR}        &       &         \\

    \rowcolor{gray!15}  \citep{fronza2013failure}    &  2013    &  Log Fail. &              &     $\bullet$      &     &     &    &         &     $\bullet$     &    &            &        &      &       &  SVM    &  SUP      &  {FPR, TPR, TNR}        &       &         \\ 

    \citep{zhang2016automated}    &  2016    &  Log Fail. &              &     $\bullet$      &     &     &    &         &     $\bullet$     &    &            &        &      &       &  LSTM    &  SUP      &  {AUC}        &       &         \\

    \rowcolor{gray!15}  \citep{islam2017predicting}    &  2017    &  Cloud &              &     $\bullet$      &  $\bullet$   &     &    &         &          &  $\bullet$  &            &        &      &   $\bullet$    &  LSTM    &  SUP      &  {F1, ACC, PREC, TPR, TNR, FPR}        &       &   $\bullet$      \\  

    \citep{pitakrat2018hora}    &  2018    &  Cloud &              &     $\bullet$      &  $\bullet$   &     &    &       $\bullet$  &          &  $\bullet$  &            &        &      &       &  Bayesian Nets, ARIMA    &  SUP      &  {AUC, PREC, REC, FPR}        &    $\bullet$   &   $\bullet$      \\  

    \rowcolor{gray!15}  \citep{lin2018predicting}    &  2018    &  Cloud  &              &     $\bullet$      &    &     &    &       $\bullet$  &          &  $\bullet$  &            &        &      &       &  LSTM, RF    &  SUP      &  {F1, PREC, REC}        &     &       \\  

    \citep{zhao2020real}    &  2020    &  Cloud  &              &     $\bullet$      &    &     &    &        &          &    &            &        &  $\bullet$    &       &  Xgboost, LIME    &  SUP      &  {F1, PREC, REC}        &     &       
  \\ \bottomrule
  
\end{tabular}}

\end{table}
\begin{table}[]
\caption{Study of particularities of AIOps Incident Prediction methods.}
\label{tab:particularities_prediction}
\renewcommand{\arraystretch}{1.0}
\resizebox{\textwidth}{!}{%
\begin{tabular}{cccccc}
\toprule
\textbf{Ref.} & \textbf{Interpretability} & \textbf{Scalability} & \textbf{Robustness} & \textbf{Temporal Evaluation}  & \textbf{Human-in-the-loop}                                           \\ \midrule
\rowcolor{gray!15} \citep{nagappan2006mining, dejaeger2012toward, ostrand2005predicting, nam2013transfer}    & $\bullet$ & & $\bullet$ & & \\
 \citep{menzies2006data}    & $\bullet$ & $\bullet$ & & &  \\
\rowcolor{gray!15} \citep{elish2008predicting}   &  & & $\bullet$ & & \\
\citep{moser2008comparative}  & $\bullet$ & & & & \\
\rowcolor{gray!15} \citep{wang2016automatically}  & & $\bullet$& $\bullet$ & & \\
\citep{li2017software}  & & & $\bullet$ & & \\
\rowcolor{gray!15} \citep{majd2020sldeep, xu2020defect}  & & $\bullet$ & & & \\ \midrule
\citep{garg1998methodology, alonso2010adaptive}  & $\bullet$ & & $\bullet$ & $\bullet$ & \\
\rowcolor{gray!15} \citep{vaidyanathan1999measurement}  & $\bullet$ & &  & $\bullet$ & \\ \midrule
\citep{zhao2010predicting}  & $\bullet$ & & & $\bullet$ & \\
\rowcolor{gray!15} \citep{wang2013online, xiao2018disk}  & $\bullet$ & & $\bullet$ & $\bullet$ & \\
\citep{xu2016health}  & & $\bullet$ & & $\bullet$ &  \\ \midrule
\rowcolor{gray!15} \citep{davis2017failuresim}  &  &  & $\bullet$ & $\bullet$ & \\
\citep{zhang2017syslog}  & $\bullet$ & $\bullet$ & & $\bullet$ & \\
\rowcolor{gray!15} \citep{sun2019system} & & $\bullet$ & & $\bullet$ & \\
\citep{khalil2020machine}  & & $\bullet$ & $\bullet$ & $\bullet$ & $\bullet$ \\ \midrule
\rowcolor{gray!15} \citep{zheng2017long, wu2018remaining}  & & & $\bullet$ & $\bullet$ & \\
\citep{ma2020deep}  & & & & $\bullet$ & \\
\rowcolor{gray!15} \citep{bellani2019towards}  & & $\bullet$ & $\bullet$ & $\bullet$ & \\
\citep{li2020data}  & &  $\bullet$ & & $\bullet$ & \\ \midrule
\rowcolor{gray!15} \citep{cohen2004correlating}    & $\bullet$ & $\bullet$ & $\bullet$  & $\bullet$ & $\bullet$ \\
\citep{chalermarrewong2012failure}  & $\bullet$ & $\bullet$ & & $\bullet$ & $\bullet$ \\
\rowcolor{gray!15} \citep{fronza2013failure}   & & $\bullet$ & $\bullet$ & & $\bullet$ \\ 
\citep{zhang2016automated}    & $\bullet$ & $\bullet$ & $\bullet$ & $\bullet$ &  \\
\rowcolor{gray!15} \citep{islam2017predicting}    & & $\bullet$ & & $\bullet$ &  \\
\citep{pitakrat2018hora}   & $\bullet$ & & & $\bullet$  & \\
\rowcolor{gray!15} \citep{lin2018predicting}  & & $\bullet$ & & & \\
\citep{zhao2020real}    & $\bullet$ & $\bullet$ & & $\bullet$ & 

\\ \bottomrule
\end{tabular}}
\end{table}

Incident prediction approaches are proactive methods designed to prevent failures (outages in extreme cases) by addressing both static aspects, such as source code, and dynamic aspects, such as the availability of computing resources. The ultimate objective is to suggest preventive measures or take immediate action as early as possible. These strategies vary extensively regarding the taxonomy we proposed (i.e., data used, area of application, etc.).

\smallbreak
\noindent \textbf{Software Defect Prediction.} SDP is an approach used to estimate the likelihood of encountering a software bug within a functional unit of code, such as a function, class, file, or module. The core assumption linking SDP to failure occurrence is that code with defects leads to errors and failures during execution. Traditionally, defect-prone software is identified using code metrics to construct defect predictors, as discussed in~\ref{subsec:data_sources}. \citet{nagappan2006mining} propose an SDP approach based on code complexity metrics. However, they highlight the challenge of multicollinearity among these metrics, making the problem more complex. To address this, they employ PCA to obtain a reduced set of uncorrelated features and use linear regression models for post-release defect prediction. They also try to employ models learned from one project when testing on different projects to evaluate cross-project applicability. This yields varied outcomes, leading to the assertion that project similarity is a crucial factor for successful transfer learning.~\citet{menzies2006data}, shift their focus from static code metrics to the choice of prediction models, advocating for the use of Naive Bayes with logarithmic features. In~\citep{elish2008predicting}, the authors examined the effectiveness of SVM in predicting defect-prone software modules, including functions in procedural software and methods in object-oriented software. They rely on code metrics, particularly McCabe metrics\citep{mccabe1976complexity} and Halstead metrics~\citep{halstead1977elements}, and compare its predictive performance with eight well-known statistical and machine learning models across four NASA datasets.~\citet{dejaeger2012toward} explored 15 different bayesian network classifiers to identify alternatives to Naive bayes that could lead to simpler networks with fewer nodes and arcs while maintaining or improving predictive performance. The study also examines the applicability of the Markov blanket principle for feature selection within the BN framework. Through the conducted evaluations, the paper demonstrates that these alternative BN classifiers can yield simpler, yet comprehensible networks with competitive predictive performance compared to the Naive Bayes classifier. The findings also emphasize the importance of balancing the interpretability and predictive power of these models. 

While the previously mentioned SDP contributions mainly centered around single-release viewpoints, another set of studies, referred to as changelog approaches, emphasize the significance of software history as a more impactful element for predicting defect density. For instance, factors like code age or the count of previous defects can serve as indicators for estimating the occurrence of new bugs~\citep{graves2000predicting}.~\citet{ostrand2005predicting} delve into the analysis of changes within extensive software systems and their correlation with previous faults, aiming to anticipate the count of defects in upcoming releases. They employ a Poisson generalized linear model, utilizing maximum likelihood estimates of model parameters to assess the significance of diverse metrics. The model is evaluated within the release cycle of an internal inventory system, encompassing various new file-level metrics such as programming language, edit flags, and age. Findings conclude that the top 20\% of files with the highest projected fault count encompassed an average of 83\% of subsequently identified faults. Conducting a comparative assessment of two sets of SDP metrics (code and change metrics),~\citet{moser2008comparative} explored the Eclipse project repository. They employed three distinct machine learning methods including Naïve Bayes, logistic regression, and decision trees. The analysis demonstrated that individual usage of change metrics is more effective than relying solely on code metrics to identify defective source files. Furthermore, a combined approach shows modest enhancements or comparable results to a change metric-oriented approach. 

~\citet{nam2013transfer} tackled the cross-project defect prediction challenge, specifically focusing on scenarios involving new or resource-limited projects. In such cases, the authors sought to leverage training data from established source projects and adapt it to target projects. To achieve this, they employed a cutting-edge transfer learning technique known as TCA (Transfer Component Analysis) to harmonize feature distributions across the source and target projects. Additionally, they introduced an extended version called \texttt{TCA+} that incorporates these aligned features into a logistic regression model for accurate prediction of faulty module files. The approach has been evaluated on a benchmark SDP dataset \texttt{AEEM} proposed by~\citet{d2012evaluating}. Critiques of traditional code metrics point out their handcrafted and simplistic nature. An alternative approach involves parsing the source code using ASTs.~\citet{wang2016automatically} question the ability of code metrics to capture semantics and distinguish between code regions with the same structure but different semantics. They propose using latent semantic representations and training a Deep Belief Network (DBN) on AST-parsed code to learn semantic features in an unsupervised fashion.~\citet{li2017software} presented \texttt{DP-CNN}, a novel framework called Defect Prediction via Convolutional Neural Network. This approach involves the extraction of a subset of AST nodes that capture diverse semantic operations during parsing. These nodes are then transformed into numerical features through mapping and word embedding processes. Subsequently, the transformed features are inputted into a 1D convolutional architecture, to automatically learn semantic and structural program features.

Unlike existing techniques that operate on coarse-grained units such as modules, classes, or files, \citet{majd2020sldeep} introduced a novel approach called \texttt{SLDeep}, which aims to identify fault-prone areas of code at a much finer-grained statement level. \texttt{SLDeep} defines a comprehensive suite of 32 statement-level metrics, encompassing factors like the utilization of binary and unary operators within a statement. The selected learning model for implementation is the LSTM model. The experiments were conducted on a diverse set of 119,989 C/C++ programs from the \texttt{Code4Bench} benchmark \citet{majd2019code4bench}. In another study, \citet{xu2020defect} proposed an approach for characterizing software defects using defect subtrees in ASTs. This approach incorporates information from fix-inducing changes and code concepts. Initially, a topic model is developed to summarize functional concepts related to defects. Each node within the defect subtrees is enriched with attributes such as types, fix-inducing changes, and code concepts. Subsequently, a GNN classifier is employed, where subtrees are represented as directed acyclic graph structures. More recently, \citet{uddin2022software} utilized a BiLSTM model in conjunction with a BERT-based semantic feature approach (\texttt{SDP-BB}) to predict defects within software. This combination captures semantic features of the code by extracting contextual information from token vectors learned by the BERT model using the BiLSTM. An attention mechanism is also integrated to capture the most crucial features for prediction. This methodology is enhanced by a data augmentation technique that generates supplementary training data. The evaluation involves both within-project defect prediction and cross-project defect prediction experiments.

\smallbreak
\noindent \textbf{Software Aging and Rejuvenation.} Software aging describes a phenomenon in which a software system experiences a gradual decline in performance and reliability as time passes. Recognized causes of software aging include memory leaks, bloats, unreleased file locks, data fragmentation, and the accumulation of numerical errors~\citep{castelli2001proactive,notaro2021survey}. Conversely, software rejuvenation pertains to proactive or corrective actions taken to eliminate accumulated error conditions and liberate system resources. These actions may include garbage collection, flushing kernel tables, re-initializing internal data structures, and similar strategies.~\citet{garg1998methodology} propose a method for estimating the time-to-exhaustion of various system resources, including free memory, file, and process table sizes, and used swap space. They utilize regression techniques and seasonal testing to identify trends and quantify the exhaustion time. In a related study,~\citet{vaidyanathan1999measurement} explore the impact of software aging resulting from the current system workload. They develop a semi-Markov reward model based on available workload and resource data, where different workload scenarios are represented as model states. The association to a specific state is determined using k-means clustering. To estimate the time-to-exhaustion of memory and swap space, a non-parametric regression technique is employed separately for each workload state. The challenge of non-linear and piece-wise linear resource consumption is tackled by~\citep{alonso2010adaptive} by utilizing an ensemble of linear regression models. These models are selected using a decision tree based on the same input features as the regression model, which consists of a combined set of hardware and software host metrics. In the study by~\citet{sudhakar2014software}, the authors advocate the utilization of a neural network architecture to capture intricate non-linear connections between resource usage and time to failure in cloud systems. 

\smallbreak
\noindent \textbf{Hardware Failures Prediction.} In large-scale computing infrastructures, ensuring hardware reliability is crucial for achieving service availability goals. However, due to the sheer number of components involved and the necessity to use commodity hardware in data centers, hardware failures pose a significant challenge. For example, Google has reported that 20-57\% of disks experience at least one sector error over a 4-6 year period~\citep{meza2015large}. Hard drives are the most frequently replaced components in large cloud computing systems, and they are a leading cause of server failure~\citep{vishwanath2010characterizing}. To address this, hard-drive manufacturers have implemented self-monitoring technologies like SMART metrics in their storage products. In the approach presented by~\citet{zhao2010predicting}, Hidden Markov and Semi-Markov Models are used to estimate likely event sequences based on SMART metric observations from a dataset of around 300 disks (with approximately two-thirds being healthy). Two models, one trained from healthy disk sequences and the other from faulty disk sequences, are used to estimate the sequence log-likelihood at test time, with the class being determined by the highest score.~\citet{wang2013online} propose a similarity-based detection algorithm that selects relevant SMART features using Minimum Redundancy Maximum Relevance (mRMR) and projects the input data into a Mahalanobis space constructed from the healthy disk population. This approach aims to detect faulty disks that deviate more from the distribution.~\citet{xu2016health} introduce the use of RNNs to model the long-term relationships in SMART data. Unlike binary classification approaches, their model is trained to predict the health status of disks, providing additional information on the remaining useful life and serving as a ranking approach. These approaches are typically used in an online setting after an offline training step. However, integrating additional data and updating the characteristics of faulty disks as new failures occur presents a challenge. To address this, the approach presented by~\citep{xiao2018disk} proposes the use of online random forests, a model that can adaptively evolve with changing data distributions through online labeling.

Regarding other hardware components, FailureSim~\citep{davis2017failuresim} presents an approach to evaluate the status of hardware in cloud data centers by employing both multi-layer perceptrons and RNNs. The proposed method concentrates on assessing 13 distinct host failure states associated with specific components such as CPU, memory, and I/O. On the other hand, addressing switch failures within datacenter networks, \citet{zhang2017syslog} introduced a novel model termed the frequent template tree (FT-tree). This model identifies frequent word combinations within the historical system log and employs them as message templates, which are then correlated with faulty behavior. Subsequently, a Hidden Semi-Markov Model is trained using the identified templates to predict failures. \citet{sun2019system} put forward a proactive hardware failure prediction scheme based on deep learning. This scheme primarily focuses on disk and memory faults, utilizing both SMART data and system logs. Overcoming the challenge of effectively handling discrete time-series data, the scheme normalizes attribute distributions from diverse vendors. The authors introduce a specialized Temporal Convolution Neural Network (TCNN) model designed to handle temporal noise, incorporating a tailored loss function for training with highly imbalanced samples. On a different note, \citet{khalil2020machine} introduced an approach aimed at anticipating potential hardware failures attributed to aging or varying conditions of Open circuits. The approach employs Fast Fourier Transform (FFT) to capture fault frequency signatures, utilizes Principal Component Analysis (PCA) to distill critical data with reduced dimensionality, and employs a CNN for learning and classifying faults. This noteworthy work stands out as the first to address fault prediction at the transistor level for hardware systems, encompassing aging, short-circuit, and open-circuit faults.

\smallbreak
\noindent \textbf{Remaining Useful Lifetime Estimation.} The Remaining Useful Life (RUL) stands as a pivotal real-time performance indicator for operating systems throughout their operational lifespan, representing the time remaining until the system ceases to function. Similar to software aging techniques, precise RUL estimation holds significant importance in planning condition-based maintenance tasks, with the aim of minimizing system downtime. A multitude of data-driven approaches have emerged to model the intricate behavior of system components. As highlighted in~\citep{berghout2022systematic}, LSTM emerges as a highly suitable tool for handling dynamic data in RUL problems. For instance, an LSTM model presented by~\citet{zheng2017long} adeptly harnesses sensor sequence information to uncover latent patterns tied to various operating conditions, faults, and degradation models. Another study by~\citet{wu2018remaining} delves into a vanilla LSTM model, a prevalent variant of LSTM often applied in language processing, for predicting the RUL of aircraft engines.~\citet{ma2020deep} conducted a novel hybrid approach called convolution-based long short-term memory (CLSTM). This approach combines CNN with LSTM networks and is designed to predict the RUL of rotating machinery, a crucial aspect of prognostics and health management (PHM). The CLSTM architecture retains the advantages of LSTM while simultaneously integrating time-frequency features. This enables the model to capture long-term dependencies and extract time-frequency domain features concurrently. Through the stacking of multiple CLSTM layers and the creation of an encoding-forecasting architecture, a deep learning model is formulated for RUL prediction. Notably, Reinforcement Learning has also found application in enhancing model adaptability through learning from experiences, even in cases of erroneous decisions. Simulation environments prove particularly advantageous in this context. For instance, a transfer learning approach developed in~\citep{bellani2019towards} leverages states, actions, and rewards to generate an optimal reward policy. These algorithms are specifically geared towards sequentially predicting RUL for a specific type of pumping system. Addressing the challenges surrounding the determination of initial prediction times for remaining useful life, \citet{li2020data} introduced a GAN approach that learns data distributions from healthy machine states. A health indicator is used to determine the initial prediction time, and adversarial training is employed to align data from various machines, thereby extracting generalized prognostic knowledge.

\smallbreak
\noindent \textbf{Software Failure Prediction.} Predicting system failures from an application perspective involves exploring potential failures that may arise across various dimensions, including jobs, tasks, processes, VMs, containers, or nodes. Existing approaches to address this issue predominantly rely on system metrics, service states, traces, and topology. For instance,~\citet{cohen2004correlating} presented an approach centered on Tree-augmented Bayesian Networks (TANs) to establish connections between observed variables and abstract service states. This strategy facilitates the anticipation and prevention of Service Level Objective (SLO) breaches and failures in the context of three-tiered Web services. The system monitors crucial system metrics such as CPU time, disk reads, and swap space, constructing a model that captures the intricate dependencies between them. By employing a heuristic selection process, the optimal graph structure, encompassing the most relevant input metrics, is determined. An approach to forecast system availability within datacenters has been introduced in a study by~\citep{chalermarrewong2012failure}. This framework employs ARMA model and fault-tree analysis. The framework identifies symptoms at the component level, such as elevated CPU temperature, malfunctioning disk sectors, and memory depletion, which function as the terminal points of the fault tree. By incorporating these symptoms and the underlying tree structure, a model encompassing dependencies in combinational logic deterministically establishes the availability state. This enables the proactive detection of errors before they escalate into failures.

In~\citep{fronza2013failure}, the authors addressed the task of predicting system failures through a log file analysis approach. This involves processing log files using the Random Indexing technique, which effectively captures intricate contextual information embedded within sequences of log messages. This ensures a nuanced representation of changes in the system's state over time. Subsequently, a weighted Support Vector Machine (SVM) approach is implemented to classify these sequences, assigning them to either impending failure scenarios or non-failure instances. Notably, the method accounts for the common challenge of imbalanced datasets, particularly emphasizing the need to maintain robust true positive rates. Similarly, the work by~\citet{zhang2016automated} unveils a method for predicting system failures through real-time analysis of console logs. This novel approach introduces log pattern extraction via clustering, which groups together log formats and content sharing similarities. Treating log patterns as analogous to words and organizing them into discrete time intervals as documents, this approach significantly simplifies the feature space, providing valuable insights into the system's current state. These identified states are then utilized as inputs to an LSTM model, facilitating the prediction of potential failures. LSTM networks also find application in an extensive characterization study carried out by~\citet{islam2017predicting} on a workload trace dataset obtained from Google. Within this study, the focus lies on predicting failures at both the job and task levels. A job comprises multiple tasks, wherein each task corresponds to a single-machine command or program. Failures are forecasted by leveraging resource usage, performance metrics, and task particulars, encompassing completion status as well as attributes linked to users, nodes, and jobs.

Drawing on a holistic strategy, the \texttt{HORA} prediction system~\citep{pitakrat2018hora} capitalizes on architectural knowledge along with live KPIs data to forecast potential Quality of Service (QoS) violations and service disruptions within distributed software systems. Central to this approach are Bayesian Networks, which play a pivotal role in constructing models that depict component interdependencies and the propagation of failures. These models establish connections between anticipated component failures, derived through auto-regressive predictors from system metrics, and larger-scale systemic challenges.~\citet{lin2018predicting} introduced an approach called MING to predict potential node failures within cloud service systems, which integrates LSTM and Random Forest models to incorporate both temporal and spatial data.  Additionally, the technique employs a ranking model and a cost-sensitive function to effectively assess and rank nodes based on their susceptibility to failure.  Demonstrating its practical effectiveness, the approach finds successful application in industrial scenarios, notably within the context of Microsoft services. \texttt{eWarn}~\citep{zhao2020real} is an approach tailored for online service systems that capitalizes on historical data and real-time alert information to forecast the likelihood of upcoming incidents. This is achieved through a combination of innovative techniques: effective feature engineering to represent pertinent alert patterns, integration of multi-instance learning to mitigate the influence of irrelevant alerts, and the generation of interpretable prediction reports using the LIME explanation technique.

\subsection{Incident Prioritization Methods} 
\label{sec:sota_review_ranking}

\begin{table}[]
\caption{Summary of reviewed AIOps Incident Priorization methods.}
\label{tab:tab_priorization}
\renewcommand{\arraystretch}{1.2}

\resizebox{\textwidth}{!}{%
\begin{tabular}{cc>{\centering}m{1.5cm}cccccccccccc>{\centering}m{2cm}>{\centering}m{1.5cm}>{\centering}m{2.5cm}cc}
\toprule
\multirow{6}{*}{\textbf{Ref.}} & \multirow{6}{*}{\textbf{Year}} & \multirow{6}{*}{\textbf{Focus}}   & \multicolumn{4}{c}{\textbf{Target Area}}  & \multicolumn{8}{c}{\textbf{Data sources}}                                                                                  & \multicolumn{1}{c}{\multirow{6}{*}{\textbf{Approach}}}    & \multirow{6}{*}{\textbf{Paradigm}}    & \multirow{6}{*}{{\textbf{\begin{tabular}[c]{c}Evaluation\\ Metrics\end{tabular}}}} 
                               & \multirow{6}{*}{\rotatebox[origin=c]{90}{\textbf{Code}}} & \multirow{6}{*}{\rotatebox[origin=c]{90}{\textbf{Dataset}}}
                                \\ \cmidrule(lr){4-7} \cmidrule(lr){8-15}
                               &        &                                          & \multicolumn{1}{c}{\rotatebox[origin=c]{90}{\textbf{Technical}}} &  \multicolumn{1}{c}{\rotatebox[origin=c]{90}{\textbf{Application}}} &  \multicolumn{1}{c}{\rotatebox[origin=c]{90}{\textbf{Functional}}} &  \multicolumn{1}{c}{\rotatebox[origin=c]{90}{\textbf{Business}}} & \multicolumn{1}{c}{\rotatebox[origin=c]{90}{\textbf{Source Code}}} &  \multicolumn{1}{c}{\rotatebox[origin=c]{90}{\textbf{Topology}}} &  \multicolumn{1}{c}{\rotatebox[origin=c]{90}{\textbf{Event Logs}}} &  \multicolumn{1}{c}{\rotatebox[origin=c]{90}{\textbf{KPIs/Metrics}}} &  \multicolumn{1}{c}{\rotatebox[origin=c]{90}{\textbf{Traffic Network}}} &  \multicolumn{1}{c}{\rotatebox[origin=c]{90}{\textbf{Reports}}} &  \multicolumn{1}{c}{\rotatebox[origin=c]{90}{\textbf{Alerts}}} & \rotatebox[origin=c]{90}{\textbf{Traces}} & \multicolumn{1}{c}{}  & \multicolumn{1}{c}{} & \multicolumn{1}{c}{} & \multicolumn{1}{c}{}                                   &                                                                                        \\ \midrule
\rowcolor{gray!15}    \citep{tian2013drone}    &  2013    &  Software Bugs       &              &         &  $\bullet$    &     $\bullet$  &   &     $\bullet$    &          &   &       &  $\bullet$   &      &       &  Lin. Reg         &  SUP      &  {F1, PREC, REC}        &       &  $\bullet$       \\

    \citep{lin2018collaborative}    &  2018    &  N/A       &              &      $\bullet$    &     &     &   &     $\bullet$    &          &   &       &     & $\bullet$     &       &  Hierarchical Bayesian Net         &  SUP      &  {ROC, PRC}        &       &       \\

\rowcolor{gray!15}    \citep{hassan2019nodoze}    &  2019    &  IDS       &              &         &  $\bullet$    &     &   &     $\bullet$    &          &   &       &     & $\bullet$     &       &  Diffusion Model         &  UNSUP      &  {ROC}        &       &       \\

    \citep{zhao2020automatically}    &  2020    &  N/A       &    $\bullet$           &      $\bullet$    &  $\bullet$    &   $\bullet$   &   &     $\bullet$    &     $\bullet$      &  $\bullet$   &       &     & $\bullet$     &       &  XGBoost   &  SUP      &  {F1, PREC, REC}        &       &       \\

\rowcolor{gray!15}    \citep{chen2020incidental}    &  2020    &  Software Bugs   &    $\bullet$           &      $\bullet$    &  $\bullet$    &   $\bullet$   &   &     $\bullet$    &     &     &       &  $\bullet$   & $\bullet$     &       &  CNN   &  SUP    &  {AUC, PREC, REC}        &    $\bullet$   &    $\bullet$   \\
    
    \citep{gupta2022improving}    &  2022    &  Software Bugs  &          &          &  $\bullet$    &   $\bullet$   & $\bullet$   &        &     &     &       &  $\bullet$   &      &       &  TOPSIS, BFOA, BAR   &  SUP    &  {F1, PREC, REC}        &      &    $\bullet$   

  \\ \bottomrule
  
\end{tabular}}

\end{table}
\begin{table}[]
\caption{Study of particularities of AIOps Incident Prioritization methods.}
\label{tab:particularities_priorization}
\renewcommand{\arraystretch}{1.0}
\resizebox{\textwidth}{!}{%
\begin{tabular}{cccccc}
\toprule
\textbf{Ref.} & \textbf{Interpretability} & \textbf{Scalability} & \textbf{Robustness} & \textbf{Temporal Evaluation}  & \textbf{Human-in-the-loop}                                           \\ \midrule

\rowcolor{gray!15} \citep{tian2013drone}    & $\bullet$ & & & & \\
\citep{lin2018collaborative}    & $\bullet$ & $\bullet$ & & $\bullet$ \\
\rowcolor{gray!15} \citep{hassan2019nodoze, zhao2020automatically}    & $\bullet$ & $\bullet$ & $\bullet$ &$\bullet$ & $\bullet$  \\
\citep{chen2020incidental}  &  &  $\bullet$ &  $\bullet$ &  $\bullet$ &  $\bullet$ \\
\rowcolor{gray!15} \citep{gupta2022improving}  & $\bullet$ & & $\bullet$ & & 

\\ \bottomrule
\end{tabular}}
\end{table}

As a large number of incidents can be reported simultaneously, it becomes time-consuming and resource-intensive to handle all of them at once. However, certain incidents require immediate attention due to their importance or severity. To address this issue, various data-driven approaches have been proposed to rank incidents or alerts based on prioritization factors. Some of these techniques can also be applied to other scenarios, as they are not solely dedicated to ranking, but also involve detection or diagnosis mechanisms. 

In their work, \citet{tian2013drone} introduced the \texttt{DRONE} framework, which provides recommendations for prioritizing bug reports. This is achieved by considering a multitude of factors, including temporal information, textual content, author details, related reports, severity, and product information. These factors are extracted as features from incident reports and subsequently used to train a discriminative model adept at handling ordinal class labels and imbalanced data. The model employs linear regression to capture the connection between the features and priority levels, followed by a thresholding approach to calibrate class labels (i.e., priority levels). In response to the challenge of characterizing and ranking diverse categorical alerts from an anomaly detection system, \citet{lin2018collaborative} introduced Collaborative Alert Ranking (CAR). CAR addresses both mining alert patterns and reducing false positives. Initially, a hierarchical Bayesian model is constructed to capture short-term and long-term dependencies within alert sequences. An entity embedding-based model is then devised to uncover content correlations among alerts based on their categorical attributes. By integrating temporal and content dependencies within a unified optimization framework, CAR provides rankings for individual alerts as well as their associated patterns. Similarly, aiming to mitigate threat alert fatigue and prioritize true attacks over false alarms, \citet{hassan2019nodoze} introduced \texttt{NODOZE}. This ranking system constructs causal dependency graphs for alert events using contextual and historical information. Anomaly scores are assigned to edges based on event frequency and are propagated using a novel network diffusion algorithm.

AlertRank, proposed by \citet{zhao2020automatically}, is an automatic and adaptive framework for identifying severe alerts. This approach leverages a variety of interpretable features, including textual and temporal alert features, as well as domain knowledge-based univariate and multivariate KPIs. The XGBoost ranking algorithm is employed to identify severe alerts, with continuous labeling utilized to obtain ground-truth labels for training and testing. \citet{chen2020incidental} conducted a large-scale empirical analysis of incidents from real-world online service systems. Their findings highlight a category of incidents labeled as \textit{Incidental Incidents}, often considered less significant and not prioritized for immediate resolution. To address this, the authors proposed \textit{DeepIP} (Deep learning-based Incident Prioritization), which employs an attention-based CNN to identify and prioritize incidental incidents using historical incident descriptions and topology information. On a distinct note, \citet{gupta2022improving} adopted a unique approach by combining the multi-criteria fuzzy Technique for Order of Preference by Similarity to Ideal Solution (TOPSIS) with the Bacterial Foraging Optimization Algorithm (BFOA) and Bar Systems (BAR) optimization. This amalgamation serves to rank software bugs and simultaneously select suitable developers.

\subsection{Incident Assignment Methods}
\label{sec:sota_review_Assignment}

\begin{table}[]
\caption{Summary of reviewed AIOps Incident Assignment methods.}
\label{tab:tab_prediction}
\renewcommand{\arraystretch}{1.2}

\resizebox{\textwidth}{!}{%
\begin{tabular}{cc>{\centering}m{1.5cm}cccccccccccc>{\centering}m{2cm}>{\centering}m{1.5cm}>{\centering}m{2.5cm}cc}
\toprule
\multirow{6}{*}{\textbf{Ref.}} & \multirow{6}{*}{\textbf{Year}} & \multirow{6}{*}{\textbf{Focus}}   & \multicolumn{4}{c}{\textbf{Target Area}}  & \multicolumn{8}{c}{\textbf{Data sources}}                                                                                  & \multicolumn{1}{c}{\multirow{6}{*}{\textbf{Approach}}}    & \multirow{6}{*}{\textbf{Paradigm}}    & \multirow{6}{*}{{\textbf{\begin{tabular}[c]{c}Evaluation\\ Metrics\end{tabular}}}} 
                               & \multirow{6}{*}{\rotatebox[origin=c]{90}{\textbf{Code}}} & \multirow{6}{*}{\rotatebox[origin=c]{90}{\textbf{Dataset}}}
                                \\ \cmidrule(lr){4-7} \cmidrule(lr){8-15}
                               &        &                                          & \multicolumn{1}{c}{\rotatebox[origin=c]{90}{\textbf{Technical}}} &  \multicolumn{1}{c}{\rotatebox[origin=c]{90}{\textbf{Application}}} &  \multicolumn{1}{c}{\rotatebox[origin=c]{90}{\textbf{Functional}}} &  \multicolumn{1}{c}{\rotatebox[origin=c]{90}{\textbf{Business}}} & \multicolumn{1}{c}{\rotatebox[origin=c]{90}{\textbf{Source Code}}} &  \multicolumn{1}{c}{\rotatebox[origin=c]{90}{\textbf{Topology}}} &  \multicolumn{1}{c}{\rotatebox[origin=c]{90}{\textbf{Event Logs}}} &  \multicolumn{1}{c}{\rotatebox[origin=c]{90}{\textbf{KPIs/Metrics}}} &  \multicolumn{1}{c}{\rotatebox[origin=c]{90}{\textbf{Traffic Network}}} &  \multicolumn{1}{c}{\rotatebox[origin=c]{90}{\textbf{Reports}}} &  \multicolumn{1}{c}{\rotatebox[origin=c]{90}{\textbf{Alerts}}} & \rotatebox[origin=c]{90}{\textbf{Traces}} & \multicolumn{1}{c}{}  & \multicolumn{1}{c}{} & \multicolumn{1}{c}{} & \multicolumn{1}{c}{}                                   &                                                                                        \\ \midrule
  \rowcolor{gray!15}  \citep{murphy2004automatic}    &  2004    &  Software Bugs     &              &         &  $\bullet$    &     $\bullet$  &   &        &          &   &       &  $\bullet$   &      &       &  Bayesian Net        &  SUP      &  {ACC}        &       &  $\bullet$       \\

    \citep{shao2008efficient}    &  2008    &  N/A    &      $\bullet$        &  $\bullet$       &  $\bullet$    &     $\bullet$  &   &        &          &   &       &    &      &    $\bullet$    &  Markov Model       &  SUP      &  {ACC}        &       &     \\

   \rowcolor{gray!15}  \citep{xuan2010automatic}    &  2010    &  Software Bugs     &            &        &  $\bullet$    &     $\bullet$  &   &        &          &   &       &  $\bullet$   &     &       &  Bayesian Net, EM      &  SemiSUP     &  {ACC}        &       &  $\bullet$    \\   

    \citep{bhattacharya2010fine}    &  2010    &  Software Bugs     &            &        &  $\bullet$    &     $\bullet$  &   &        &          &   &       &  $\bullet$   &     &  $\bullet$     &  Naive Bayes      &  SUP     &  {ACC}        &       &  $\bullet$    \\   

    \rowcolor{gray!15} \citep{alenezi2013efficient}    &  2013    &  Software Bugs     &            &        &  $\bullet$    &     $\bullet$  &   &        &          &   &       &  $\bullet$   &     &      &  Naive Bayes      &  SUP     &  {F1, PREC, REC}        &       &  $\bullet$    \\   

    \citep{wang2014fixercache}    &  2014    &  Software Bugs     &            &        &  $\bullet$    &     $\bullet$  &   &        &          &   &       &  $\bullet$   &     &  $\bullet$    &  Similarity Computation      &  UNSUP     &  {ACC, Diversity}        &       &  $\bullet$    \\

   \rowcolor{gray!15}  \citep{xi2019bug}    &  2019    &  Software Bugs     &            &        &  $\bullet$    &     $\bullet$  &   &        &          &   &       &  $\bullet$   &     &  $\bullet$    &    GRU    &  SUP     &  {ACC}      &       &  $\bullet$    \\

    \citep{DBLP:conf/sigsoft/LeeHLKJ17}    &  2017    &  Software Bugs     &            &        &  $\bullet$    &     $\bullet$  &   &        &          &   &       &  $\bullet$   &     &     &    CNN    &  SUP     &  {ACC}      &       &    \\ 

    \rowcolor{gray!15} \citep{chen2019continuous}    &  2019    &  N/A    &      $\bullet$      &   $\bullet$     &  $\bullet$    &     $\bullet$  &   &   $\bullet$     &          &   &       &  $\bullet$   &     &     &    CNN, GRU    &  SUP     &  {ACC}      &       &    \\ 

    \citep{DBLP:conf/kdd/PhamJDOJ20}    &  2020    &  N/A   &      $\bullet$      &   $\bullet$     &  $\bullet$    &     $\bullet$  &   &   $\bullet$     &          &   &       &  $\bullet$   &     &     &    CNN   &  SUP     &  {ACC}      &       &    \\ 

    \rowcolor{gray!15} \citep{DBLP:conf/dsaa/RemilBPRK21}    &  2021    &  N/A   &      $\bullet$      &   $\bullet$     &  $\bullet$    &     $\bullet$  &   &   $\bullet$     &          &  $\bullet$  &       &  $\bullet$   &  $\bullet$    &     &    Subgroup Discovery   &  SUP     &  {F1, Fidelity}      &   $\bullet$     &

  \\ \bottomrule
  
\end{tabular}}

\end{table}
\begin{table}[]
\caption{Study of particularities of AIOps Incident Assignment methods.}
\label{tab:particularities_assignment}
\renewcommand{\arraystretch}{1.0}
\resizebox{\textwidth}{!}{%
\begin{tabular}{cccccc}
\toprule
\textbf{Ref.} & \textbf{Interpretability} & \textbf{Scalability} & \textbf{Robustness} & \textbf{Temporal Evaluation}  & \textbf{Human-in-the-loop}                                           \\ \midrule

\rowcolor{gray!15} \citep{murphy2004automatic}  & $\bullet$  &  &  &  &  \\
\citep{shao2008efficient,DBLP:conf/dsaa/RemilBPRK21}    & $\bullet$ & $\bullet$ & $\bullet$ & & \\
\rowcolor{gray!15} \citep{xuan2010automatic}    & $\bullet$ & & & $\bullet$ & \\
\citep{bhattacharya2010fine}    & $\bullet$ & $\bullet$ & $\bullet$ & $\bullet$ & $\bullet$ \\
\rowcolor{gray!15} \citep{alenezi2013efficient}    & $\bullet$ & $\bullet$ & & &  \\
\citep{wang2014fixercache}    & $\bullet$ & $\bullet$ & & $\bullet$ & \\
\rowcolor{gray!15} \citep{xi2019bug}    & & $\bullet$ & & $\bullet$ & $\bullet$ \\
\citep{DBLP:conf/sigsoft/LeeHLKJ17,DBLP:conf/kdd/PhamJDOJ20}    & & $\bullet$ & & & \\
\rowcolor{gray!15} \citep{chen2019continuous}    & & $\bullet$ & $\bullet$ & $\bullet$ & \\
 \bottomrule
\end{tabular}}
\end{table}

Numerous data-driven approaches have been proposed to optimize the process of incident assignment (referred also to as routing, or triaging) by automatically assigning incidents to the appropriate service team and individual. Typically, these approaches involve training a classifier using historical incident reports that contain textual information, topology data, or prioritization scores. The trained classifier is then used to assign new incidents. 

The preceding works have primarily relied on text preprocessing methods and supervised classification models within traditional machine learning and statistical frameworks. For instance, \citet{murphy2004automatic} employed a Bayesian learning approach to incident reports, representing them as a bag of words from a predefined vocabulary, which conveniently adapts to multi-class classification. However, this method's accuracy in predicting report assignments to developers was found to be limited, achieving only 30\% correct predictions. In~\citep{shao2008efficient}, a unique perspective is taken by solely utilizing incident resolution sequences without accessing the incident ticket content. This is achieved by developing a Markov model that captures the decision-making process behind successful problem resolution paths. The order of the model is thoughtfully chosen based on conditional entropy derived from ticket data. Addressing the challenge of limited labeled incident reports, \citet{xuan2010automatic} presented a semi-supervised text classification technique. They combine a Naive Bayes classifier with expectation-maximization, leveraging both labeled and unlabeled incident reports. An iterative process involves initial classifier training with labeled bug reports, followed by iterative labeling of unlabeled bug reports to enhance classifier training. A weighted recommendation list further refines performance, utilizing multiple developer weights during classifier training.

\citet{bhattacharya2010fine} tackled the task by employing various techniques, including refined classification with additional attributes, intra-fold updates during training, a precise ranking function for recommending developers in tossing graphs, and multi-feature tossing graphs. Results not only exhibit a noteworthy accuracy but also demonstrate a substantial reduction in tossing path lengths. Similarly,~\citet{alenezi2013efficient} employed a series of selection term techniques to allocate software bugs to experienced developers. The process is initiated with the application of a conventional text processing methodology, transforming textual data into a coherent and meaningful representation. Subsequently, a bug-term matrix was crafted, incorporating term frequency weights. Various term selection methods were then employed to effectively mitigate both data dimensionality and sparsity issues. These included techniques like Log Odds Ratio, Term Frequency Relevance Frequency, and Mutual Information. Following this, a Naive Bayes classifier was trained on the resulting representations.~\citet{wang2014fixercache} introduced \texttt{FixerCache}, an unsupervised approach that leverages developers' component-specific activeness to create dynamic developer caches for each product component. When a new bug report arises, \texttt{FixerCache} recommends highly active fixers from the developer cache to participate in resolving the bug. The developer cache is dynamically updated following the verification and resolution of each bug report.~\citet{xi2019bug} emphasized the role of tossing sequence paths in predicting suitable service teams or individuals for handling software bugs. Their proposal, iTriage, employs a sequence-to-sequence model to jointly capture features from textual content and tossing sequences, integrating them through an encoder-decoder classification model.

In recent times, the focus has shifted towards advanced natural language processing techniques combined with sophisticated deep learning approaches. For example, \citet{DBLP:conf/sigsoft/LeeHLKJ17} pioneered the use of a convolutional neural network and pre-trained Word2Vec embeddings for incident assignment. \texttt{DeepCT}~\citep{chen2019continuous} presents an approach that views incident triage as a continuous process involving intensive discussions among engineers. It employs a GRU-based model with an attention-based mask strategy and a revised loss function to learn knowledge from discussions and update triage results incrementally. \texttt{DeepTriage}~\citep{DBLP:conf/kdd/PhamJDOJ20} was devised to address incident assignment challenges such as imbalanced incident distribution, diverse input data formats, scalability, and building engineers' trust. This approach employs CNN models to recommend the responsible team for each incident. Focusing on interpretability, \citet{DBLP:conf/dsaa/RemilBPRK21} proposed a framework that includes preprocessing steps like lemmatization using a pre-trained transformer model. Incident textual information is vectorized using TF-IDF and fed into an LSTM-based attention model for predicting the suitable service team. Furthermore, a Subgroup Discovery approach groups predicted incident tickets and labels into subgroups supporting the same explanation of the model's decision-making process. 

\subsection{Incident Classification Methods}
\label{sec:sota_review_classification}

\begin{table}[]
\caption{Summary of reviewed AIOps Incident Classification methods.}
\label{tab:tab_classification}
\renewcommand{\arraystretch}{1.2}

\resizebox{\textwidth}{!}{%
\begin{tabular}{cc>{\centering}m{1.5cm}cccccccccccc>{\centering}m{2cm}>{\centering}m{1.5cm}>{\centering}m{2.5cm}cc}
\toprule
\multirow{6}{*}{\textbf{Ref.}} & \multirow{6}{*}{\textbf{Year}} & \multirow{6}{*}{\textbf{Focus}}   & \multicolumn{4}{c}{\textbf{Target Area}}  & \multicolumn{8}{c}{\textbf{Data sources}}                                                                                  & \multicolumn{1}{c}{\multirow{6}{*}{\textbf{Approach}}}    & \multirow{6}{*}{\textbf{Paradigm}}    & \multirow{6}{*}{{\textbf{\begin{tabular}[c]{c}Evaluation\\ Metrics\end{tabular}}}} 
                               & \multirow{6}{*}{\rotatebox[origin=c]{90}{\textbf{Code}}} & \multirow{6}{*}{\rotatebox[origin=c]{90}{\textbf{Dataset}}}
                                \\ \cmidrule(lr){4-7} \cmidrule(lr){8-15}
                               &        &                                          & \multicolumn{1}{c}{\rotatebox[origin=c]{90}{\textbf{Technical}}} &  \multicolumn{1}{c}{\rotatebox[origin=c]{90}{\textbf{Application}}} &  \multicolumn{1}{c}{\rotatebox[origin=c]{90}{\textbf{Functional}}} &  \multicolumn{1}{c}{\rotatebox[origin=c]{90}{\textbf{Business}}} & \multicolumn{1}{c}{\rotatebox[origin=c]{90}{\textbf{Source Code}}} &  \multicolumn{1}{c}{\rotatebox[origin=c]{90}{\textbf{Topology}}} &  \multicolumn{1}{c}{\rotatebox[origin=c]{90}{\textbf{Event Logs}}} &  \multicolumn{1}{c}{\rotatebox[origin=c]{90}{\textbf{KPIs/Metrics}}} &  \multicolumn{1}{c}{\rotatebox[origin=c]{90}{\textbf{Traffic Network}}} &  \multicolumn{1}{c}{\rotatebox[origin=c]{90}{\textbf{Reports}}} &  \multicolumn{1}{c}{\rotatebox[origin=c]{90}{\textbf{Alerts}}} & \rotatebox[origin=c]{90}{\textbf{Traces}} & \multicolumn{1}{c}{}  & \multicolumn{1}{c}{} & \multicolumn{1}{c}{} & \multicolumn{1}{c}{}                                   &                                                                                        \\ \midrule
\rowcolor{gray!15}     \citep{yang2014towards}    &  2014    &  Software Bugs     &              &         &  $\bullet$    &     $\bullet$  &   &        &          &   &       &  $\bullet$   &      &       &  LDA        &  UNSUP      &  {F1, PREC, REC, MRR}        &       &  $\bullet$       \\

    \citep{lim2014identifying}    &  2014    &  N/A   &              &       $\bullet$  &      &      &   &        &          &  $\bullet$ &       &     &      &       &  HMRF        &  UNSUP      &  {PREC, REC}        &       &  $\bullet$       \\

\rowcolor{gray!15}    \citep{xia2016improving}    &  2016    &  Software Bugs     &              &         &  $\bullet$    &     $\bullet$  &   &     $\bullet$   &          &   &       &  $\bullet$   &      &       &  MTM        &  UNSUP      &  {ACC}        &       &  $\bullet$       \\  

    \citep{zeng2017knowledge}    &  2017   &  N/A    &    $\bullet$          &     $\bullet$    &  $\bullet$    &     $\bullet$  &   &      &          &   &       &  $\bullet$   &      &    $\bullet$    &  Hierarchical Classification   &  SUP      &  {F1, PREC, REC}        &       &     \\   

\rowcolor{gray!15}    \citep{zhao2020understanding}    &  2020   &  N/A    &    $\bullet$          &     $\bullet$    &     &    &   &   $\bullet$    &          &   &       &    &   $\bullet$    &       &  EVT, DBSCAN   &  UNSUP      &  {F1, PREC, REC}        &       &     

  \\ \bottomrule
  
\end{tabular}}

\end{table}
\begin{table}[]
\caption{Study of particularities of AIOps Incident Classification methods.}
\label{tab:particularities_classif}
\renewcommand{\arraystretch}{1.0}
\resizebox{\textwidth}{!}{%
\begin{tabular}{cccccc}
\toprule
\textbf{Ref.} & \textbf{Interpretability} & \textbf{Scalability} & \textbf{Robustness} & \textbf{Temporal Evaluation}  & \textbf{Human-in-the-loop}                                           \\ \midrule

\rowcolor{gray!15} \citep{yang2014towards}    & $\bullet$ & & $\bullet$ & & \\ 
\citep{lim2014identifying}    & $\bullet$ & $\bullet$ &  & $\bullet$ & \\
\rowcolor{gray!15} \citep{xia2016improving}    & $\bullet$ & $\bullet$ & $\bullet$ & $\bullet$ &  \\
\citep{zeng2017knowledge}    & $\bullet$ & $\bullet$ & &  &  $\bullet$ \\
\rowcolor{gray!15} \citep{zhao2020understanding}    & $\bullet$ & $\bullet$ &  $\bullet$ &  & 
\\ \bottomrule
\end{tabular}}
\end{table}

As discussed earlier, the primary objective of incident classification is to enhance the diagnosis of incidents, thereby centralizing the efforts of the maintenance team. However, this category has often been overlooked in the review process and is generally associated with deduplication, triage, or prioritization. Nevertheless, there exist few research works that align perfectly with this category, offering approaches to organize a large volume of incidents into representative sets of issues or topics. 

\citet{yang2014towards} employed LDA to extract topics from bug reports and identify relevant bug reports corresponding to each topic. Their methodology involves initially determining the topics of a new bug report and subsequently employing multiple features (such as component, product, priority, and severity) to identify reports with shared features as the new bug report. Another notable contribution, presented by~\citep{lim2014identifying}, proposes a Hidden Markov Random Field (HMRF) based approach for automatically identifying recurrent performance issues in large-scale software systems. Their approach formulates the problem as an HMRF-based clustering problem, which involves learning metric discretization thresholds and optimizing the clustering process. \citet{xia2016improving} introduced an innovative bug-triaging framework utilizing a specialized topic modeling algorithm known as a multi-feature topic model (MTM). MTM extends LDA by incorporating product and component information from bug reports, effectively mapping the term space to the topic space. Additionally, they introduced an incremental learning method named \texttt{TopicMiner}, which exploits the topic distribution of a new bug report to assign an appropriate fixer based on their affinity with the topics.

In another context, \citet{zeng2017knowledge} developed a methodology to identify the underlying categories of IT problems from ticket descriptions. To achieve this, they employed a hierarchical multi-label classification technique for categorizing monitoring tickets, introducing a novel contextual hierarchy. A new greedy algorithm, \texttt{GLabel}, is employed to address the optimization challenge. The approach also leverages domain expert knowledge alongside ticket instances to guide the hierarchical multi-label classification process. In addressing the issue of alert storms, \citet{zhao2020understanding} presented a two-stage approach: alert storm detection and alert storm summary. During the alert storm summary phase, an alert denoising method filters out irrelevant alerts by learning patterns from the system's normal states. Subsequently, alerts indicating service failures are clustered based on textual and topological similarities. From each cluster, the most representative alert is selected, thereby forming a concise set of alerts for further investigation.

\subsection{Incident Deduplication Methods}
\label{sec:sota_review_deduplication}

\begin{table}[]
\caption{Summary of reviewed AIOps Incident Deduplication methods.}
\label{tab:tab_deduplication}
\renewcommand{\arraystretch}{1.2}

\resizebox{\textwidth}{!}{%
\begin{tabular}{cc>{\centering}m{1.5cm}cccccccccccc>{\centering}m{2cm}>{\centering}m{1.5cm}>{\centering}m{2.5cm}cc}
\toprule
\multirow{6}{*}{\textbf{Ref.}} & \multirow{6}{*}{\textbf{Year}} & \multirow{6}{*}{\textbf{Focus}}   & \multicolumn{4}{c}{\textbf{Target Area}}  & \multicolumn{8}{c}{\textbf{Data sources}}                                                                                  & \multicolumn{1}{c}{\multirow{6}{*}{\textbf{Approach}}}    & \multirow{6}{*}{\textbf{Paradigm}}    & \multirow{6}{*}{{\textbf{\begin{tabular}[c]{c}Evaluation\\ Metrics\end{tabular}}}} 
                               & \multirow{6}{*}{\rotatebox[origin=c]{90}{\textbf{Code}}} & \multirow{6}{*}{\rotatebox[origin=c]{90}{\textbf{Dataset}}}
                                \\ \cmidrule(lr){4-7} \cmidrule(lr){8-15}
                               &        &                                          & \multicolumn{1}{c}{\rotatebox[origin=c]{90}{\textbf{Technical}}} &  \multicolumn{1}{c}{\rotatebox[origin=c]{90}{\textbf{Application}}} &  \multicolumn{1}{c}{\rotatebox[origin=c]{90}{\textbf{Functional}}} &  \multicolumn{1}{c}{\rotatebox[origin=c]{90}{\textbf{Business}}} & \multicolumn{1}{c}{\rotatebox[origin=c]{90}{\textbf{Source Code}}} &  \multicolumn{1}{c}{\rotatebox[origin=c]{90}{\textbf{Topology}}} &  \multicolumn{1}{c}{\rotatebox[origin=c]{90}{\textbf{Event Logs}}} &  \multicolumn{1}{c}{\rotatebox[origin=c]{90}{\textbf{KPIs/Metrics}}} &  \multicolumn{1}{c}{\rotatebox[origin=c]{90}{\textbf{Traffic Network}}} &  \multicolumn{1}{c}{\rotatebox[origin=c]{90}{\textbf{Reports}}} &  \multicolumn{1}{c}{\rotatebox[origin=c]{90}{\textbf{Alerts}}} & \rotatebox[origin=c]{90}{\textbf{Traces}} & \multicolumn{1}{c}{}  & \multicolumn{1}{c}{} & \multicolumn{1}{c}{} & \multicolumn{1}{c}{}                                   &                                                                                        \\ \midrule
   \rowcolor{gray!15} \citep{hiew2006assisted}    &  2006    &  Software Bugs     &              &         &  $\bullet$    &     $\bullet$  &   &        &          &   &       &  $\bullet$   &      &       &  K-Means        &  UNSUP      &  {F1, PREC, REC}        &       &  $\bullet$       \\

    \citep{runeson2007detection}    &  2007    &  N/A    &  $\bullet$      &     $\bullet$    &  $\bullet$    &     $\bullet$  &   &   $\bullet$      &          &   &       &  $\bullet$   &      &       &  Cosine Similarity        &  UNSUP      &  {RR@k}        &       &      \\

    \rowcolor{gray!15} \citep{sureka2010detecting}    &  2010    &  Software Bugs   &       &        &  $\bullet$    &     $\bullet$  &   &        &          &   &       &  $\bullet$   &      &       &  N-gram similarity        &  UNSUP      &  {RR@k}        &       &   $\bullet$   \\  

    \citep{sun2011towards}    &  2011    &  Software Bugs   &       &        &  $\bullet$    &     $\bullet$  &   &  $\bullet$      &          &   &       &  $\bullet$   &      &       &  BM25F Measure    &  SUP      &  {RR@k}      &       &   $\bullet$   \\  

    \rowcolor{gray!15} \citep{banerjee2012automated}    &  2012    &  Software Bugs   &       &        &  $\bullet$    &     $\bullet$  &   &     &          &   &    &  $\bullet$   &      &      &  LCS Measure    &  UNSUP      &  {RR@k}      &      &   $\bullet$   \\  

    \citep{zhou2012learning}    &  2012    &  Software Bugs   &       &        &  $\bullet$    &     $\bullet$  &   &  $\bullet$   &          &   &    &  $\bullet$   &      &      &  LTR    &  SemiSUP      &  {MRR, RR@k}      &      &   $\bullet$   \\  

   \rowcolor{gray!15}  \citep{xie2018detecting}    &  2018    &  Software Bugs   &       & $\bullet$   &  $\bullet$    &     $\bullet$  &   &     &          &   &    &  $\bullet$   &      &      &  CNN    &  SUP      &  {F1, ACC}      &      &   $\bullet$   \\ 

    \citep{he2020duplicate}    &  2020    &  Software Bugs   &       & $\bullet$   &  $\bullet$    &     $\bullet$  &   &     &          &   &    &  $\bullet$   &      &      &  BERT    &  SUP      &  {F1, ACC, AUC}      &      &   $\bullet$   \\ 

   \rowcolor{gray!15}  \citep{messaoud2022duplicate}    &  2022    &  Software Bugs   &       & $\bullet$   &  $\bullet$    &       &   &     &          &   &    &  $\bullet$   &      &      &  BERT, MLP    &  SUP      &  {F1, PREC, REC}      &      &   $\bullet$   

  \\ \bottomrule
  
\end{tabular}}

\end{table}
\begin{table}[]
\caption{Study of particularities of AIOps Incident Assignment methods.}
\label{tab:particularities_assignment}
\renewcommand{\arraystretch}{0.7}
\resizebox{\textwidth}{!}{%
\begin{tabular}{cccccccc}
\toprule
\textbf{Ref.} & \textbf{Interpretability} & \textbf{Scalability} & \textbf{Robustness} & \textbf{Temporal Evaluation}                                           \\ \midrule

\rowcolor{gray!15} \citep{hiew2006assisted, runeson2007detection, sun2011towards}    & $\bullet$ & $\bullet$ &  & $\bullet$   \\

\citep{sureka2010detecting}   & $\bullet$ & & $\bullet$ & $\bullet$  \\
\rowcolor{gray!15} \citep{banerjee2012automated, zhou2012learning}    & $\bullet$ & & & $\bullet$  

\\ \bottomrule
\end{tabular}}
\end{table}

Incident deduplication aims to identify the most similar incidents among a set of historical incidents, which exhibit slight differences but primarily address the same problem. This work can be categorized into two main categories. The first category of research is centered around techniques for detecting duplicate incident reports based on their descriptions and characteristics. 

In~\citep{hiew2006assisted}, an initial attempt was made to identify duplicate bug reports using textual information. Their approach involved building a model that clustered similar reports, represented as document vectors via TF-IDF, into representative centroids. When a new incident report is submitted, the method calculates cosine similarity to each centroid, looking for instances of high similarity to identify potential duplicates. \citet{runeson2007detection} adopted a similar approach, enhancing it by considering additional textual features such as software versions, testers, and submission dates. Effective preprocessing steps like stemming and stop word removal were also incorporated. \citet{sureka2010detecting} proposed a distinctive N-gram-based model, setting it apart from the previous word-based methods. This study explored the utility of low-level character features, offering benefits like resilience against noisy data and effective handling of domain-specific term variations. Incorporating a comprehensive similarity assessment, \citet{sun2011towards} introduced a retrieval function (REP). This function considered textual content similarity in summary and description fields as well as the similarity of non-textual attributes such as product, component, and version. The paper extended the frequently used BM25F similarity measure in information retrieval.

Recognizing the limitations of word-based approaches that may identify reports discussing different problems as duplicates due to shared common words, \citet{banerjee2012automated} introduced \texttt{FactorLCS} approach that employed common sequence matching when calculating the textual similarity between incident reports. \citet{zhou2012learning} introduced \texttt{BugSim}, leveraging a learning-to-rank concept by incorporating both textual and statistical features. A similarity function for bug reports was established based on these features. The model was trained using a set of duplicate and non-duplicate report pairs, adjusting feature weights with the stochastic gradient descent algorithm. Candidate duplicate reports for a new bug report were retrieved using this trained model. 

Recently, deep learning techniques have gained traction. \citet{xie2018detecting} introduced \texttt{DBR-CNN}, utilizing a CNN model and word embedding techniques to assess semantic similarities between pairs of incident reports. This approach departed from prior methods that primarily relied on common words or sequences of words for lexical similarity computation. Similarly, \citet{he2020duplicate} incorporated a CNN while introducing a unique bug report pair representation known as the dual-channel matrix. This matrix was formed by concatenating two single-channel matrices, each representing a bug report. These pairs were then fed into a CNN model named \texttt{DC-CNN}, designed to capture interconnected semantic relationships. In \citet{messaoud2022duplicate}, BERT-MLP was proposed. This approach utilized a pre-trained language model, BERT, to process unstructured bug report data and capture contextual relationships. The BERT model's output was then input to a multilayer perceptron classifier to predict duplicate bug reports.

The second category primarily relies on designing similarity metrics that can reflect the semantic crash similarity between execution reports, specifically for stack traces, which we discuss briefly.~\citet{DBLP:conf/csmr/LerchM13} employed the TF-IDF-based scoring function from Lucene library~\citep{lucene}.~\citet{DBLP:conf/qrs/SaborHL17} proposed DURFEX system which uses the package name of the subroutines and then segment the resulting stack traces into N-grams to compare them using the Cosine similarity. Some alternative techniques propose to compute the similarity using derivatives of the Needleman-Wunsch algorithm~\citep{needleman1970general}. In~\citep{DBLP:conf/icac/BrodieMLMMWCS05}, the authors suggested adjusting the similarity based on the frequency and the position of the matched subroutines.~\citet{DBLP:conf/icse/DangWZZN12} proposed a new similarity measure called PDM in their framework Rebucket to compute the similarity based on the offset distance between the matched frames
and the distance to the top frame. More recently, TraceSim~\citep{DBLP:conf/sigsoft/VasilievKCKLP20} has been proposed to take into consideration both the frame position and its global inverse frequency.~\citet{DBLP:conf/seke/MorooAH17} present an approach that combines TF-IDF coefficient with PDM. Finally, we outline some earlier approaches that used edit distance, as it is equivalent to optimal global alignment~\citep{DBLP:conf/osdi/BartzSPKGCL08,DBLP:conf/icde/ModaniGLSM07}.

\subsection{Root Cause Analysis Methods}
\label{sec:sota_review_rca}

\begin{table}[]
\caption{Summary of reviewed AIOps Root Cause Analysis methods.}
\label{tab:tab_classification}
\renewcommand{\arraystretch}{1.2}

\resizebox{\textwidth}{!}{%
\begin{tabular}{cc>{\centering}m{1.5cm}cccccccccccc>{\centering}m{2cm}>{\centering}m{1.5cm}>{\centering}m{2.5cm}cc}
\toprule
\multirow{6}{*}{\textbf{Ref.}} & \multirow{6}{*}{\textbf{Year}} & \multirow{6}{*}{\textbf{Focus}}   & \multicolumn{4}{c}{\textbf{Target Area}}  & \multicolumn{8}{c}{\textbf{Data sources}}                                                                                  & \multicolumn{1}{c}{\multirow{6}{*}{\textbf{Approach}}}    & \multirow{6}{*}{\textbf{Paradigm}}    & \multirow{6}{*}{{\textbf{\begin{tabular}[c]{c}Evaluation\\ Metrics\end{tabular}}}} 
                               & \multirow{6}{*}{\rotatebox[origin=c]{90}{\textbf{Code}}} & \multirow{6}{*}{\rotatebox[origin=c]{90}{\textbf{Dataset}}}
                                \\ \cmidrule(lr){4-7} \cmidrule(lr){8-15}
                               &        &                                          & \multicolumn{1}{c}{\rotatebox[origin=c]{90}{\textbf{Technical}}} &  \multicolumn{1}{c}{\rotatebox[origin=c]{90}{\textbf{Application}}} &  \multicolumn{1}{c}{\rotatebox[origin=c]{90}{\textbf{Functional}}} &  \multicolumn{1}{c}{\rotatebox[origin=c]{90}{\textbf{Business}}} & \multicolumn{1}{c}{\rotatebox[origin=c]{90}{\textbf{Source Code}}} &  \multicolumn{1}{c}{\rotatebox[origin=c]{90}{\textbf{Topology}}} &  \multicolumn{1}{c}{\rotatebox[origin=c]{90}{\textbf{Event Logs}}} &  \multicolumn{1}{c}{\rotatebox[origin=c]{90}{\textbf{KPIs/Metrics}}} &  \multicolumn{1}{c}{\rotatebox[origin=c]{90}{\textbf{Traffic Network}}} &  \multicolumn{1}{c}{\rotatebox[origin=c]{90}{\textbf{Reports}}} &  \multicolumn{1}{c}{\rotatebox[origin=c]{90}{\textbf{Alerts}}} & \rotatebox[origin=c]{90}{\textbf{Traces}} & \multicolumn{1}{c}{}  & \multicolumn{1}{c}{} & \multicolumn{1}{c}{} & \multicolumn{1}{c}{}                                   &                                                                                        \\ \midrule
   \rowcolor{gray!15}  \citep{bahl2007towards}    &  2007    &  Network Faults    &    $\bullet$      &      &     &    &    &   $\bullet$      &          &       $\bullet$    &  $\bullet$   &   &   &       &  Inference Graph     &  UNSUP      &  {Qualitative Eval.}        &       &       \\

    \citep{attariyan2012x}    &  2012    &  Software Faults    &          &  $\bullet$    &  $\bullet$   &    &  $\bullet$  &         &          &       $\bullet$    &   &   &   &       &  DIFT       &  UNSUP      &  {FPR}        &       &       \\

   \rowcolor{gray!15} \citep{nguyen2013fchain}    &  2013    &  Cloud     &  $\bullet$      &  $\bullet$    &   &    &    &   $\bullet$      &          &       $\bullet$    &   &   &   &       &  Makrov Model      &  UNSUP      &  {PRC}        &       &       \\

    \citep{chen2014causeinfer}    &  2014    &  Cloud    &     &  $\bullet$    &   &    &    &   $\bullet$      &          &       $\bullet$    &   &   &   &       &  Causality Graph      &  UNSUP      &  {PREC, REC}        &       &       \\

   \rowcolor{gray!15} \citep{lin2016idice}    &  2016    &  Cloud   &     &  $\bullet$    &   &    &    &   $\bullet$      &     &       $\bullet$    &   &   &   &       &  FPM     &  UNSUP      &  {F1, PREC, REC}        &       &       \\

    \citep{sun2018hotspot}    &  2018    &  Cloud   &     &  $\bullet$    &   &    &    &   $\bullet$      &     &       $\bullet$    &   &   &   &       &  MCTS      &  UNSUP      &  {F1, PREC, REC}        &       &       \\

   \rowcolor{gray!15} \citep{li2019generic}    &  2019    &  Cloud   &     &  $\bullet$    &   &    &    &   $\bullet$      &  $\bullet$    &       $\bullet$    &   &   &   &       &  FPM      &  UNSUP      &  {F1}   &    $\bullet$    &   $\bullet$     \\

    \citep{samir2019controller}    &  2019    &  Cluster   &     &  $\bullet$    &   &    &    &        &      &       $\bullet$    &   &   &   &       &  HHMM      &  UNSUP      &  {REC, FNR, FPR}   &     &   $\bullet$     \\

   \rowcolor{gray!15} \citep{jeyakumar2019explainit}    &  2019    &  Datacenters   &  $\bullet$    &    &   &    &    &        &      &       $\bullet$    &   &   &   &       &   Causality Graph      &  UNSUP      &  {Ranking ACC, Success rate}   &     &       \\

    \citep{ma2020diagnosing}    &  2020    &  Databases   &    & $\bullet$   &  $\bullet$ &    &    &        &      &       $\bullet$    &   &   &   &   $\bullet$    &   Bayesian Net      &  UNSUP      &  {F1, PREC, REC}    &     &       \\

    \rowcolor{gray!15} \citep{DBLP:conf/kbse/RemilBMCK21}    &  2021    &  Databases   &    & $\bullet$   &  $\bullet$ &    &    &     $\bullet$   &      &       $\bullet$    &   &   & $\bullet$  &   $\bullet$    &   Subgroup Discovery  &  SUP  &  {PREC, Confidence}    &  $\bullet$   &  $\bullet$     \\   

    \citep{li2022actionable}    &  2022    &  Cloud   &  $\bullet$  & $\bullet$   &   &    &    &     $\bullet$   &      &       $\bullet$    &   &   &   &      &   CNN-GRU  &  SUP  &  {MAR, RR@k}    &  $\bullet$   &  $\bullet$     \\   

    \rowcolor{gray!15} \citep{li2022causal}    &  2022    &  Cloud   &    & $\bullet$   &   &    &    &     $\bullet$   &      &       $\bullet$    &   &   &   &      &   Bayesian Net  &  UNSUP  &  {RR@k}    &  $\bullet$   &  $\bullet$     \\  \midrule

    \citep{renieres2003fault}    &  2003    &  SFL   &    &   &  $\bullet$ &  $\bullet$   &  $\bullet$   &        &      &           &   &   &   &      &   K-NN  &  SUP  &  {$\text{Score}_{\text{U}}$, $\text{Score}_{\text{NN}}$}   &    &      \\  
    
    \rowcolor{gray!15} \citep{wong2013dstar}    &  2013    &  SFL   &    &   &  $\bullet$ &  $\bullet$   &  $\bullet$   &        &      &           &   &   &   &      &   Suspiciousness Score  &  SUP  &  {EXAM, Wilcoxon Signed-Rank Test}   &  $\bullet$  &   $\bullet$   \\

    \citep{xuan2014learning}    &  2014    &  SFL   &    &   &  $\bullet$ &  $\bullet$   &  $\bullet$   &        &      &           &   &   &   &      &   RankBoost  &  SUP  &  {Wasted effort}   &   &  $\bullet$    \\

    \rowcolor{gray!15}\citep{liu2006statistical}    &  2006    &  SFL   &    &   &  $\bullet$ &  $\bullet$   &  $\bullet$   &        &      &           &   &   &   &      &   Statistical Debugging  &  SUP  &  {T-Score}   & 
    & $\bullet$    \\

    \citep{abreu2009spectrum}    &  2009    &  SFL   &    &   &  $\bullet$ &  $\bullet$   &  $\bullet$   &        &      &           &   &   &   &      &   Bayesian Net  &  SUP  &  {Wasted effort}   & $\bullet$  &   $\bullet$   \\

    \rowcolor{gray!15}\citep{denmat2005data}    &  2005    &  SFL   &    &   &  $\bullet$ &  $\bullet$   &  $\bullet$   &        &      &           &   &   &   &      &   Association Rules  &  SUP  &  {$-$}   &   &     \\

    \citep{cellier2008formal}    &  2008    &  SFL   &    &   &  $\bullet$ &  $\bullet$   &  $\bullet$   &        &      &           &   &   &   &      &   FCA  &  SUP  &  {$-$}   &   &     \\

    \rowcolor{gray!15}\citep{zhang2014software}    &  2014    &  SFL   &    &   &  $\bullet$ &  $\bullet$   &  $\bullet$   &        &      &           &   &   &   &      &   Markov Logic  &  SUP  &  {ACC}   &   &  $\bullet$    \\

    \citep{sohn2017fluccs}    &  2017    &  SFL   &    &   &  $\bullet$ &  $\bullet$   &  $\bullet$   &        &      &           &   &   &   &      &   Genetic Prog. SVM  &  SUP  &  {ACC, Waster Effort, PREC}   &   &     \\

    \rowcolor{gray!15}\citep{li2019deepfl}    &  2019    &  SFL   &    &   &  $\bullet$ &  $\bullet$   &  $\bullet$   &        &      &           &   &   &   &      &   RNN, MLP  &  SUP  &  {MAR, RR@k, MFR}   & $\bullet$  &  $\bullet$    \\

    \citep{li2021fault}    &  2021    &  SFL   &    &   &  $\bullet$ &  $\bullet$   &  $\bullet$   &        &      &           &   &   &   &      &   CNN-GRU  &  SUP  &  {MAR, RR@k, MFR}   & $\bullet$  &  $\bullet$    

  \\ \bottomrule
  
\end{tabular}}

\end{table}
\begin{table}[]
\caption{Study of particularities of AIOps Root Cause Analysis methods.}
\label{tab:particularities_rca}
\renewcommand{\arraystretch}{1.0}
\resizebox{\textwidth}{!}{%
\begin{tabular}{cccccccc}
\toprule
\textbf{Ref.} & \textbf{Interpretability} & \textbf{Scalability} & \textbf{Robustness} & \textbf{Temporal Evaluation}  & \textbf{Human-in-the-loop}                                           \\ \midrule

\rowcolor{gray!15} \citep{bahl2007towards,nguyen2013fchain,sun2018hotspot,li2019generic} & $\bullet$ & $\bullet$ & $\bullet$ &  & \\
\citep{attariyan2012x} & $\bullet$ & $\bullet$ & $\bullet$ & $\bullet$ &  \\
\rowcolor{gray!15} \citep{lin2016idice} & $\bullet$ & & $\bullet$ &  &  $\bullet$ \\
\citep{samir2019controller,DBLP:conf/kbse/RemilBMCK21} & $\bullet$ & $\bullet$ & & & $\bullet$ \\
\rowcolor{gray!15} \citep{ma2020diagnosing, li2022actionable,li2022causal,jeyakumar2019explainit, chen2014causeinfer} & $\bullet$ & $\bullet$ & $\bullet$  & & $\bullet$ 
  \\
\midrule
\citep{renieres2003fault,sohn2017fluccs,cellier2008formal,denmat2005data,wong2013dstar} & $\bullet$ &  \\
\rowcolor{gray!15} \citep{xuan2014learning,abreu2009spectrum,zhang2014software} & $\bullet$ & $\bullet$ & & &\\
\citep{liu2006statistical} & $\bullet$ & $\bullet$ & $\bullet$ & \\ 
\rowcolor{gray!15} \citep{li2019deepfl} & & $\bullet$ &  & & 
\\ \bottomrule
\end{tabular}}
\end{table}

Root cause analysis approaches aim to determine the underlying faults that give rise to software bugs, errors, anomalies, or hardware failures. More concretely, root cause analysis is a diagnostic task that needs to be performed when reporting incidents, either after or in parallel with the triage process. In complex systems, it is necessary to first isolate and restrict the analysis to the faulty component or functionality, a process known as Fault Localization in many research communities. Specifically, Fault Localization involves identifying a set of components (devices, hosts, software modules, etc.) that serve as the initial trigger for an error within the system. It is important to note that fault localization can operate at different layers, including the technical, application, functional, and business layers. To simplify, we can distinguish between two types of fault localization. The first is technical and applicative fault localization or troubleshooting, which spans network, hardware, and application layers (e.g., faulty components in cloud environments).  The second is software fault localization, which revolves around the functional and business layers, essentially addressing flaws that lead to bugs. In the context of software fault localization, the central focus lies in analyzing the source code, whether the software is deployed across numerous machines or a more limited set.

\smallbreak
\noindent \textbf{Technical and Application Troubleshooting.} The study conducted by~\citet{bahl2007towards} introduced the \texttt{Sherlock} approach, which focuses on localizing performance issues within enterprise networks. This is achieved by constructing probabilistic inference graphs based on packet exchange observations in the network infrastructure. These graphs consist of three node types: root cause nodes (representing internal IP entities), observation nodes (representing clients), and meta-nodes that capture dependencies between these types. Each node is associated with a categorical random variable indicating its state (up, troubled, down). The state is influenced by other nodes through probability dependencies. The learning process of the inference graph involves monitoring packet exchanges between nodes during regular network operations. Once established, the graph enables the utilization of data from observation nodes to derive sets of state-node assignment vectors, reflecting the anticipated network operational state. \texttt{X-Ray}~\citep{attariyan2012x} addresses the challenge of troubleshooting performance issues by providing insights into the reasons behind specific events during performance anomalies. The key contribution is a performance summarization technique, achieved by instrumenting binaries while applications execute. This technique attributes performance costs to each basic block and utilizes dynamic information flow tracking to estimate the likelihood that a block was executed due to each potential root cause. The overall cost of each potential root cause is then summarized by aggregating the per-block cost multiplied by the cause-specific likelihood across all basic blocks. This technique can also be differentially applied to explain performance differences between two similar activities. 

\texttt{FChain}~\citep{nguyen2013fchain} is a fault localization system designed for identifying faulty components in online cloud environments. It operates as a black-box solution, leveraging low-level system metrics to detect performance anomalies. Anomalies are sorted based on manifestation time, and a discrete Markov model is used to sequentially examine these components. Techniques such as analyzing interdependencies between components and studying the propagation trend are employed to filter out spurious correlations. Similarly, \texttt{CauseInfer}~\citep{chen2014causeinfer} is a black-box cause inference system that constructs a two-layered hierarchical causality graph to aid in identifying performance problem causes. The system employs statistical methods, including a novel Bayesian change point detection method, to infer potential causes along the causal paths present in the graph.

\citet{lin2016log} proposed \texttt{LogCluster}, where a knowledge base is used to expedite the retrieval of logs associated with recurring issues, thus streamlining the process of problem identification. The methodology treats logs as sequences consisting of discrete log events, which are then subjected to an initial transformation into a vector format through the application of Inverse Document Frequency (IDF), where these individual events are considered as terms. Following this, an agglomerative hierarchical clustering technique is employed to group these logs, resulting in the selection of representative instances for each cluster. During the system's operational phase, queries are directed towards this condensed set of representatives, significantly alleviating the effort required for log retrieval based on similarity. 

\citet{lin2016idice} proposed \texttt{iDice}, a pattern mining approach based on frequent pattern mining and change detection within time series data. The approach is applied to identify effective attribute combinations associated with emerging issues in topology data. Given a volume of customer issue reports over time, the goal is to identify an attribute combination that divides the multi-dimensional time series dataset into two partitions: one with a significant increase in issue volume and the other without such an increase. Similarly,~\citet{sun2018hotspot} introduced \texttt{HotSpot}, which employs Monte Carlo Tree Search (MCTS) to efficiently explore attribute combinations and measure their correlation with sudden changes in the Page View metric. The approach proposes a potential score based on the ripple effect, quantifying how anomalies propagate through different attribute combinations. A hierarchical pruning strategy is used to narrow down the search space, focusing on attribute combinations with the highest potential to be the root cause. Another akin approach, \texttt{Squeeze} by~\citet{li2019generic}, applies a pattern mining approach to structured logs for identifying attribute combinations responsible for abnormal behaviors. \texttt{Squeeze} utilizes a bottom-up, then top-down searching strategy, efficiently exploring the multi-dimensional search space by first identifying potentially relevant attributes and then pinpointing the root cause.

Another work conducted by~\citep{samir2019controller} employed Hierarchical Hidden Markov Models (HHMM) to associate resource anomalies with root causes in clustered resource environments. Markov models are constructed on different levels and trained with the Baum-Welch algorithm using response time sequences as observations. \citet{jeyakumar2019explainit} introduced \texttt{ExplainIt}, an approach for unsupervised root-cause analysis in complex systems like data centers, utilizing KPIs data. This system enables operators to articulate causal hypotheses systematically ranked to identify root causes behind significant events. Using a declarative language akin to SQL, \texttt{ExplainIt} facilitates generating hypotheses that probe the complex probabilistic graphical causal model in the system.

The \texttt{iSQUAD} framework, introduced by \citet{ma2020diagnosing}, addresses the detection and diagnosis of Intermittent Slow Queries (iSQs) in cloud databases. These queries, arising from external intermittent performance issues, can pose significant risks to users. Unlike typical slow queries, iSQs are less predictable and present more intricate diagnostic challenges. Employing a machine learning-based approach, \texttt{iSQUAD} leverages Anomaly Extraction, Dependency Cleansing, Type-Oriented Pattern Integration Clustering (TOPIC), and Bayesian Case Model components. In their work, \citet{DBLP:conf/kbse/RemilBMCK21} focused on SQL workload analysis for identifying schema issues and automatically pinpointing subsets of queries that share specific properties while targeting certain performance measures. These measures encompass slow execution times, concurrency issues, high I/O communications, and more. The approach involves parsing queries to extract key attributes (such as tables, fields, and predicates), and augmenting queries with pertinent information such as performance metrics, environmental features, and anomaly alerts. Utilizing a pattern mining approach dubbed Subgroup Discovery, the authors uncover a subset of queries exhibiting anomalies within patterns formed by conjunctions of conditions on these attributes. Furthermore, the integration of a visual tool allows iterative and interactive learning from the obtained outcomes.

The proposed methodology by \citet{li2022actionable} introduces \texttt{DejaVu}, an interpretable approach designed to localize recurring failures within online service systems. These recurring failures manifest as repeated instances of the same type across various locations. Engineers can identify indicative metrics for each failure type, aiding in recognizing underlying issues and guiding mitigation actions based on domain expertise. These sets of indicative metrics facilitate the identification of candidate failure units associated with recurring issues. To capture intricate system dependencies, a failure dependency graph (FDG) is established, linking failure units with interdependencies. When a failure occurs, the monitoring system triggers \texttt{DejaVu}, which is trained on historical failures. DejaVu then utilizes the latest FDG and metric data to recommend suspicious failure units from the candidate set on the FDG, streamlining the fault localization process.~\citet{li2022causal} proposed Causal Inference-based Root Cause Analysis (\texttt{CIRCA}), which frames the root cause analysis challenge as intervention recognition within a novel causal inference task. The core principle centers around a sufficient condition for monitoring variables to serve as root cause indicators. This involves a probability distribution change conditioned on parents within a Causal Bayesian Network (CBN). Specifically tailored to online service systems, CIRCA constructs a graph of monitoring metrics based on system architecture knowledge and a set of causal assumptions.

\smallbreak
\noindent \textbf{Software Fault Localization.} A conventional software fault localization strategy generally yields a collection of statements or source code blocks that could potentially be linked to bugs. Unlike the SDP strategy, this approach relies on failure patterns observed during production runs and unit tests, rather than predictions regarding the probability of a code component transitioning into a faulty state. Numerous approaches have been developed to tackle the problem of software fault localization. One prominent category is program spectrum-based techniques, which rely on the similarity between program execution profiles obtained from execution traces. These profiles represent both successful and faulty runs of programs. For example,~\citet{renieres2003fault} employs nearest neighbor search to compare a failed test with a similar successful test, using the Hamming distance as a measure of similarity. Another well-known technique, \texttt{Tarantula}~\citep{jones2002visualization}, utilizes coverage and execution results to compute the suspiciousness score of each statement. This score is based on the number of failed and successful test cases covering the statement, as well as the total number of successful and failed test cases. Subsequent research in this field has proposed refinements to the suspiciousness scoring also known as ranking metrics, such as \texttt{Ochiai}~\citep{abreu2006evaluation}, \texttt{Crosstab}~\citep{wong2011towards} and \texttt{DStar}~\citep{wong2013dstar}.~\citet{xuan2014learning} addressed the challenge of finding an optimal ranking metric for fault identification by combining multiple existing metrics. The approach dubbed MULTRIC employs a two-phase process involving learning and ranking: it learns from faulty and non-faulty code elements to build a ranking model and then applies this model to compute the suspiciousness score for program entities when new faults emerge.  
Statistical debugging approaches have also been explored.~\citet{liu2006statistical} introduce a statistical debugging method called \texttt{SOBER}, which analyzes predicate evaluations in failing and passing runs. By estimating the conditional probability of observing a failure given the observation of a specific predicate, the approach identifies predicates with higher probabilities, indicating their potential involvement in software bugs or their proximity to them.~\citet{abreu2009spectrum} later proposes a Bayesian reasoning approach known as \texttt{BARINEL}, which incorporates a probabilistic framework for estimating the health probability of components. This model, based on propositional logic, captures the interaction between successful and failed components. Furthermore, data mining techniques have shown promise in fault localization due to their ability to unveil hidden patterns in large data samples.

~\citet{denmat2005data} introduced an approach that reinterprets Tarantula as a data-mining challenge. In this method, association rules denoting the connection between an individual statement and a program failure are extracted using coverage information and test suite execution outcomes. The significance of these rules is assessed through two well-known classical data mining metrics, namely, \textit{confidence} and \textit{lift}. These values can be understood as indicators of the potential for a statement to be suspicious, hinting at the presence of bugs.~\citet{cellier2008formal} discussed a data mining approach that relies on a combination of association rules and Formal Concept Analysis (FCA) as a mean to assist in fault localization. This technique aims to identify rules that associate statement coverage with corresponding execution failures, measuring the frequency of each rule. A threshold is set to determine the minimum number of failed executions covered by a selected rule. The generated rules are then partially ranked using a rule lattice, and the ranking is examined to locate the fault. In~\citep{zhang2014software}, the authors proposed an approach that utilizes multi-relational data mining techniques for fault localization. Specifically, this technique utilizes Markov logic, combining first-order logic and Markov random fields with weighted satisfiability testing for efficient inference, alongside a voted perceptron algorithm for discriminative learning. When applied to fault localization, Markov logic integrates various information sources including statement coverage, static program structure details, and prior bug knowledge to enhance the accuracy of fault localization efforts.
 
Recently, there has been a notable shift towards employing machine and deep learning approaches to tackle the software fault localization problem. For instance,~\citet{sohn2017fluccs} extended spectrum-Based Fault Localization with code and change metrics, to enhance fault localization precision. They applied Genetic Programming (GP) and linear rank Support Vector Machines for learning to rank, utilizing both suspiciousness values and additional metrics.~\citet{li2019deepfl} introduced  \texttt{DeepFL}, a deep learning approach designed for learning-based fault localization. It addresses challenges posed by increasing feature dimensions in advanced fault localization techniques, by automatically identifying effective existing and latent features. \texttt{DeepFL} is demonstrated using suspiciousness-value-based, fault-proneness-based, and textual-similarity-based features collected from fault localization, defect prediction, and information retrieval domains. \texttt{DeepRL4FL}~\citep{li2021fault} is another deep learning approach that treats fault localization as an image pattern recognition problem. It locates buggy code at statement and method levels by introducing code coverage representation learning and data dependencies representation learning for program statements. These dynamic information types in a code coverage matrix are combined with static code representation learning of suspicious source code. Inspired by crime scene investigation, the approach simulates analyzing crime scenes (failed test cases and statements), related individuals (statements with dependencies), and usual suspects (similar buggy code). \texttt{DeepRL4FL} organizes test cases, marks error-exhibiting statements, and leverages data dependencies for comprehensive fault identification. A Convolutional Neural Network classifier is employed, utilizing fused vector representations from multiple sources to detect faulty statements/methods effectively. 

\subsection{Incident Correlation Methods}
\label{sec:sota_review_correlation}

\begin{table}[]
\caption{Summary of reviewed AIOps Incident Correlation methods.}
\label{tab:tab_correlation}
\renewcommand{\arraystretch}{1.2}

\resizebox{\textwidth}{!}{%
\begin{tabular}{cc>{\centering}m{1.5cm}cccccccccccc>{\centering}m{2cm}>{\centering}m{1.5cm}>{\centering}m{2.5cm}cc}
\toprule
\multirow{6}{*}{\textbf{Ref.}} & \multirow{6}{*}{\textbf{Year}} & \multirow{6}{*}{\textbf{Focus}}   & \multicolumn{4}{c}{\textbf{Target Area}}  & \multicolumn{8}{c}{\textbf{Data sources}}                                                                                  & \multicolumn{1}{c}{\multirow{6}{*}{\textbf{Approach}}}    & \multirow{6}{*}{\textbf{Paradigm}}    & \multirow{6}{*}{{\textbf{\begin{tabular}[c]{c}Evaluation\\ Metrics\end{tabular}}}} 
                               & \multirow{6}{*}{\rotatebox[origin=c]{90}{\textbf{Code}}} & \multirow{6}{*}{\rotatebox[origin=c]{90}{\textbf{Dataset}}}
                                \\ \cmidrule(lr){4-7} \cmidrule(lr){8-15}
                               &        &                                          & \multicolumn{1}{c}{\rotatebox[origin=c]{90}{\textbf{Technical}}} &  \multicolumn{1}{c}{\rotatebox[origin=c]{90}{\textbf{Application}}} &  \multicolumn{1}{c}{\rotatebox[origin=c]{90}{\textbf{Functional}}} &  \multicolumn{1}{c}{\rotatebox[origin=c]{90}{\textbf{Business}}} & \multicolumn{1}{c}{\rotatebox[origin=c]{90}{\textbf{Source Code}}} &  \multicolumn{1}{c}{\rotatebox[origin=c]{90}{\textbf{Topology}}} &  \multicolumn{1}{c}{\rotatebox[origin=c]{90}{\textbf{Event Logs}}} &  \multicolumn{1}{c}{\rotatebox[origin=c]{90}{\textbf{KPIs/Metrics}}} &  \multicolumn{1}{c}{\rotatebox[origin=c]{90}{\textbf{Traffic Network}}} &  \multicolumn{1}{c}{\rotatebox[origin=c]{90}{\textbf{Reports}}} &  \multicolumn{1}{c}{\rotatebox[origin=c]{90}{\textbf{Alerts}}} & \rotatebox[origin=c]{90}{\textbf{Traces}} & \multicolumn{1}{c}{}  & \multicolumn{1}{c}{} & \multicolumn{1}{c}{} & \multicolumn{1}{c}{}                                   &                                                                                        \\ \midrule
\rowcolor{gray!15}    \citep{luo2014correlating}    &  2014    &   N/A     &             &     $\bullet$     &   &     &   &        &          &   $\bullet$  &     &     &   $\bullet$   &       &  K-NN        &  UNSUP      &  {F1}        &       &       \\ 

    \citep{su2019coflux}    &  2019    &   N/A     &             &     $\bullet$     &   &     &   &        &          &   $\bullet$  &     &     &      &       &  Cross-Correlation      &  UNSUP      &  {F1, PREC, REC, PRC}        &       &       \\ 

\rowcolor{gray!15}    \citep{tan2013system}    &  2013    &   IDS    &     $\bullet$        &          &   &     &   &        &          &    &  $\bullet$    &     &      &       &  Mahalanobis Distance      &  UNSUP      &  {REC, FPR, ACC}        &       &   $\bullet$    \\ 

    \citep{bateni2013time}    &  2013    &   IDS    &     $\bullet$        &          &   &     &   &        &          &    &  $\bullet$    &     &      &       &  CART      &  UNSUP      &  {Completeness, Soundness, FCR}        &       &   $\bullet$    

  \\ \bottomrule
  
\end{tabular}}

\end{table}
\begin{table}[]
\caption{Study of particularities of AIOps Incident Correlation methods.}
\label{tab:particularities_correlation}
\renewcommand{\arraystretch}{1.0}
\scalebox{0.9}{%
\begin{tabular}{cccc}
\toprule
\textbf{Ref.} & \textbf{Interpretability} & \textbf{Scalability} & \textbf{Robustness}                                      \\ \midrule

\rowcolor{gray!15} \citep{luo2014correlating,bateni2013time}    &  $\bullet$ & $\bullet$ &   \\
\citep{su2019coflux, tan2013system}      & $\bullet$ & $\bullet$ & $\bullet$   \\
 \bottomrule
                               \end{tabular}}
\end{table}

Incident correlation research typically concentrates on analyzing correlations among alerting signals, occurring incidents, or the associations between alerting signals and incidents. Existing correlation algorithms primarily evaluate raw key performance indicators, or they transform these KPIs into events and then analyze their correlations.

~\citet{luo2014correlating} focused on correlating events with KPIs data. They formulated the correlation problem as a two-sample problem to assess the correlation between KPI time series and event sequences in online service systems. They employ the nearest neighbors method to evaluate the existence of the correlation and analyze temporal relationships and monotonic effects. \texttt{CoFlux}~\citep{su2019coflux} is an unsupervised approach for correlating KPIs in internet service operations management. It introduced the concept of KPI flux-correlation, which involves identifying interactions between KPIs through fluctuations, especially under anomalous situations. The study emphasized the challenge of accurately distinguishing fluctuations from normal variations in KPIs with various structural characteristics. The proposed approach automatically determines flux-correlation between two KPIs, including temporal order and direction of fluctuations, without manual algorithm selection or parameter tuning through robust feature engineering and cross-correlation.

\citet{tan2013system} proposed a multivariate correlation analysis system for attack detection by extracting geometric correlations between characteristics of network traffic. Their solution utilizes the Mahalanobis distance to measure the similarity between traffic records.~\citet{bateni2013time} presented an Enhanced Random Directed Time Window (ERDTW) alert selection policy based on sliding time windows analysis. ERDTW classifies time intervals into relevant (safe) and irrelevant (dangerous) based on attributes described in mathematical logic rules. For example, if a time interval contains numerous alerts with the same IP address, it is more likely to be flagged as dangerous.

\subsection{Incident Mitigation Methods}
\label{sec:sota_review_mitigation}

\begin{table}[]
\caption{Summary of reviewed AIOps Incident Mitigation methods.}
\label{tab:tab_mitigation}
\renewcommand{\arraystretch}{1.2}

\resizebox{\textwidth}{!}{%
\begin{tabular}{cc>{\centering}m{1.5cm}cccccccccccc>{\centering}m{2cm}>{\centering}m{1.5cm}>{\centering}m{2.5cm}cc}
\toprule
\multirow{6}{*}{\textbf{Ref.}} & \multirow{6}{*}{\textbf{Year}} & \multirow{6}{*}{\textbf{Focus}}   & \multicolumn{4}{c}{\textbf{Target Area}}  & \multicolumn{8}{c}{\textbf{Data sources}}                                                                                  & \multicolumn{1}{c}{\multirow{6}{*}{\textbf{Approach}}}    & \multirow{6}{*}{\textbf{Paradigm}}    & \multirow{6}{*}{{\textbf{\begin{tabular}[c]{c}Evaluation\\ Metrics\end{tabular}}}} 
                               & \multirow{6}{*}{\rotatebox[origin=c]{90}{\textbf{Code}}} & \multirow{6}{*}{\rotatebox[origin=c]{90}{\textbf{Dataset}}}
                                \\ \cmidrule(lr){4-7} \cmidrule(lr){8-15}
                               &        &                                          & \multicolumn{1}{c}{\rotatebox[origin=c]{90}{\textbf{Technical}}} &  \multicolumn{1}{c}{\rotatebox[origin=c]{90}{\textbf{Application}}} &  \multicolumn{1}{c}{\rotatebox[origin=c]{90}{\textbf{Functional}}} &  \multicolumn{1}{c}{\rotatebox[origin=c]{90}{\textbf{Business}}} & \multicolumn{1}{c}{\rotatebox[origin=c]{90}{\textbf{Source Code}}} &  \multicolumn{1}{c}{\rotatebox[origin=c]{90}{\textbf{Topology}}} &  \multicolumn{1}{c}{\rotatebox[origin=c]{90}{\textbf{Event Logs}}} &  \multicolumn{1}{c}{\rotatebox[origin=c]{90}{\textbf{KPIs/Metrics}}} &  \multicolumn{1}{c}{\rotatebox[origin=c]{90}{\textbf{Traffic Network}}} &  \multicolumn{1}{c}{\rotatebox[origin=c]{90}{\textbf{Reports}}} &  \multicolumn{1}{c}{\rotatebox[origin=c]{90}{\textbf{Alerts}}} & \rotatebox[origin=c]{90}{\textbf{Traces}} & \multicolumn{1}{c}{}  & \multicolumn{1}{c}{} & \multicolumn{1}{c}{} & \multicolumn{1}{c}{}                                   &                                                                                        \\ \midrule
\rowcolor{gray!15}    \citep{zhou2016resolution}    &  2016    &  -       &        $\bullet$       &    $\bullet$      &  $\bullet$    &     $\bullet$  &   &     $\bullet$    &          &   &       &  $\bullet$   &      &     &  K-NN, LDA    &  UNSUP      &  {ACC, MAP, avgSim}        &       &       \\ 

    \citep{wang2017constructing}    &  2017    &  -       &        $\bullet$       &    $\bullet$      &  $\bullet$    &     $\bullet$  &   &     $\bullet$    &          &   &       &  $\bullet$   &      &     &  Ontology Model    &  SUP      &  {F1, PREC, REC}        &       &       \\ 

\rowcolor{gray!15}    \citep{lin2018hardware}    &  2018    &  -       &        $\bullet$       &    $\bullet$      &  $\bullet$    &     $\bullet$  &   &     $\bullet$    &   $\bullet$       &   &       &  $\bullet$   &      &     &  FastText    &  UNSUP      &  {$-$}        &       &       \\ 

    \citep{ding2012healing}    &  2012    &  -       &             &    $\bullet$      &  $\bullet$    &      &   &        &   $\bullet$       &   &       &    &      &   $\bullet$   &  FCA    &  UNSUP      &  {REC, ROC}        &       &     

  \\ \bottomrule
  
\end{tabular}}

\end{table}
\begin{table}[]
\caption{Study of particularities of AIOps Incident Mitigation methods.}
\label{tab:particularities_mitigation}
\renewcommand{\arraystretch}{1.0}
\resizebox{\textwidth}{!}{%
\begin{tabular}{ccccccc}
\toprule
\textbf{Ref.} & \textbf{Year} & \textbf{Interpretability} & \textbf{Scalability} & \textbf{Robustness} & \textbf{Temporal Evaluation}  & \textbf{Human-in-the-loop}                                           \\ \midrule

\rowcolor{gray!15} \citep{zhou2016resolution}    &  2016 & $\bullet$ & & $\bullet$ & & \\
\citep{wang2017constructing}    &  2017 & $\bullet$ & $\bullet$ & $\bullet$ & $\bullet$ & $\bullet$ \\
\rowcolor{gray!15} \citep{ding2012healing}    &  2012 & $\bullet$ & & &$\bullet$ & 

\\ \bottomrule
                               \end{tabular}}
\end{table}

Through the triage and diagnosis steps of incident management, valuable knowledge is gained, including the identification of incident scope, retrieval of historical duplicates, and analysis of root causes. This knowledge enables the initiation of automatic repair actions known as mitigation or remediation actions. Incident mitigation has received less attention compared to reporting and diagnosis tasks, as it is often a consequence of the outcomes of those processes. Once the underlying problem is clarified through diagnosis, the recovery steps become readily identifiable and attainable without the need for complex models. However, our commitment extends to providing a list of research works that focus on resolution tasks, even when triage or diagnosis are involved.

In a study conducted by~\citep{zhou2016resolution}, similarity-based algorithms are proposed to suggest resolutions for recurring problems based on incident tickets. The approach retrieves k suggestions for ticket resolution using a k-NN approach. The similarity between tickets is evaluated using a combination of numerical, categorical, and textual data, with individual and aggregate similarity measures defined. The solution is further extended to address false-positive tickets in both historical and incoming data. This is achieved by classifying tickets using a binary classifier and weighing ticket importance based on the prediction outcome. The final solution recommendation considers both importance and similarity. The paper also explores ideas for improving feature extraction, such as topic discovery and metric learning.~\citet{wang2017constructing} propose a cognitive framework based on ontologies to construct domain-specific knowledge and suggest recovery actions for IT service management tickets. The approach involves analyzing free-form text in ticket summaries and resolution descriptions. Domain-specific phrases are extracted using language processing techniques, and an ontology model is developed to define keywords, classes, relations, and a hierarchy. This model is then utilized to recommend resolution actions by matching concept patterns extracted from incoming and historical tickets using similarity functions like the Jaccard distance.~\citet{lin2018hardware} employed natural language processing techniques to predict repair actions for hardware failures based on closed incident tickets. Through the analysis of raw text logs, up to five repair actions are recommended.~\citet{ding2012healing} proposed an automated mining-based approach for suggesting appropriate healing actions. The method involves generating signatures of an issue using transaction logs, retrieving historical issues based on these signatures, and suggesting a suitable healing action by adapting actions used for similar past issues.

\begin{table}[]
\caption{Publicly Available Datasets and Benchmarks for AIOps Methods.}
\label{tab:datasets}
\renewcommand{\arraystretch}{1.0}
\resizebox{\textwidth}{!}{
\begin{tabular}{lllll}
\toprule
\multicolumn{1}{l}{\textbf{Application Area}}  & \multicolumn{1}{l}{\textbf{Dataset/Bench}}                                    & \multicolumn{1}{l}{\textbf{Data Source}}                                     & \multicolumn{1}{l}{\textbf{Refs.}} & \multicolumn{1}{l}{\textbf{Link for data and description}} \\ \midrule
\rowcolor{gray!15}Intrusion Detection Systems          & ABILENE                                                              & Network Traffic                                                     &   \cite{lakhina2004diagnosing,lakhina2005mining}                        & \url{https://tinyurl.com/2p8d3ufb}                      \\
Intrusion Detection Systems      & SNDLIB                                                               & Network Traffic                                                     &         \cite{lakhina2005mining}       & \url{https://tinyurl.com/yc6dfdkd}                      \\
\rowcolor{gray!15}Intrusion Detection Systems           & SWAT/WADI                                                            & Network Metrics                                                     &     \cite{deng2021graph,audibert2020usad,li2021multivariate}                      & \url{https://tinyurl.com/5n87td8c}                      \\
Intrusion Detection Systems           & CIC-IDS2017                                                          & Network Traffic                                                     &         N/A                   & \url{https://tinyurl.com/bdezufhc}                      \\
\rowcolor{gray!15}Denial of Service Attacks              & CIC-DDoS2019                                                         & Network Traffic                                                     &             N/A              & \url{https://tinyurl.com/8ywj34sf}                      \\
Packet Injection Attacks              & NETRESEC                                                             & Network Traffic                                                     &         N/A                  & \url{https://tinyurl.com/cf3a42m9}                      \\ \midrule
\rowcolor{gray!15}Edge Streams                          & SNAP                                                                 & Network Traffic                                                     &       \cite{yu2018netwalk}                    & \url{https://tinyurl.com/2nv4h37s}                      \\
Edge Streams                          & DBLP                                                                 & Network Traffic                                                     &              \cite{yu2018netwalk,chang2021f}             & \url{https://tinyurl.com/34n576wp}                      \\
\rowcolor{gray!15}Edge Streams                          & DARPA                                                                & Network Traffic                                                     &    \cite{chang2021f,yoon2019fast}                       & \url{https://tinyurl.com/55ev547r}                      \\
Edge Streams                          & ENRON                                                                & Network Traffic                                                     &    \cite{chang2021f,yoon2019fast}                       & \url{https://tinyurl.com/3k96xfat }                     \\ \midrule
\rowcolor{gray!15}Traffic Anomaly Detection             & MAWI                                                                 & Network Metrics                                                     &    \cite{siffer2017anomaly}                       & \url{https://tinyurl.com/4n2uc2te}                      \\
Anomaly Detection                     & Yahoo                                                                & Univariate TS                                                       &      \cite{ren2019time}                     & \url{https://tinyurl.com/5awjuj85}                      \\ \midrule
\rowcolor{gray!15}Software Changes                      & Alibaba                                                              & Univariate TS                                                       &      \cite{zhao2021identifying}                     & \url{https://tinyurl.com/mrjddhvk}                      \\
Software Changes                      & AIOps2018                                                            & Univariate TS                                                       &      \cite{wang2022identifying}                     & \url{https://tinyurl.com/bdzzeuwz}                      \\ \midrule
\rowcolor{gray!15} Anomaly Detection                     & NASA-SMAP                                                            & Multivariate TS                                                     &      \cite{su2019robust,audibert2020usad}                     & \url{https://tinyurl.com/m8pnwvkf}                      \\
Anomaly Detection                     & SMD                                                                  & Multivariate TS                                                     &      \cite{su2019robust,audibert2020usad,li2021multivariate}                     & \url{https://tinyurl.com/yj5au5me}                      \\
\rowcolor{gray!15} Anomaly Detection                     & ASD                                                                  & Multivariate TS                                                     &      \cite{li2021multivariate}                     & \url{https://tinyurl.com/yj5au5me}                      \\
Anomaly Detection                     & NASA-MSL                                                             & Multivariate TS                                                     &      \cite{su2019robust}                     & \url{https://tinyurl.com/ypcafk99}                      \\
\rowcolor{gray!15} Anomaly Detection                     & Power Plant                                                          & Multivariate TS                                                     &      \cite{zhang2019deep}                     & \url{https://tinyurl.com/ynuz78s8}                      \\ \midrule
Log Anomaly Detection                 & OpenStack                                                            & Log Events                                                          &      \cite{du2017deeplog}                     & \url{https://tinyurl.com/yhhckue3}                      \\
\rowcolor{gray!15} Log Anomaly Detection                 & HFDS/ Hadoop                                                         & Log Events                                                          &      All log methods                    & \url{https://tinyurl.com/2j8ebupx}                      \\
Log Anomaly Detection                 & BGL                                                                  & Log Events                                                          &      \cite{guo2021logbert,xia2021loggan,meng2019loganomaly}                     & \url{https://tinyurl.com/2j8ebupx}                      \\
\rowcolor{gray!15} Log Anomaly Detection                 & Thunderbird                                                          & Log Events                                                          &      \cite{guo2021logbert}                     & \url{https://tinyurl.com/2j8ebupx}                      \\
Log Anomaly Detection                 & Spark/Apache/HPC                                                     & Log Events                                                          & N/A                       & \url{https://tinyurl.com/2j8ebupx}                      \\ \midrule
\rowcolor{gray!15} Software Defect Prediction            & PROMISE                                                              & Code Metrics                                                        &      \cite{menzies2006data,uddin2022software,li2017software,wang2016automatically}                     & \url{https://tinyurl.com/bd9zz34m}                      \\
Software Defect Prediction            & Eclipse                                                              & Code Metrics                                                        &      \cite{dejaeger2012toward,moser2008comparative}                     & \url{https://tinyurl.com/2p899sdd}                    \\
\rowcolor{gray!15} Software Defect Prediction            & NASA                                                                 & Code Metrics                                                        &      \cite{elish2008predicting,dejaeger2012toward}                     & \url{https://tinyurl.com/2a5869vz}                      \\
Software Defect Prediction            & Code4BENCH                                                           & \begin{tabular}[c]{@{}l@{}}Defect Code\\ Benchmark\end{tabular}     &      \cite{majd2019code4bench}                     & \url{https://tinyurl.com/3bk7fnu3}                      \\
\rowcolor{gray!15} Software Defect Prediction            & AEEM                                                                 & \begin{tabular}[c]{@{}l@{}}Code and \\ Process Metrics\end{tabular} &      \cite{nam2013transfer}                     & \url{https://tinyurl.com/zr6cmsb9}                      \\
Software Defect Prediction            & GHPR                                                                 & \begin{tabular}[c]{@{}l@{}}Java Code and\\ Metrics\end{tabular}     &      \cite{xu2020defect}                     & \url{https://tinyurl.com/3zpp4d7a}                      \\
\rowcolor{gray!15} Software Defect Prediction            & JIRA                                                                 & \begin{tabular}[c]{@{}l@{}}Open Source \\ Project\end{tabular}      & N/A                       & \url{https://tinyurl.com/zr6cmsb9}                      \\ \midrule
Disk Failures                         & SMART                                                                & Multivariate TS                                                     &      \cite{zhao2010predicting,wang2013online,xu2016health}                     & \url{https://tinyurl.com/2vexmjmu}                      \\
\rowcolor{gray!15} Disk Failures                         & Backblaze                                                            & Multivariate TS                                                     &      \cite{xiao2018disk}                     & \url{https://tinyurl.com/rfp4j8jw}                      \\
DRAM Failures                         & DRAM                                                                 & Multivariate TS                                                     &      \cite{sun2019system}                     & \url{https://tinyurl.com/3yyz7e67}                      \\ \midrule
\rowcolor{gray!15} Remaining Useful Lifetime             & C-MAPSS                                                              & Multivariate TS                                                     &      \cite{zheng2017long,wu2018remaining,ma2020deep}                     & \url{https://tinyurl.com/sbyu5584}                      \\
Remaining Useful Lifetime             & Milling                                                              & Multivariate TS                                                     &      \cite{zheng2017long}                     & \url{https://tinyurl.com/2p8zjevy}                      \\
\rowcolor{gray!15} Remaining Useful Lifetime             & PHM2008                                                              & Multivariate TS                                                     &      \cite{zheng2017long,li2020data}                     & \url{https://tinyurl.com/2r2955ss}                      \\
Remaining Useful Lifetime             & XJTU-SY                                                              & Multivariate TS                                                     &      \cite{li2020data}                     & \url{https://tinyurl.com/272nxd3v}                      \\
\rowcolor{gray!15} Remaining Useful Lifetime             & PRONOSTICA                                                           & Multivariate TS                                                     & N/A                       & \url{https://tinyurl.com/4c8exsk9}                      \\ \midrule
Software Prediction                   & Google Cluster                                                       & Traces and Metrics                                                  &      \cite{islam2017predicting}                     & \url{https://tinyurl.com/4abzuz8m}                      \\ \midrule
\rowcolor{gray!15} Bug Triage                            & Eclipse                                                              & Bug Reports                                                         &      \cite{tian2013drone,gupta2022improving,murphy2004automatic,xuan2010automatic,bhattacharya2010fine}                     & \url{https://tinyurl.com/352v7ddc}                      \\

\rowcolor{gray!15}                         &                                                             &                                                     &      \cite{alenezi2013efficient,wang2014fixercache,xi2019bug,yang2014towards}                     &                       \\

\rowcolor{gray!15}                         &                                                             &                                                     &      \cite{xia2016improving,hiew2006assisted,sureka2010detecting,sun2011towards}                     &                      \\
\rowcolor{gray!15}                         &                                                             &                                                     &      \cite{banerjee2012automated,zhou2012learning,he2020duplicate,messaoud2022duplicate}                     &                     \\                         

Bug Triage                            & Mozilla                                                              & Bug Reports                                                         &      \cite{chen2020incidental,bhattacharya2010fine,wang2014fixercache,xi2019bug,yang2014towards}                     & \url{https://tinyurl.com/352v7ddc}                       \\

                       &                                                               &                                                         &      \cite{xia2016improving,sun2011towards,messaoud2022duplicate, sureka2010detecting}                     &                     \\

\rowcolor{gray!15} Bug Triage                            & HFDS/ Hadoop                                                         & Bug Reports                                                         &      \cite{xie2018detecting}                     & \url{https://tinyurl.com/352v7ddc}                      \\
Bug Triage                            & Thunderbird                                                          & Bug Reports                                                         &      \cite{messaoud2022duplicate}                     & \url{https://tinyurl.com/352v7ddc}                      \\
\rowcolor{gray!15} Bug Triage                            & NetBeans                                                             & Stack Traces                                                        &      \cite{DBLP:conf/sigsoft/VasilievKCKLP20}                     & \url{https://tinyurl.com/ycxxyjfd}                      \\
Bug Triage                            & \begin{tabular}[c]{@{}l@{}}Cassandra/Mesos/\\ JDT/Spark\end{tabular} & Bug Reports                                                         & N/A                       & \url{https://tinyurl.com/352v7ddc}                      \\ \midrule

\rowcolor{gray!15} Database Issues     & Infologic                       & \begin{tabular}[c]{@{}l@{}}SQL Queries \\ Alerts, ASH \end{tabular} &  \cite{DBLP:conf/kbse/RemilBMCK21}                       & \url{https://tinyurl.com/3fyb2s76}                      \\ \midrule

Software Fault Localization           & Defect4JS                                                            & Bug Reports                                                         &      \cite{li2021fault}                     & \url{https://tinyurl.com/4uvcu24y}                      \\
\rowcolor{gray!15} Software Fault Localization           & MEGA                                                                 & Bug Reports                                                         &      \cite{li2019deepfl}                     & \url{https://tinyurl.com/mpas9yan}                      \\
Intrusion Detection Systems & KDDCup1999                                                           & Network Metrics                                                     &      \cite{tan2013system}                     & \url{https://tinyurl.com/mrxxfftu}                      \\ \bottomrule
\end{tabular}}
\end{table}
\section{Publicly Available Datasets and Benchmarks for AIOps Methods}
\label{sec:reviewDatasets}

In this section, we present a comprehensive overview of the key publicly available datasets and benchmarks relevant to the field of AIOps, specifically in the context of incident management procedures across various application areas. Table~\ref{tab:datasets} serves as a valuable resource, not only listing datasets used in prior research methodologies but also highlighting additional datasets that have remained untapped in previous studies, yet hold significance for the respective research application domains. We organize these datasets by incident task or application area, specifying their data sources, cross-referencing them with the methodologies they have been utilized in, and, importantly, providing direct links to access the datasets along with detailed descriptions. It's worth noting that datasets designed for incident triage can also serve incident prioritization purposes and sub-categories, including assignment, classification, and deduplication.

We acknowledge the paramount importance of this information, both for newcomers venturing into the domain of AIOps and for practitioners seeking to replicate existing research findings, make informed comparisons among similar techniques within the same research domain, and evaluate their own contributions. While some previous surveys have offered lists of datasets tailored to specific domains, our contribution, to the best of our knowledge, represents the first comprehensive compilation spanning multiple application areas.

\section{Conclusion and Open Challenges}
\label{sec:conclusion}
\begin{table}[]
\caption{Summary of all reviewed AIOps methods for Incident Management Procedure.}
\label{tab:tab_all}
\renewcommand{\arraystretch}{1.0}

\resizebox{\textwidth}{!}{%
\begin{tabular}{ccll}
\toprule
\textbf{Incident Task}                                                          & \textbf{Application Area}                                                                   & \multicolumn{2}{c}{\textbf{Type of AI Techniques}}                                                                                                                                                                                                                                                                                                                     \\ \midrule
\multicolumn{2}{c}{\begin{tabular}[c]{c}\textbf{Incident} \\ \textbf{Detection}\end{tabular}}                                                                                             & \begin{tabular}[c]{l}Auto-Encoders (Others)~\cite{yu2018netwalk,xu2018unsupervised,li2021multivariate}\\ Auto-Regressive Models~\cite{pena2017anomaly}\\ Clustering~\cite{karami2015fuzzy,yu2018netwalk,sharma2013cloudpd,li2018robust,he2018identifying}\\ CNNs~\cite{ren2019time,zhang2019deep}\\ Combinatorial Optimization~\cite{karami2015fuzzy}\\ Conventional Classifiers~\cite{liu2015opprentice}\\ Dimensionality Reduction~\cite{lakhina2004diagnosing,pascoal2012robust,xu2009detecting}\\ Fourier Transform~\cite{ren2019time} \end{tabular} & \begin{tabular}[c]{l} FSM~\cite{fu2009execution} \\GANs~\cite{audibert2020usad,xia2021loggan}\\ Graph Models~\cite{deng2021graph,lou2010mining}\\ Language Models~\cite{meng2019loganomaly,zhang2019robust,guo2021logbert}\\ Markov Models~\cite{bang2017anomaly,sharma2013cloudpd}\\ Nearest Neighbors Search~\cite{sharma2013cloudpd}\\ RNNs~\cite{zhao2021identifying,wang2022identifying,su2019robust,zhang2019deep,du2017deeplog,meng2019loganomaly,zhang2019robust,xia2021loggan}\\ Statistical Models~\cite{lakhina2005mining,pena2017anomaly,xie2016distributed,yoon2019fast} \\ Continued ...~\cite{chang2021f,vallis2014novel,siffer2017anomaly,xu2018unsupervised}\end{tabular} \\ \midrule

\multirow{7}{*}{\begin{tabular}[c]{c}\textbf{Incident} \\ \textbf{Prediction}\end{tabular}} & SDP                                                                                         
& \begin{tabular}[c]{l}Bayesian Models~\cite{menzies2006data,dejaeger2012toward,moser2008comparative,wang2016automatically}\\ CNNs~\cite{li2017software}\\ Conventional Classifiers~\cite{elish2008predicting,nam2013transfer}\\ Convention Regressors~\cite{nagappan2006mining,ostrand2005predicting}\\ Dimensionality Reduction~\cite{nagappan2006mining}\end{tabular}                                                              & \begin{tabular}[c]{l}Graph Models~\cite{xu2020defect}\\ Language Models~\cite{uddin2022software}\\ RNNs~\cite{majd2020sldeep,uddin2022software}\\ Topic Modeling~\cite{xu2020defect}\end{tabular}   \\ \cmidrule(l){2-4}

                                                            & \begin{tabular}[c]{c}Hardware Failures \\ Aging/Rejuv.\\ RUL\end{tabular} & \begin{tabular}[c]{l}Clustering~\cite{vaidyanathan1999measurement}\\ CNNs~\cite{sun2019system,khalil2020machine,ma2020deep}\\ Conventional Classifiers~\cite{xiao2018disk,davis2017failuresim}\\ Conventional Regressors~\cite{garg1998methodology,vaidyanathan1999measurement,alonso2010adaptive,sudhakar2014software,araujo2014software}\\ Dimensionality Reduction~\cite{khalil2020machine}\\ Fourier Transform~\cite{khalil2020machine}\end{tabular}                                             & \begin{tabular}[c]{l}GANs~\cite{li2020data}\\ Markov Models~\cite{vaidyanathan1999measurement,zhao2010predicting,zhang2017syslog}\\ Pattern Mining~\cite{zhang2017syslog}\\ RL~\cite{bellani2019towards}\\ RNNs~\cite{xu2016health,davis2017failuresim,zheng2017long,wu2018remaining,ma2020deep}\\ Statistical Models~\cite{garg1998methodology,wang2013online}\end{tabular}                                             \\ \cmidrule(l){2-4}

                                                & Software Failures   
                                                & \begin{tabular}[c]{l}Auto-Regressive Models~\cite{chalermarrewong2012failure}\\ Bayesian Models~\cite{cohen2004correlating,pitakrat2018hora}\end{tabular}                                             & \begin{tabular}[c]{l}Conventional Classifiers~\cite{fronza2013failure,lin2018predicting,zhao2020real}\\ RNNs~\cite{zhang2016automated,islam2017predicting,lin2018predicting}\end{tabular}                                                                                    \\ \midrule

\multicolumn{2}{c}{\begin{tabular}[c]{c}\textbf{Incident} \\ \textbf{Prioritization}\end{tabular}}     & \begin{tabular}[c]{l}Bayesian Models~\cite{lin2018collaborative,hassan2019nodoze}\\ CNNs~\cite{chen2020incidental} \\Ranking Models~\cite{zhao2020automatically} \end{tabular}    &    \begin{tabular}[c]{l}Combinatorial Optimization~\cite{gupta2022improving}\\ Conventional Regressors~\cite{tian2013drone}\end{tabular}                                                               \\ \midrule

\multicolumn{2}{c}{\begin{tabular}[c]{c}\textbf{Incident} \\ \textbf{Assignment}\end{tabular}}   & \begin{tabular}[c]{l}Bayesian Models~\cite{murphy2004automatic,xuan2010automatic,bhattacharya2010fine,alenezi2013efficient}\\ CNNs~\cite{DBLP:conf/sigsoft/LeeHLKJ17,chen2019continuous,DBLP:conf/kdd/PhamJDOJ20}\\ Markov Models~\cite{shao2008efficient} \end{tabular}                                                                                                           & \begin{tabular}[c]{l}Pattern Mining~\cite{DBLP:conf/dsaa/RemilBPRK21} \\ RNNs~\cite{xi2019bug,chen2019continuous,DBLP:conf/dsaa/RemilBPRK21}\\ Similarity-based~\cite{wang2014fixercache}\end{tabular}                                                                            \\ \midrule

\multicolumn{2}{c}{\begin{tabular}[c]{c}\textbf{Incident} \\ \textbf{Classification}\end{tabular}}        & \begin{tabular}[c]{l}Clustering~\cite{zhao2020understanding}\\ Conventional Classification~\cite{zeng2017knowledge}\end{tabular}                                                                                                                          & \begin{tabular}[c]{l}Markov Models~\cite{lim2014identifying}\\ Topic Modeling~\cite{yang2014towards,xia2016improving}\end{tabular}                                                                                     \\ \midrule

\multicolumn{2}{c}{\begin{tabular}[c]{c}\textbf{Incident} \\ \textbf{Deduplication}\end{tabular}}  & \begin{tabular}[c]{l}Clustering~\cite{hiew2006assisted}\\ CNNs~\cite{xie2018detecting}\\ Language Models~\cite{he2020duplicate,messaoud2022duplicate}\end{tabular}                                                                                                                               & \begin{tabular}[c]{l}Ranking Models~\cite{zhou2012learning}\\ Similarity-based~\cite{runeson2007detection,sureka2010detecting,sun2011towards,banerjee2012automated} \\ Other Similarity-based~\cite{DBLP:conf/csmr/LerchM13,DBLP:conf/qrs/SaborHL17,DBLP:conf/icac/BrodieMLMMWCS05,DBLP:conf/icse/DangWZZN12} \\ Continued ...~\cite{DBLP:conf/sigsoft/VasilievKCKLP20,DBLP:conf/seke/MorooAH17,DBLP:conf/osdi/BartzSPKGCL08,DBLP:conf/icde/ModaniGLSM07}\end{tabular}                                                                                  \\ \midrule

\multirow{5}{*}{\begin{tabular}[c]{c}\textbf{Root Cause} \\ \textbf{Analysis}\end{tabular}} & \begin{tabular}[c]{c}Technical \\ Troubleshooting\end{tabular}                        & \begin{tabular}[c]{l}Bayesian Models~\cite{ma2020diagnosing,li2022causal}\\ CNNs~\cite{li2022actionable}\\ Graph Models~\cite{bahl2007towards,attariyan2012x,chen2014causeinfer,jeyakumar2019explainit}\end{tabular}                                                                                                                             & \begin{tabular}[c]{l}Markov Models~\cite{nguyen2013fchain,samir2019controller}\\ Pattern Mining~\cite{lin2016idice,sun2018hotspot,li2019generic,DBLP:conf/kbse/RemilBMCK21}\\ RNNs~\cite{li2022actionable}\end{tabular}                                                                              \\ \cmidrule(l){2-4}

                                                                        & SFL                                                                                         & \begin{tabular}[c]{l} Bayesian Models~\cite{abreu2009spectrum} \\ CNNs~\cite{li2021fault} \\Combinatorial Optimization~\cite{sohn2017fluccs}\\ Conventional Classifiers~\cite{sohn2017fluccs,li2019deepfl}\\ Markov Models~\cite{zhang2014software}\end{tabular}        & \begin{tabular}[c]{l}Nearest Neighbors Search~\cite{renieres2003fault} \\ Pattern Mining~\cite{denmat2005data,cellier2008formal}\\ Ranking Models~\cite{xuan2014learning}\\ RNNs~\cite{li2019deepfl,li2021fault}\\ Statistical Models~\cite{wong2013dstar,liu2006statistical}\end{tabular}                                                        \\ \midrule
\multicolumn{2}{c}{\begin{tabular}[c]{c}\textbf{Incident} \\ \textbf{Correlation}\end{tabular}}     & \begin{tabular}[c]{l}Conventional Classifiers~\cite{bateni2013time}\\ Nearest Neighbors Search~\cite{luo2014correlating}\end{tabular}                                                                                                                      & \begin{tabular}[c]{l} Similarity-based~\cite{su2019coflux,tan2013system} \end{tabular}                                                                                                                                           \\ \midrule
                                                                    
\multicolumn{2}{c}{\begin{tabular}[c]{c}\textbf{Incident} \\ \textbf{Mitigation}\end{tabular}}     & \begin{tabular}[c]{l}Language Models~\cite{lin2018hardware}\\ Nearest Neighbors Search~\cite{zhou2016resolution}\end{tabular}                                                                                                                        & \begin{tabular}[c]{l}Pattern Mining~\cite{wang2017constructing,ding2012healing}\\ Topic Modeling~\cite{zhou2016resolution}\end{tabular}                                                                                    \\ \bottomrule
\end{tabular}}
\end{table}

As stated in the introduction, we have observed a notable surge of interest in the AIOps domain in recent years, both within the research and industry sectors. However, it is imperative to acknowledge that this field still lacks centralization and a structured framework. This arises from the need to combine diverse specialized disciplines, encompassing software engineering, machine learning, big data analysis, optimization, and more. AIOps, characterized by its novelty and inherently interdisciplinary nature, has yet to establish a distinct and cohesive identity as a clearly defined area of study. This presents substantial challenges for both practitioners and researchers to fully comprehend the current state of the art and identify potential limitations and gaps. The absence of standardized terminology and conventions for aspects such as data management, problem targeting, focus areas, implementation details, requirements, and capabilities further exacerbates the situation. This absence not only result in the absence of technical guidelines and a coherent roadmap for implementing effective AIOps solutions but also makes it difficult for new researchers entering the field to discover and compare contributions from various disciplines addressing the same problems.

As a result, this survey serves as an introductory resource to the AIOps domain, specifically tailored for incident management procedures. Our primary goal was to establish a foundational knowledge base for AIOps, which can benefit both industries looking to transition to AIOps infrastructures and future research endeavors. In addition to presenting the fundamentals, including the essential building blocks required for a systematic AIOps approach in intelligent and effective incident management, while addressing pain points and desired outcomes, we offer a comprehensive exploration of existing approaches designed to handle various tasks within the defined incident management process.


\begin{figure}
	\centering
	\includegraphics[width=0.9\linewidth]{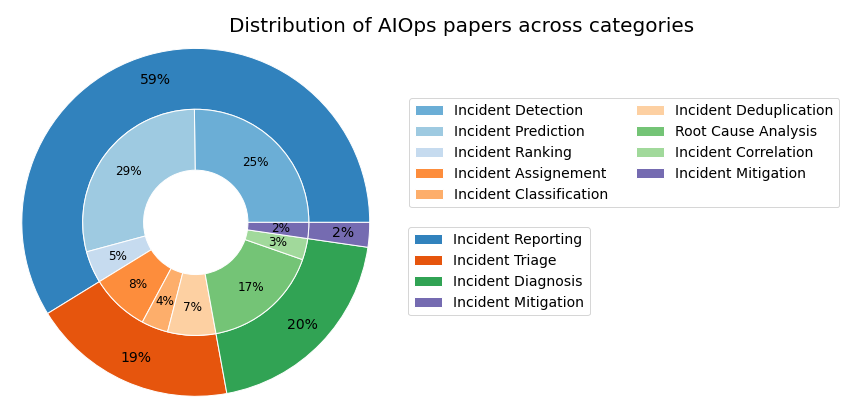}
    \captionsetup{font=footnotesize}
	\caption{Distribution of analyzed AIOps papers across incident tasks and their subcategories.}%
	\label{fig:distribution_papers}
\end{figure}

In Figure~\ref{fig:distribution_papers}, we analyze the distribution of AIOps papers reviewed in this survey across incident tasks and their subcategories. It is evident that certain research categories have garnered more attention than others, resulting in an imbalance in contributions across each phase. This disparity can be attributed to several factors. For instance, the incident prediction and detection phases collectively constitute more than half of the papers analyzed in this survey, sourced from prestigious conferences and journals. Conversely, categories such as incident classification, correlation, and mitigation have received relatively less attention. This discrepancy aligns logically with the fundamental requirements of an AIOps framework, where detection, prediction, and root cause analysis are of paramount importance. Furthermore, as discussed earlier, some categories may not be as critical to pursue. For example, if the root cause can be efficiently identified by referencing a past duplicate incident, there may be no need for an automated mitigation technique. Similarly, some categories may become less essential based on specific use cases. For instance, if an incident is already routed to a qualified individual with no competing priorities, the ranking process may be of reduced significance. However, it remains beneficial to encourage diverse contributions employing various techniques and the latest algorithms to ensure that tasks, when required, can be executed accurately and swiftly.

In relation to each incident task, we observed specific aspects that have garnered considerable research focus. This includes identifying prevalent contexts within the related research, extensively utilized data for those contexts, as well as the most commonly employed model types and learning paradigms, among other factors. For example, when examining incident detection methods, a predominant focus emerges on addressing technical aspects, including hardware and network layers, as well as application layers. This is typically accomplished by analyzing Key Performance Indicators (KPIs), system metrics, network traffic data, and event logs. The prevailing models tend to be unsupervised, with significant attention placed on statistical models, dimensionality reduction approaches, and general auto-encoders like Recurrent Neural Networks (RNNs). In incident prediction tasks, research efforts have also addressed failures occurring in the functional strata. Many of these methods are supervised and leverage patterns from past failures or outages. They frequently make use of forecasting methods, especially for time series data, particularly in contexts like software aging and estimating remaining useful lifetimes.

When it comes to incident prioritization, topology and alerting signals data play a prominent role, often using supervised techniques. In contrast, incident reports serve as the primary source of information for tasks related to assignment and deduplication. However, for these latter two tasks, most methods predominantly focus on addressing software-related issues, particularly software bugs. In terms of root cause analysis methods, interpretability is a key consideration. For software fault localization, the approach predominantly involves analyzing source code through supervised methods. On the other hand, technical fault localization methods tend to be unsupervised and draw from descriptive models in pattern mining, often incorporating topology data to contextualize faults.

In addition to the significant challenge addressed in this survey, which involves organizing the body of knowledge within the AIOps research area, we have identified several other concerns, pain points, and open challenges that require attention. These include aspects related to the design, unique characteristics, availability, and reproducibility of AIOps models.  

To begin, our initial observation highlights that despite the myriad challenges faced by predictive models, including issues like the absence of clear ground truth labels, the need for manual efforts to obtain high-quality data, extremely imbalanced datasets, and the complexity of dependencies and relationships among components and services, AIOps continues to place a significant emphasis on the development of predictive models, particularly for incident detection and prediction. However, there exists a lesser-known yet highly valuable approach that has been underutilized. This approach involves employing descriptive models, such as pattern mining in general, and specifically, techniques like supervised rule discovery~\cite{DBLP:journals/widm/Atzmueller15,DBLP:conf/pkdd/Wrobel97} and formal concept analysis~\cite{DBLP:conf/ijcai/KaytoueKN11,cellier2008formal}. These models harness the power of data mining to extract informative patterns from data, which can be instrumental in detecting, diagnosing, and resolving issues. Descriptive models offer distinct advantages when it comes to tackling challenges related to data diversity, complexity, and quality. This makes them particularly valuable in scenarios involving tasks like deduplication of incidents and dealing with intricate dependencies. Therefore, it is imperative to shift our focus towards enhancing descriptive models in conjunction with predictive models, recognizing the unique strengths they bring to the AIOps landscape.

Additionally, it is crucial to emphasize the importance of evolving the evaluation methods for these models. While the majority of research examined in this survey has centered on contingency table metrics, only a few have considered the contextual and temporal aspects outlined in Sections~\ref{subsec:eval_metrics} and~\ref{subsec:desiderata}. For instance, the challenge also is to ensure that the behavior of the model during the training phase is consistent with its performance in the testing and production phases. Traditional metrics used to assess models are susceptible to the contamination zone phenomenon~\cite{fourure2021anomaly}, which may lead to erroneous assessments. Indeed,~\citet{fourure2021anomaly}, highlight that by parameterizing the proportion of data between training and testing sets, the F1-score of anomaly detection models can be artificially increased.

When it comes to addressing interpretability concerns, many research efforts that employ black box models to carry out their tasks acknowledge the importance of incorporating an interpretation layer into their methods. They claim that the output of their models is accompanied by explanations regarding the reasons behind their predictions. However, several critical questions persist in this regard. It remains unclear how these interpretations remain internally consistent when the model encounters new data, or how they compare externally when different models produce the same results but with differing interpretations. Additionally, the stability of model interpretations over time, particularly with updates and improvements, raises important questions (see Section~\ref{subsec:desiderata} for further details). Furthermore, only a limited number of methods involve human expertise in the process. In reality, practitioners' insights can significantly guide both the learning and implementation of these models. Interestingly, in the context of pattern mining approaches, several techniques, such as Subjective Interestingness~\cite{DBLP:journals/datamine/Bie11}, can facilitate the process of updating a user's knowledge about the most interesting data over time. Scalability is another major concern. In AIOps environments, it is essential not only to focus on the efficiency of the model but also on its overall performance. While optimizing (TTx) times (including detection, engagement, and mitigation), which have received comparatively less attention, is crucial for the successful implementation of automated incident management procedures, performance evaluation is often overlooked when comparing different models addressing the same research field. In practical scenarios, a model that takes less time to execute while maintaining a 90\% F-Score for detection may be preferred over a model with a 95\% F-Score that requires a longer execution time. This emphasizes the need to strike a balance between model performance and efficiency in real-world AIOps applications.

Finally, to advance this field significantly, it is imperative to advocate for a closer partnership between academia and industry. Furthermore, open-sourcing initiatives should be encouraged. While it is totally understandable that publishing datasets may raise potential rights or confidentiality concerns, and sharing models employed in production scenarios could be susceptible to reverse-engineering, there is a critical need to find common ground, especially in the research community. Facilitating the reproducibility of certain results can be a significant catalyst for advancing research in this field. This collaborative approach, where academia and industry work together and contribute to open-source initiatives, can pave the way for the development of more robust and impactful AIOps solutions.


\newpage
\bibliographystyle{ACM-Reference-Format}
\bibliography{biblio}

\newpage

\appendix

\section{List of Abbreviations}
\begin{table}[h]
  \centering
  \scalebox{0.85}{
  \begin{tabularx}{\textwidth}{lL}
    \toprule
    \textbf{Abbreviation} & \textbf{Meaning} \\
    \midrule
    ACID & Atomicity, Consistency, Isolation, Durability \\
    ACODS & Ant Colony Optimization for Digital Signature \\
    AE & Auto Encoder \\
    AIOPS & Artificial Intelligence for IT Operations \\
    ARIMA & Auto Regressive Integrated Moving Average \\
    ASH & Active Session History \\
    AST & Abstract Syntax Tree \\
    AUC & Area Under the Curve \\
    BERT & Bidirectional Encoder Representations from Transformers \\
    BFOA & Bacterial Foraging Optimization Algorithm \\
    CAR & Collaborative Alert Ranking \\
    CART & Classification and Regression Trees \\
    CBN & Causal Bayesian Network \\
    CBO & Coupling Between Objects \\
    CNN & Convolutional Neural Network \\
    CPM & Correlational Paraconsistent Machine \\
    ConvLSTM & Convolutional Long Short-Term Memory \\
    DBMS & Database Management System \\
    DBN & Deep Belief Network \\
    DBSCAN & Density-Based Spatial Clustering of Applications with Noise \\
    DIFT & Dynamic Information Flow Tracking \\
    DIT & Depth of Inheritance Tree \\
    DRAM & Dynamic Random Access Memory \\
    DSL & Domain-Specific Language \\
    DSNSF & Digital Signatures of Network Segment using Flow Analysis \\
    DTW & Dynamic Time Warping \\
    DoS & Denial of Service \\
    EM & Expectation-Maximization \\
    ETL & Extract, Transform, Load \\
    FCA & Formal Concept Analysis \\
    FF & Frequency-Factorization \\
    FFT & Fast Fourier Transform \\
    FPM & Frequent Pattern Mining \\
    FSA & Finite State Automaton \\
    FSM & Finite State Machine \\
    FT-Tree & Frequent Template Tree \\
    GLM & Generalized Linear Model \\
    GNN & Graph Neural Network \\
    GP & Genetic Programming \\
    GRU & Gated Recurrent Unit \\
    HDFS & Hadoop Distributed File System \\
    HMM & Hidden Markov Model \\
    HMRF & Hidden Markov Random Field \\
    HsMM & Hierarchical state-based Markov Model \\
    IDC & Internet Data Center \\
    IDS & Intrusion Detection System \\
    IPS & Intrusion Prevention System \\
    ITOA & Information Technology Operations Analytics \\
    \bottomrule
  \end{tabularx}}
\end{table}

\begin{table}
  \centering
  \scalebox{0.85}{
  \begin{tabularx}{\textwidth}{lL}
    \toprule
    \textbf{Abbreviation} & \textbf{Meaning} \\
    \midrule
    K-NN & k-Nearest Neighbors \\
    KPI & Key Performance Indicator \\
    LCOM & Lack of Cohesion in Methods \\
    LDA & Latent Dirichlet Allocation \\
    LIME & Local Interpretable Model-Agnostic Explanations \\
    LOC & Lines of Code \\
    LSTM & Long Short-Term Memory \\
    MLP & Multi-Layer Perceptron \\
    MRMR & Minimum Redundancy Maximum Relevance \\
    MTM & Multi-Feature Topic Model \\
    MTTx & Mean Time To X \\
    NOM & Number of Methods \\
    OLAP & Online Analytical Processing \\
    PCA & Principal Component Analysis \\
    PDF & Probability Density Functions \\
    PL & Paraconsistent Logic \\
    PRC & Precision-Recall Curve \\
    PSO & Particle Swarm Optimization \\
    Qos & Quality of Service \\
    RCA & Root Cause Analysis \\
    RFC & Response for a Class \\
    RNN & Recurrent Neural Network \\
    ROC & Receiver Operating Characteristic \\
    RUL & Remaining Useful Life \\
    SDP & Software Defect Prediction \\
    SLO & Service Level Objective \\
    SMART & Self-Monitoring, Analysis and Reporting Technology \\
    SMM & Semi Markov Model \\
    SR & Spectral Residual \\
    TAN & Tree Augmented Naive Bayes \\
    TCNN & Temporal Convolutional Neural Network \\
    TF-IDF & Term Frequency-Inverse Document Frequency \\
    TL & Transfer Learning \\
    TOPSIS & Technique for Order of Preference by Similarity to Ideal Solution \\
    WMC & Weighted Methods per Class \\
    WSN & Wireless Sensor Network \\
    XAI & Explainable Artificial Intelligence \\
    \bottomrule
  \end{tabularx}}
\end{table}

\end{document}